\documentclass[twocolumn,10pt,a4paper]{article}
\usepackage{palatino}
\usepackage{amsmath}
\usepackage{amssymb}
\usepackage{graphicx}
\usepackage{graphics}

\title{Concentric Characterization and Classification of Complex Network
Nodes: Theory and Application to Institutional Collaboration}

\author{Luciano da Fontoura Costa\thanks{} , Marilza A
Rodrigues Tognetti, Filipi Nascimento Silva}

\begin{document}

\bibliographystyle{unsrt}
\twocolumn[
    \begin{@twocolumnfalse}
\maketitle
\abstract{Differently from theoretical scale-free networks, most of real
networks present multi-scale behavior with nodes structured in
different types of functional groups and communities. While the
majority of approaches for classification of nodes in a complex
network has relied on local measurements of the topology/connectivity
around each node, valuable information about node functionality can be
obtained by Concentric (or Hierarchical) Measurements. In this paper
we explore the possibility of using a set of Concentric Measurements
and agglomerative clustering methods in order to obtain a set of
functional groups of nodes. Concentric clustering coefficient and
convergence ratio are chosen as segregation parameters for the
analysis of a institutional collaboration network including various
known communities (departments of the University of S\~ao Paulo). A
dendogram is obtained and the results are analyzed and discussed.
Among the interesting obtained findings, we emphasize the scale-free
nature of the obtained network, as well as the identification of
different patterns of authorship emerging from different areas
(e.g. human and exact sciences). Another interesting result concerns
the relatively uniform distribution of hubs along the concentric
levels, contrariwise to the non-uniform pattern found in theoretical
scale free networks such as the BA model.}
\end{@twocolumnfalse}
\vspace{1cm}
  ]
 {
    \renewcommand{\thefootnote}%
      {\fnsymbol{footnote}}
    \footnotetext[1]{Cybernetic Vision Research
 Group, GII-IFSC, Universidade de S\~ ao Paulo, S\~{a}o Carlos, SP,
 Caixa Postal 369, 13560-970, Brasil, luciano@if.sc.usp.br.}
  }

\section{Introduction}

One of the inherent features of complex networks concerns their
structured patterns of connectivity, which depart from the largely
uniform degree distribution found in random graphs
~\cite{erdos:1959er,Albert:2002ys,Costa:2007yq,Newman:2003vn}.It is
such a complex connectivity, found in some real and theoretical
networks, that gives rise to interesting structural elements like
communities and scale-free node degree
distributions~\cite{Barabasi:1999rt}. Though such patterns can be
sometimes identified by considering only simple features such as the
node degrees, more information can be obtained by considering
additional measurements~\cite{Newman:2007lr}. Indeed, some types of
communities can be overlooked while considering only such
measurements.  Even more information about the heterogeneity of
networks connectivity can be provided by the consideration of
concentric (or hierarchical) measurements, obtained by taking into
account successive neighborhoods around each node ~\cite{Costa:2004fj,
Costa:2006uq, Fontoura-Costa:2007qy}. This possibility has been
preliminary explored.  In~\cite{Costa:2006lr}, those measurements were
used in order to obtain interesting information about the topological
features of the networks as a whole. The results showed distinct
behaviors for real and grown networks, with the latter often
exhibiting a mixture of features typical to different models. That
paper also illustrated the possibility of clustering of groups of
nodes with similar concentric connectivity.

The current work extends in a more systematic and formal way such
preliminary investigations. More specifically, we adopt a sound way to
measure the similarity between the distributions of concentric
measurements, namely by calculating the Spearman correlation between
those features. Compared to the previously adopted consideration of
the Euclidean distances, such an approach accounts for less
sensitivity to the absolute values of the measurements. The potential
of this approach is illustrated with respect to the important problem
of scientific collaboration, as quantified by co-authorship, between
the staff of the largest Brazilian university, namely the University
of S\~ao Paulo -- USP. A dataset of scientific publication covering
from 2003 and 2004\footnote{The authors are thankful to Adriana Cybele
Ferrari, Edna Knorich and SIBi-USP(Integrated System of Libraries of
University of S\~ao Paulo), for providing the dataset used in this
paper.} was considered in order to build a collaborative network where
each node corresponds to a member of staff, while the links are
provided by co-authorships in publications indexed by forty libraries integrated with SIBi-USP.
Interestingly, the original dataset also included the
respective affiliations of each author, so that a preliminary
identification of possible communities (departments of USP) was
available for use as a reference.

A series of concentric measurements were calculated from this network
and had their average and standard deviation values compared to
theoretical models (Erd\H{o}s-R\'enyi -- ER, and Barab\'asi-Albert --
BA). Subsequently, the new methodology for node classification was
applied in order to organize the network nodes into clusters, possibly
corresponding to the communities existing in the network. An
average-based concentric clustering algorithm
\cite{Duda:2001fk,Fontoura-Costa:2001lr} was used for obtaining such a
clusterization. A series of interesting results was obtained. First,
as could be expected, we found that the collaborative network exhibits
a scale-free like distribution of node degrees. Among the several
considered concentric measurements, the convergence ratio and
concentric clustering coefficient were found to contribute to
particularly to the discrimination between the nodes, and were
consequently adopted for the c clustering of nodes. When the obtained
clusters were compared with the original institutional departments, a
more definite correspondence was found in the case of exact
sciences. A less clear adherence with the original departments was
found for humanities and biological sciences. Such findings suggest a
more localized pattern of co-authorship in the case of exact
sciences. Another interesting finding regards the several deviations
of specific properties between the collaborative network and the BA
theoretical model. For instance, the real network was found to have
larger values of average shortest paths, indicating higher network
sizes when compared to the BA counterpart. In addition, the
information provided by the convergence ratio, suggests that the large
size of this network is a consequence of the uniform distribution of
hubs along concentric levels, where the hubs tend to be connected to
low degree nodes, while hubs are almost connected one another in the
BA case.

\section{Basic Concepts and Models of Networks.}

Consider a undirected and weighted network $\Gamma$ defined by $N$
nodes and a set of $K$ weighted edges connecting those nodes. $\Gamma$
can be completely specified by an adjacency matrix $G$ with elements
$G_{ij} = G_{ji}$ (i.e. a symmetric matrix) where the strength of a
connection between node $i$ and node $j$ is $G_{ij}$(i.e. the value of
matrix at $i$-th line and $j$-th column), and a null value represents
no connection.

Nodes can be characterized by the traditional immediate neighborhood
features, i.e. the node degree and clustering coefficient. For
weighted networks, the node degree of a node $i$, represented by
$k_i$, is defined as the sum of all weight values of edges that
connects $i$ to any other nodes. More specifically, considering the
adjacency matrix representation $G_{ij}$, the node degree can be
calculated by:

\begin{equation}
k_i=\sum_{j=1}^{n}G_{ij} \label{eq:knormal}
\end{equation}

The clustering coefficient of a node $i$, abbreviated as $cc_i$, is
defined as the number of connections, $e_1(i)$, among the nodes in the
immediate neighbors of $i$ divided by he maximum possible number of
connections of those nodes. Let $n_1(i)$ be the number of nodes at
the immediate neighbor of $i$. The clustering coefficient can then be
calculated as:

\begin{equation}
cc_d(i)= 2 \frac{e_1(i)}{n_1(i)(n_1(i)-1)} \label{eq:ccnormal}
\end{equation}

This paper considers two theoretical network models for comparison
purposes, namely: Erd\'os-R\'enyi (random networks) and
Barab\'asi-Albert(a scale-free model). The Erd\'os-R\'enyi (ER) model
is defined by a network created connecting the pairs of nodes with a
constant probability~\cite{erdos:1959er,Albert:2002ys,Costa:2007yq},
resulting in a binomial distribution of node degrees. A
Barab\'asi-Albert (BA) network is created by starting which a
non-connected network with $m_0$ nodes and then adding new nodes
progressively with $k$ new connections between each new node and those
already in the network. The probability of the new connections is
proportional to the node degree of the existing nodes. This procedure
results in a scale-free network, where the node degree distribution
follows a power law and shows the presence of hubs.

\section{Concentric Measurements}

\begin{figure}
\begin{center}
\includegraphics[scale=0.5]{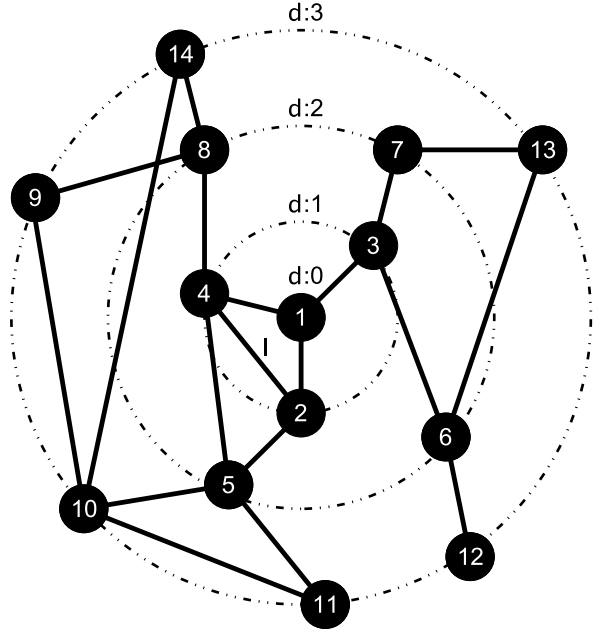}
\caption{A small network with 3 concentric levels considering
node $i$ as reference~\label{fig:hierdemo}}
\end{center}
\end{figure}

A ring $R_{d}(i)$ representing the nodes that are at the
\emph{concentric level} $d$ centered at node $i$, is defined as the
sub-graph containing all nodes whose shortest path value, starting at
node $i$, is $d$. A network with three concentric levels is
illustrated in Figure~\ref{fig:hierdemo}, where the rings $R_{d}(1)$
(i.e. centered at node $1$) of levels $d=1$, $d=2$ and $d=3$ are
represented by concentric circles, i.e. with $R_{0}(1)$=\{1\},
$R_{1}(1)$=\{2, 3, 4\}, $R_{2}(1)$=\{5, 6, 8\} and $R_{3}(1)$=\{9, 10,
11, 12, 13, 14\}.

The concept of concentric levels allows the complementation of the
traditional measurements, focused not on the local topological
properties of nodes, but taking into account its successive
neighborhoods. In general, the concentric measurements are calculated
considering relationships between the nodes and edges at two or more
concentric levels. The following measurements can be naturally
generalized for weighted networks by performing some modifications.

The \emph{concentric node degree}, $k_d(i)$, of a reference node $i$
at the concentric level $d$ is defined as corresponding to the number
of edges connecting the nodes in $R_d(i)$ and $R_{d+1}(i)$. As an
example, we have for Figure~\ref{fig:hierdemo} $k_0(1)=3$, $k_1(1)=5$
and $k_2(1)=7$. Note that the concentric node degree is not an average
value taken among the number of nodes in $R_d(i)$. This measurement is
the direct extension of the well known node degree where the reference
node is understood as the nodes inside the ball $B_d(i)$ (i.e. the
ball containing the nodes in rings $0$ to $d$). The concentric node
degree can be extended for weighted networks by taking the sum of the
weight values for every connection between these nodes and the nodes
at the next level.

The \emph{concentric clustering coefficient}, $cc_d(i)$, is the
immediate generalization of the traditional clustering coefficient and
considers the only the nodes and connections of the ring $R_{d}(i)$.
It is defined as in Equation~\ref{eq:cc}, where the number of edges in
$R_d(i)$ is expressed as $e_d(i)$, and the number of elements is
represented as $n_d(i)$.

\begin{equation}
cc_d(i)= 2 \frac{e_d(i)}{n_d(i)(n_d(i)-1)} \label{eq:cc}
\end{equation}

For node $i=1$ in the network shown in Figure~\ref{fig:hierdemo},
we have that $cc_1(1)=1/3$, $cc_2(1)=0$, and $cc_3(1)=1/5$.

Other interesting concentric measurements which can be obtained with
respect to the reference node $i$ and used in order to complete the
characterization of complex networks include the following:

{\bf Convergence ratio ($C_d(i)$):} Corresponds to the ratio between
the concentric node degree of node $i$ at distance $d$ and the
number of nodes in the ring at next ring, i.e.

\begin{equation}
C_d(i)=\frac{k_{d}(i)}{n_{d+1}(i)}.
\end{equation}

This measurement quantifies the average number of edges received by
each node in the ring $d+1$. We have necessarily that $C_0(i)=1$ for
whatever node selected as the reference $i$. In the case illustrated
in Figure~\ref{fig:hierdemo}, we have $C_0(1)=1$, $C_1(1)=5/4$ and
$C_2(1)=1$.

{\bf Intra-ring degree ($A_d(i)$):} This measurement is obtained by
taking the average among the degrees of the nodes in the subnetwork
$\gamma_d(i)$. Observe that only those edges between the nodes in such
a subnetwork are considered, therefore overlooking the connections
established by such nodes within the nodes in the rings at levels
$d-1$ and $d+1$. For instance, we have for the situation in
Figure~\ref{fig:hierdemo} that $A_1(1)=1/3$, $A_2(1)=0$ and
$A_3(1)=1/2$. For weighted networks the value of intra-ring is the
average of the weights of all nodes at the rings $R\{d-1\}$ and
$R\{d+1\}$.

{\bf Inter-ring degree ($E_d(i)$):} This measurement corresponds to
the average of the number of connections between each node in the ring
$R_d(i)$ and those in $R_{d+1}(i)$. For instance, for
Figure~\ref{fig:hierdemo} we have $E_0(1)=3$, $E_1(1)=5/3$ and
$E_2(1)=3/2$. Observe that $E_d(i)=k_d(i)/n_d(i)$.

{\bf concentric common degree ($H_d(i)$):} Equal to the average node
degree among the nodes in $R_d(i)$, considering all edges in the
original network. For Figure~\ref{fig:hierdemo} we have $H_0(1)=1$,
$H_1(1)=10/3$ and $H_2(1)=16/7$. The concentric common degree
expresses the average node degree at each concentric level, indicating
how the network node degrees are distributed along the network
hierarchies.

Table~\ref{tab:meas} summarizes the concentric measurements to be used
in this paper, all of which are defined with respect to one of the
network nodes, identified by $i$, taken as a reference and at a
distance $d$ from that node.

\begin{table}
\begin{center}
\vspace{1cm}
\begin{tabular}{cr} \hline
$e_d(i)$ & conc. number of edges among \\
& the nodes in the ring $R_d(i)$ \\ \hline
$n_d(i)$ & conc. number of nodes \\
& in the ring $R_d(i)$ \\ \hline
$k_d(i)$ & concentric degree \\
& of node $i$ at distance $d$ \\ \hline
$A_d(i)$ & intra-ring node degree \\
& of node $i$ at distance $d$ \\ \hline
$E_d(i)$ & inter-ring node degree \\
& of node $i$ at distance $d$ \\ \hline
$H_d(i)$ & concentric common degree \\
& $i$ of node at distance $d$ \\ \hline
$cc_d(i)$ & conc. clustering coefficient \\
& of node $i$ at distance $d$ \\ \hline
$C_d(i)$ & convergence rate at \\
& concentric level $d$ \\ \hline

\end{tabular}
\caption{The concentric measurements considered in the current
article.~\label{tab:meas}}
\end{center}
\end{table}

\section{Statistical Concepts}

Two statistical methodologies are used in the present paper in order
to obtain groups of nodes with similar concentric measurements. The
first step is to choose a distance
measurement~\cite{Duda:2001fk,Fontoura-Costa:2001lr}(i.e. how similar
two nodes are, in terms of a set of measurements). Because of the
varying forms of the concentric measurement distributions, a
non-parametric distance such as the Spearman rank correlation
coefficient, should be adopted.

\subsection{Spearman Rank Correlation.}

The Spearman rank correlation coefficient is a statistical measurement
quantifying how strong is the tendency of two random variables to vary
together. Unlike the Pearson correlation coefficient, this measurement
is not restricted to linear joint variations and can be used to
quantify the similarity between the form of two curves (or data
sets). In fact, the Spearman rank correlation is a special case of the
Pearson correlation, where every value of the curve is ranked before
calculating the coefficient.

Given two normalized distributions of two random discrete variables, X
and Y, the Pearson correlation coefficient between them is defined by
the covariance of those two variables divided by their respective
standard deviations, i.e.:
\begin{equation}
\label{eq:pearson}
r_{X Y}=\frac{cov(X,Y)}{\sigma_X \sigma_Y}
\end{equation}

Given $n$ samples of the random variables $X$ and $Y$, henceforth
expressed $x_i$ and $y_i$, the respective Pearson Correlation
Coefficient can be estimated as:

\begin{equation}
\label{eq:pearson2}
r_{X Y}=\frac{\sum (x_i - \bar{x})(y_i - \bar{y})}{(n-1)\sigma_X \sigma_Y}
\end{equation}

The Spearman rank correlation coefficient $\rho$ can be obtained by
replacing $X$ and $Y$ values by their ranked version $X^*$ and
$Y^*$. For example, considering a data set with $X=\{1.1, 3, 0.5,
100\}$ and $Y=\{2, 0.8, 1,0.1\}$, the ranked data set will be
$X^*=\{2,3,1,4\}$ $Y^*=\{4,2,3,1\}$, and the Spearman rank coefficient
will be $\rho=r_{X^* Y^*}$

\section{Methodology}

In order to illustrate the use of concentric measurements and node
classification, we consider a collaboration network obtained from real
data, which will be compared to Barabasi-Albert(BA) and Erd\"os and
R\'enyi(ER) counterparts.

The collaboration network, presented in this work for the first time,
was created by collecting data about co-authorship in published
articles, where each node represents an author and each undirected
edge represents a paper written by the respective two nodes
(authors). Because of the possible existence of more than one paper by
the same two authors, those edges are weighted with values
representing the number of papers that those authors wrote
together. This network was obtained from the library database of
"Universidade de S\~ao Paulo". In addition to the collaborative
information, every node was labeled with the author corresponding
department. The network resulted with 5630 nodes and an average
topological node degree of $\left< k_t \right> \approx 15$ and average
strength of $\left< k \right> \approx 40$ .

The simulated networks, of type BA and ER, were obtained by the
classical methods~\cite{Costa:2007yq}. Random networks(ER) were
generated by selecting edges with uniform probability $p$, while the
BA networks were grown by starting with $m0$ randomly interconnected
nodes and adding new nodes with $m$ edges which are attached to the
existing nodes with probability proportional to their respective node
degrees.  The networks were created with 5000 nodes and average node
degree of $\left< k \right> \approx 16$ for BA and $\left< k \right>
\approx 15$ for ER.

We started the analysis of the characteristics of the real and
theoretical networks by considering the traditional node degree.
Next, a collection of concentric measurements were obtained for all
the networks, considering every node of as the center (reference), and
then taking the average values and average $\pm$ standard
deviations. The considered measurements were the concentric node
degree, concentric clustering coefficient, intra-ring degree,
inter-ring degree, common node degree and convergence ratio. The
distributions of such measurements obtained for the three networks
were compared as discussed in the next section.

While distributions of the concentric measurements supply subsidies
for a global characterization of the networks, they do not convey
information about the individual node concentric characteristics. This
information can be obtained in terms of the individual node concentric
measurements among the several levels centered at this node. In order
to classify those nodes into groups with similar concentric features,
the agglomerative hierarchical clustering algorithm, using spearman
rank coefficient as distance measure, was applied over the individual
node concentric clustering coefficients and convergence ratios. This
data was obtained only for the collaboration network. The resulting
tree (called dendogram) was truncated so as to yield eight groups of
nodes with similar properties.

Because the nodes in collaborative network are labeled with the
respective department of the corresponding author, the effectiveness
in the segregation of those groups can be quantified in the sense of
the percentage of nodes common to departments and communities.

\section{Results and Discussions}

This section begins by presenting the results obtained for the
concentric level measurements described in the methodology section and
then discusses such results while comparing the collaborative network
with the other two models. Finally, the agglomerative concentric
clustering results are shown as dendograms and average concentric
measurements distributions obtained for each group. The effectiveness
of the segregation of labeled groups is presented in the form of pie
charts.

\begin{figure*}
\begin{center}
\includegraphics[scale=0.3]{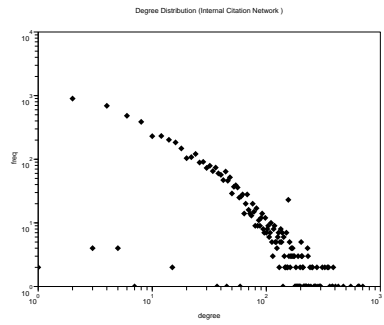}

(a)\\

\begin{tabular}{cc}
\includegraphics[scale=0.3]{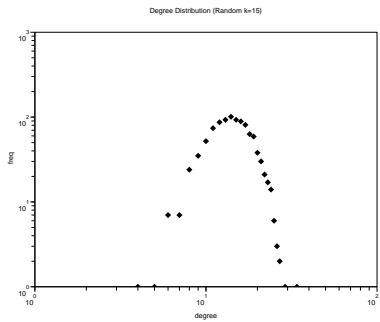}&
\includegraphics[scale=0.3]{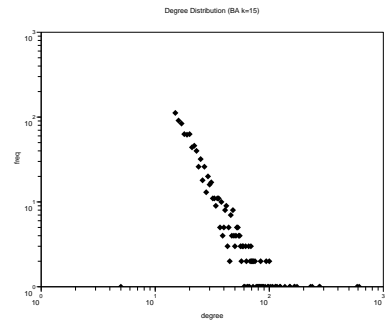}\\
(b) & (c)\\
\end{tabular}
\caption{Average degree distributions for the three considered
networks: (a)Collaborative network, (b) random ER model and (c)
Barab\'asi-Albert BA model. ER and BA networks with average degree
$\left< k \right>=4$. ~\label{fig:distrib}}
\end{center}
\end{figure*}

By obtaining the traditional degree distribution of the considered
networks, as can be seen in the loglog curves of Figure
~\ref{fig:distrib}, the collaboration network (a) can be understood as
a scale-free network, like the BA model (b), because of the well-known
power law behavior of those curves. It is interesting to note that the
power coefficients of the two scale free networks are distinct.

\begin{figure}
\includegraphics[scale=0.4]{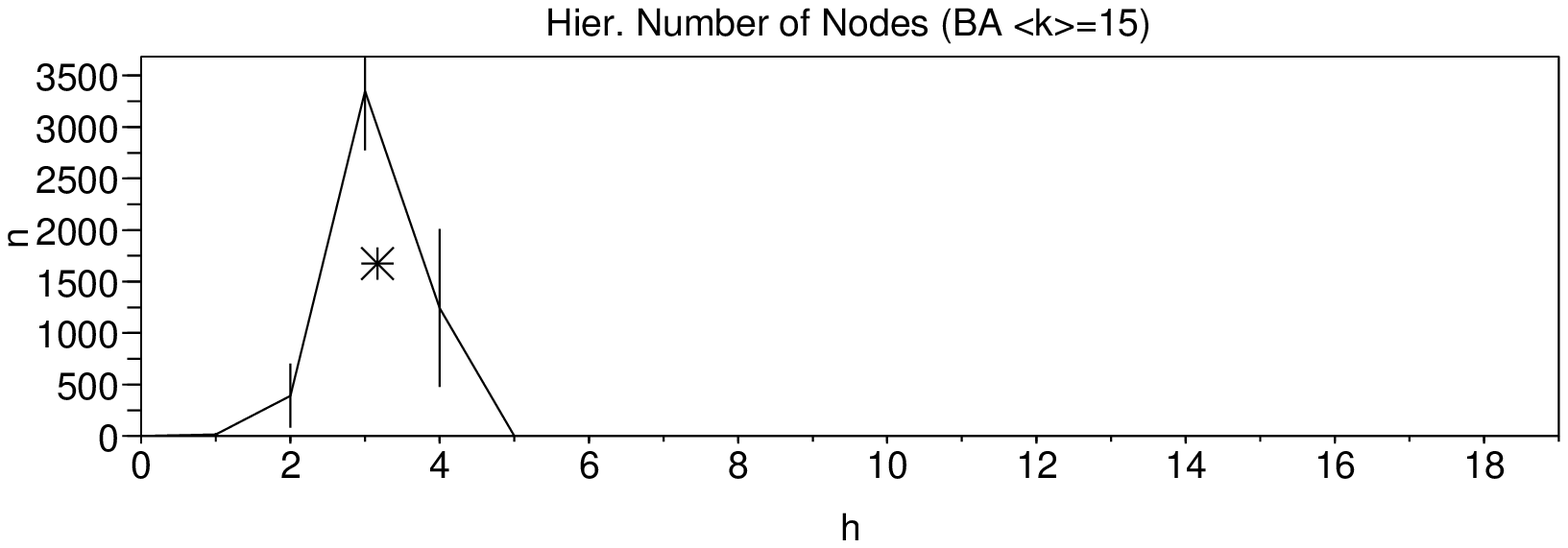}
\includegraphics[scale=0.4]{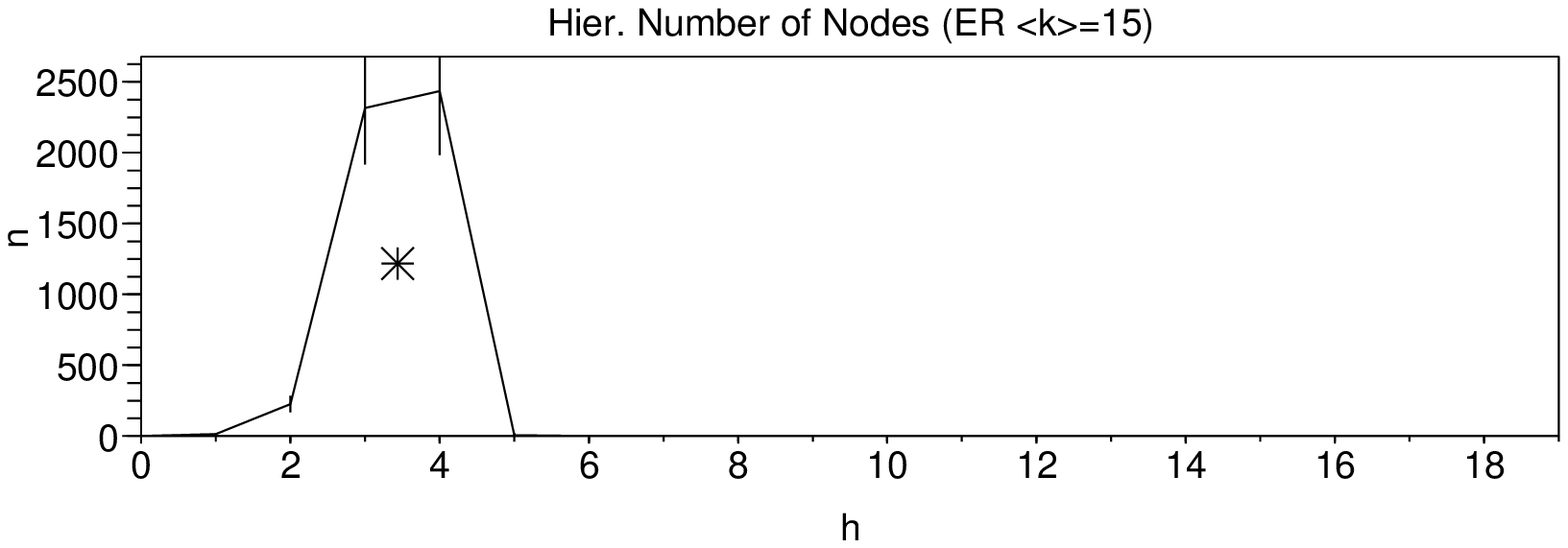}
\includegraphics[scale=0.4]{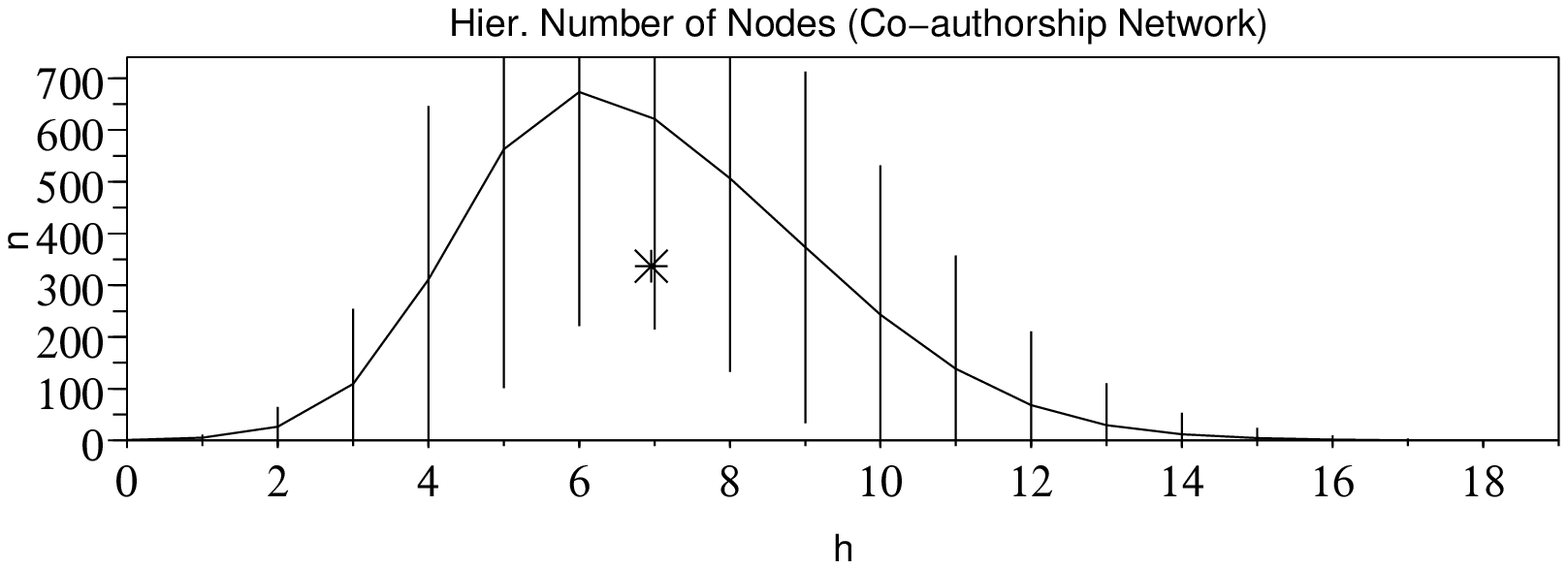}
\caption{Hierarchical number of nodes (average $\pm$ standard
deviation) for all considered networks, which are identified above
each graph. Observe that most curves are characterized by a
peak. The average value of the shortest path between any two nodes
is marked by an asterisk. ~\label{fig:sim1}}
\end{figure}

\begin{figure}
\includegraphics[scale=0.4]{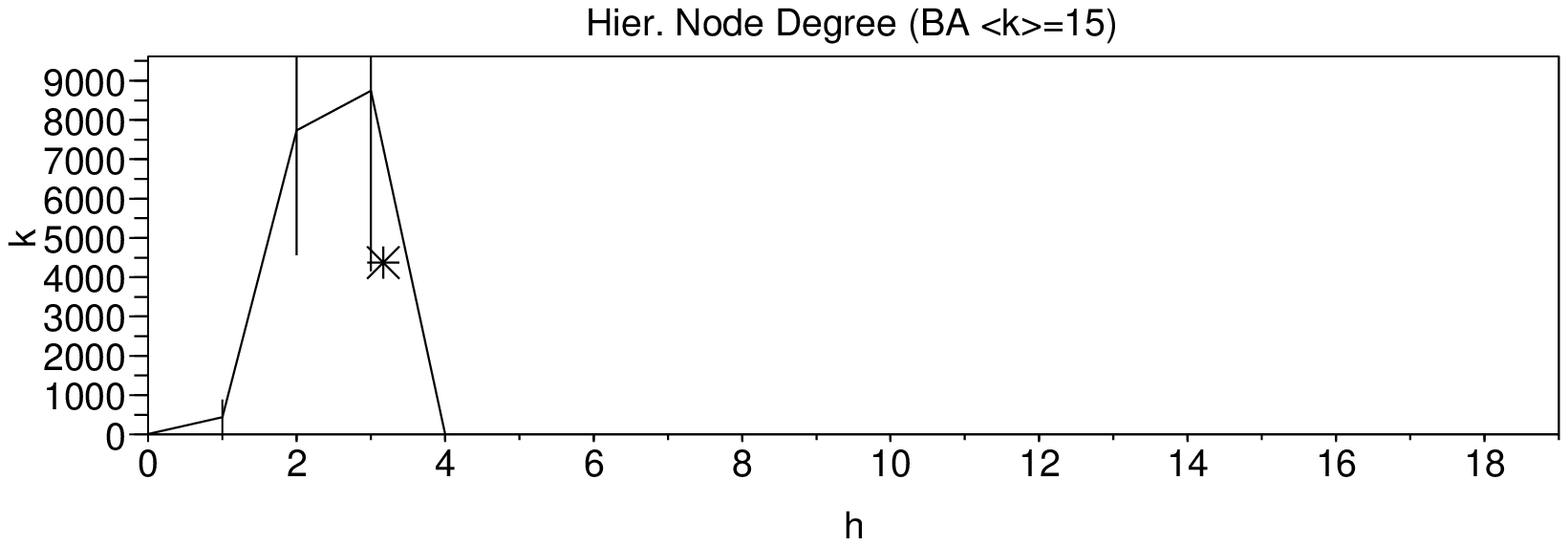}
\includegraphics[scale=0.4]{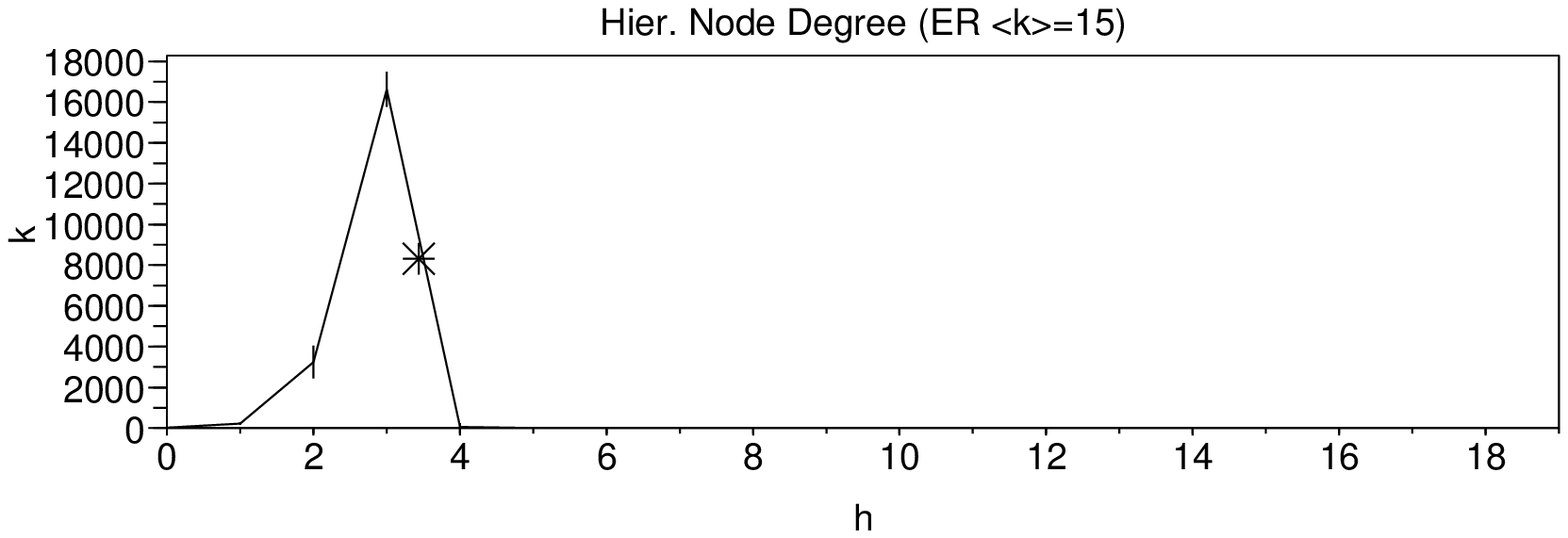}
\includegraphics[scale=0.4]{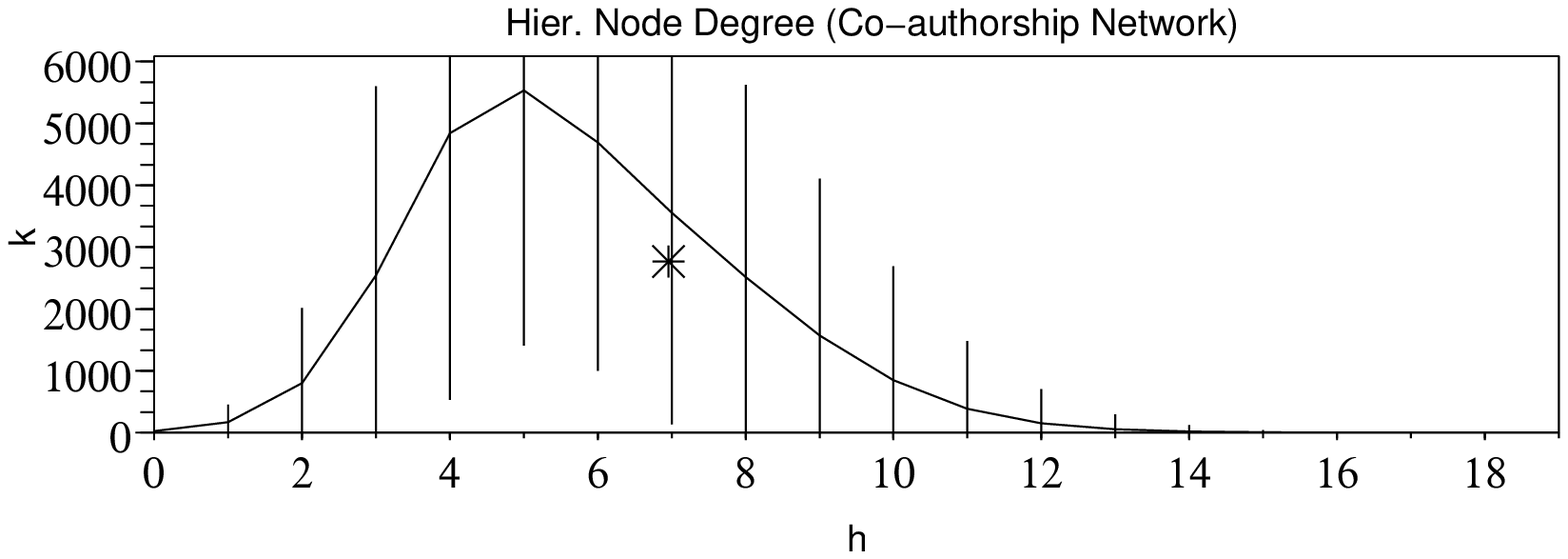}
\caption{Hierarchical node degrees obtained for all the considered
network models. The curves are similar to those obtained for the
concentric number of nodes, except for a expected offset at one
level.~\label{fig:sim2}}
\end{figure}

\begin{figure}
\includegraphics[scale=0.4]{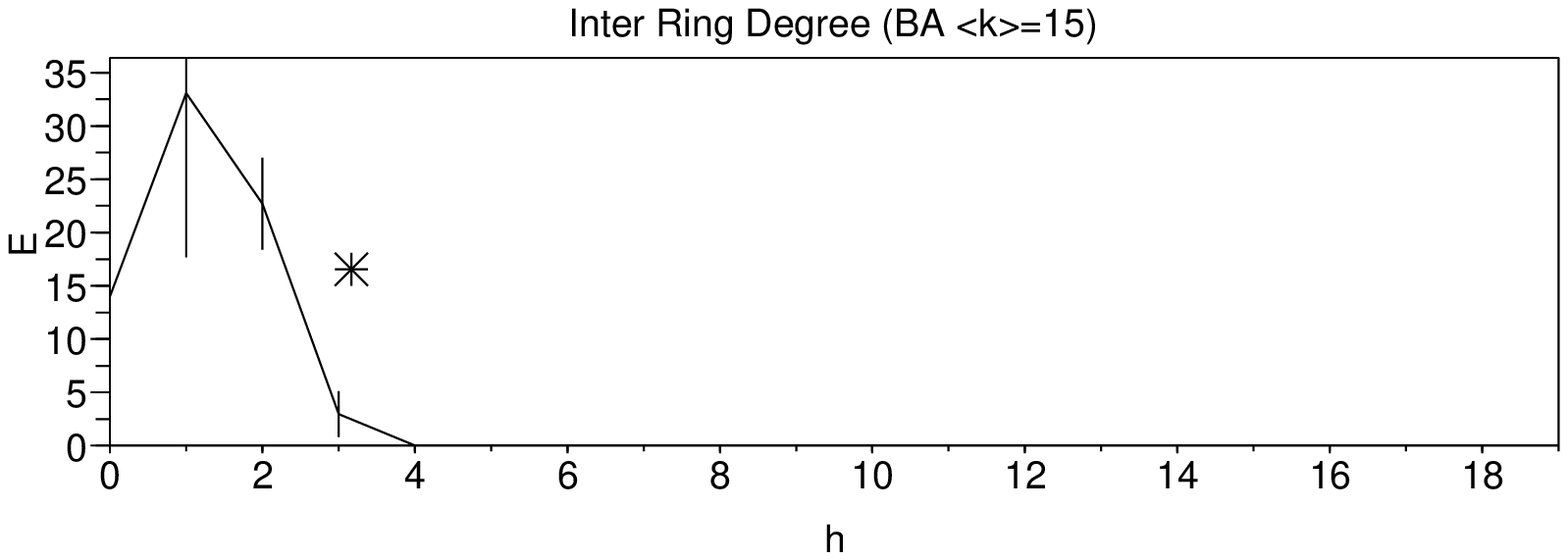}
\includegraphics[scale=0.4]{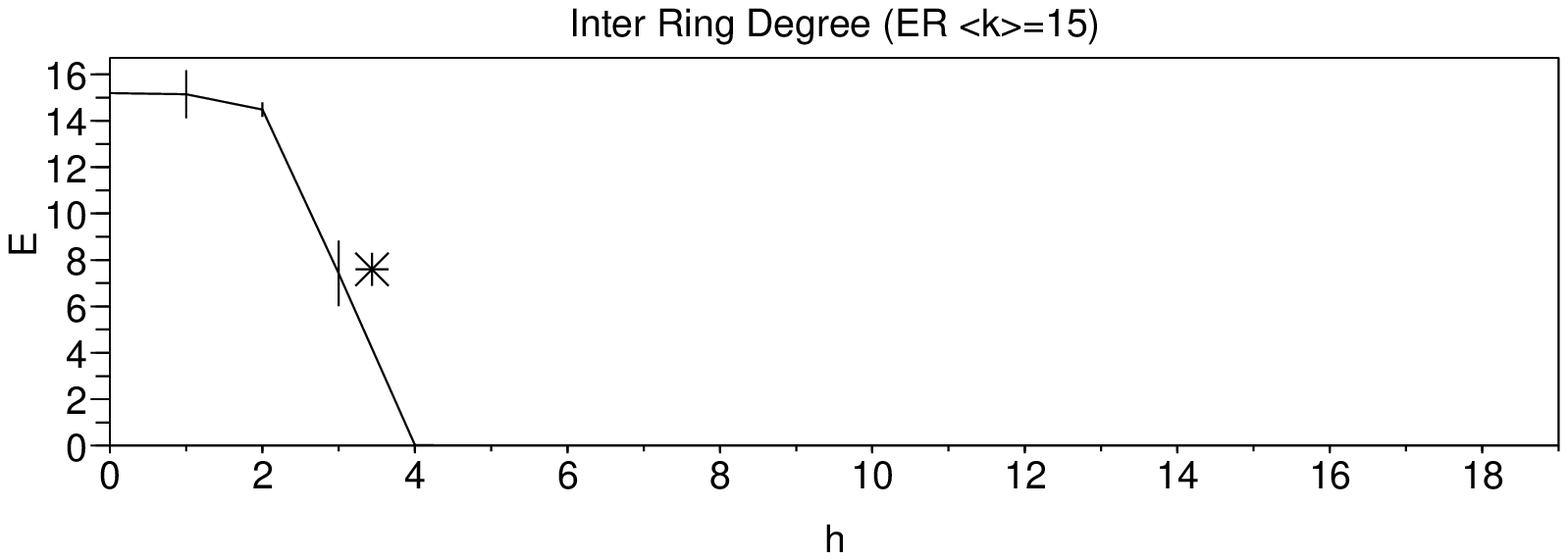}
\includegraphics[scale=0.4]{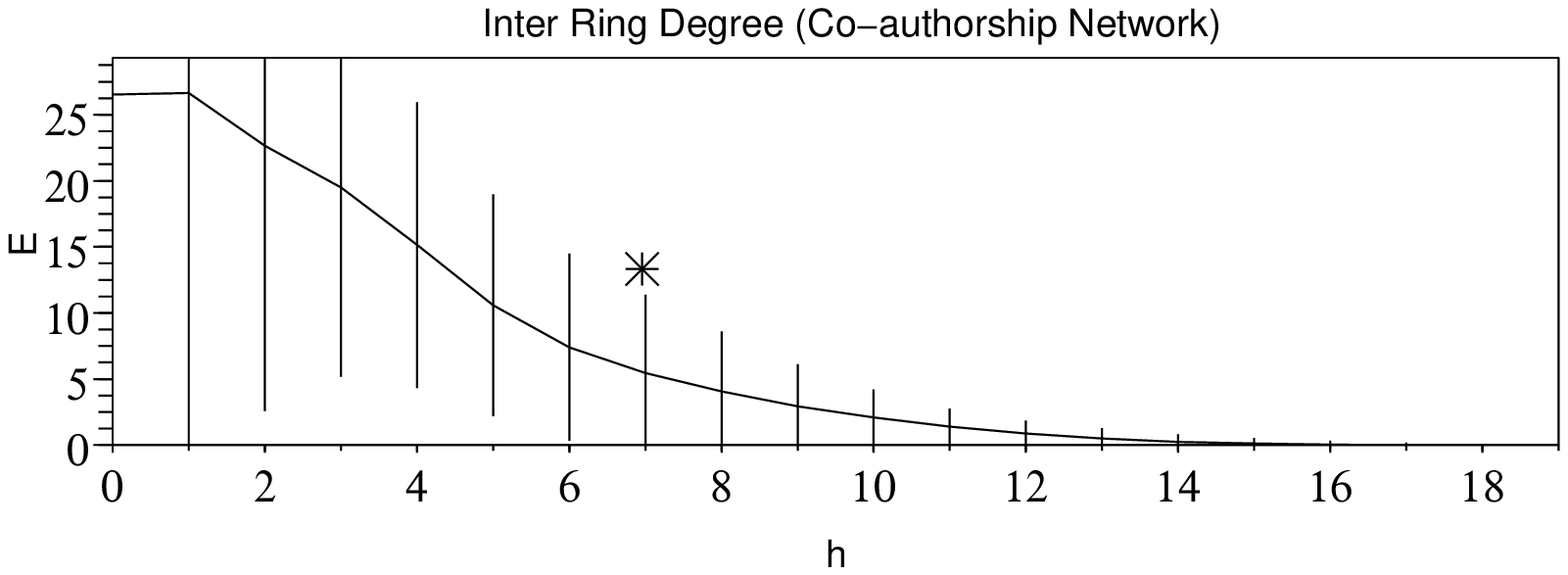}
\caption{Inter ring degree values for the considered
network models.~\label{fig:sim3}}
\end{figure}

\begin{figure}
\includegraphics[scale=0.4]{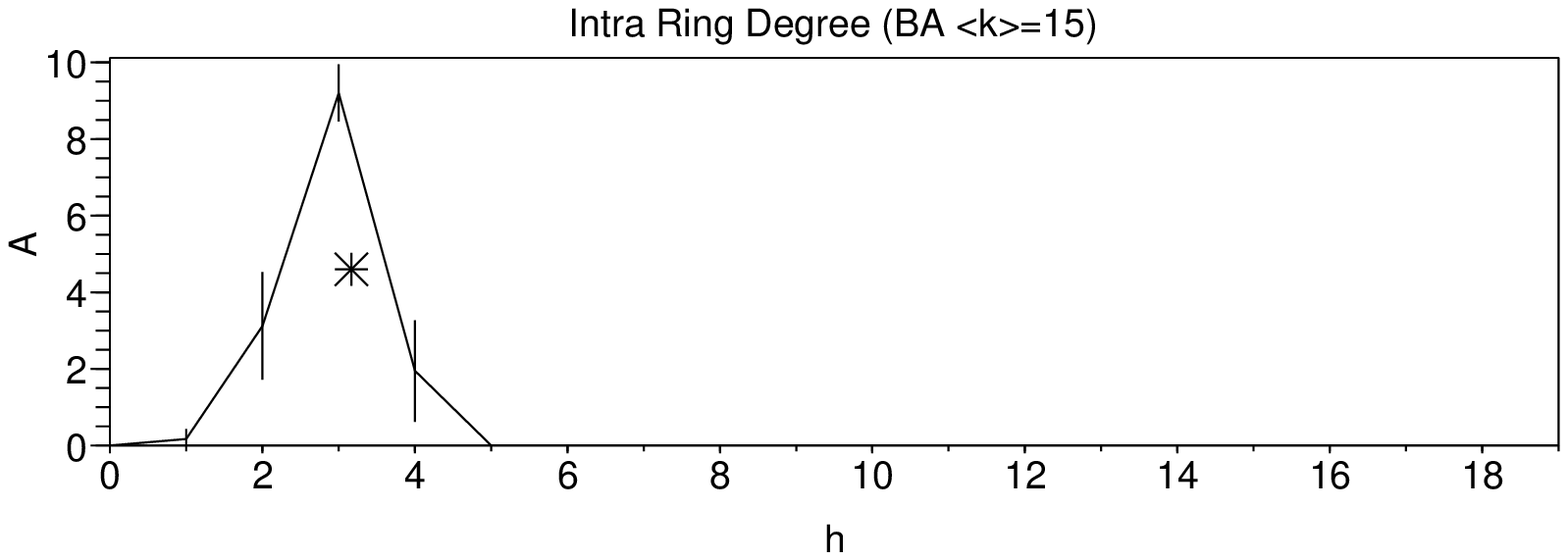}
\includegraphics[scale=0.4]{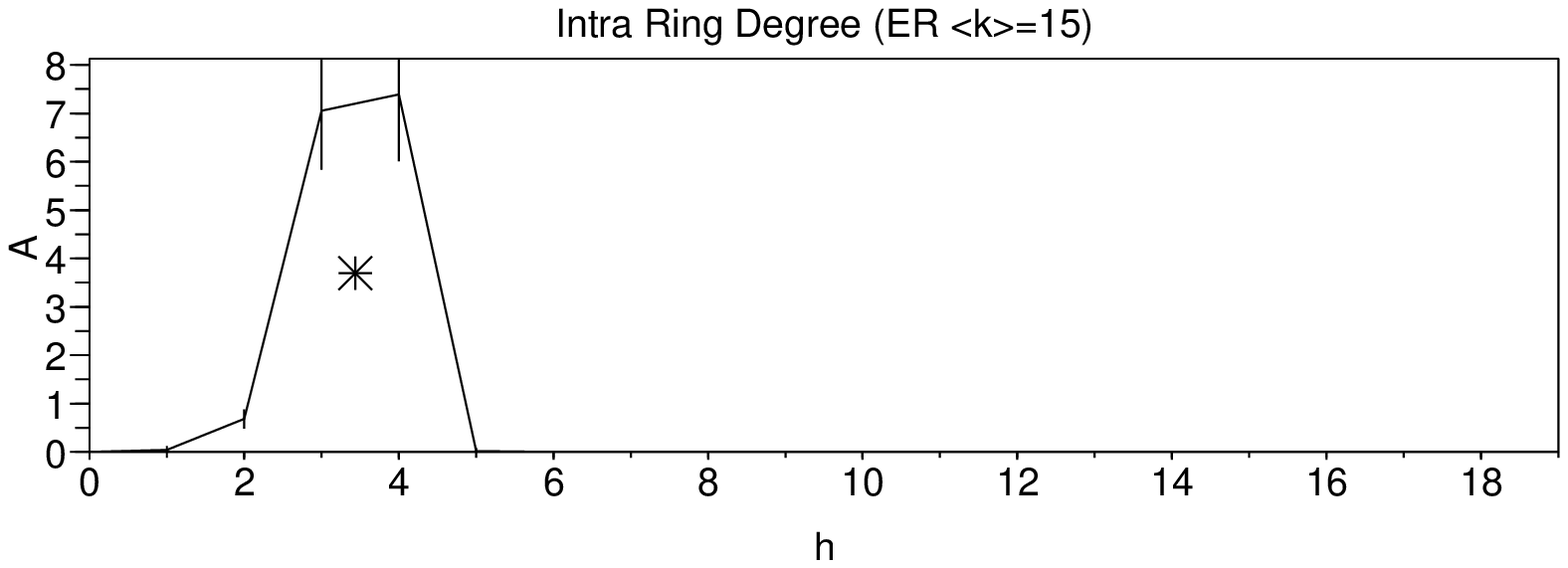}
\includegraphics[scale=0.4]{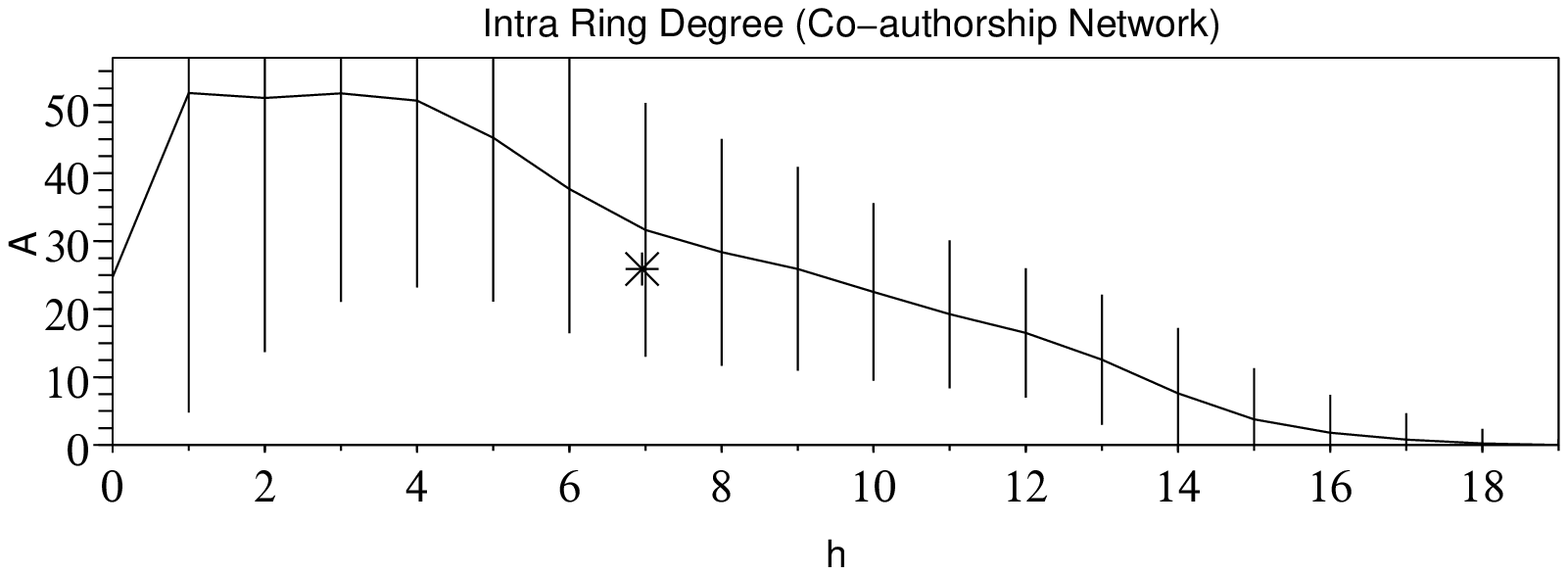}
\caption{Intra Ring Degree values for the considered network
models.~\label{fig:sim4}}
\end{figure}

\begin{figure}
\includegraphics[scale=0.4]{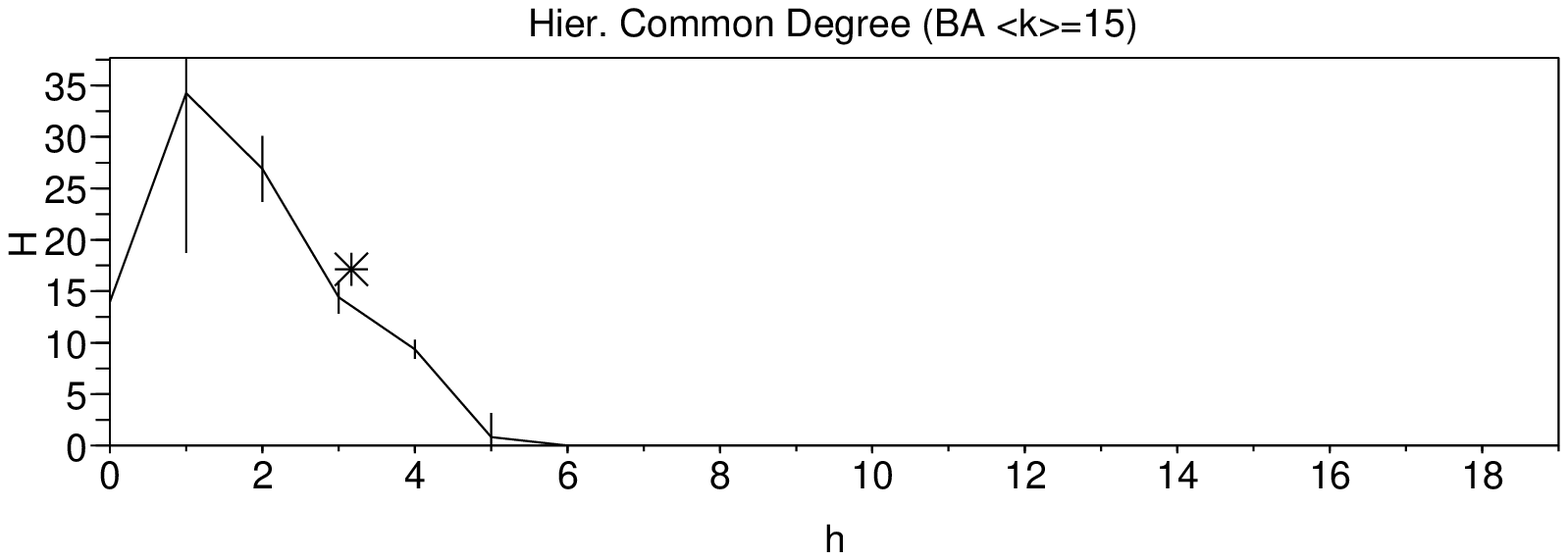}
\includegraphics[scale=0.4]{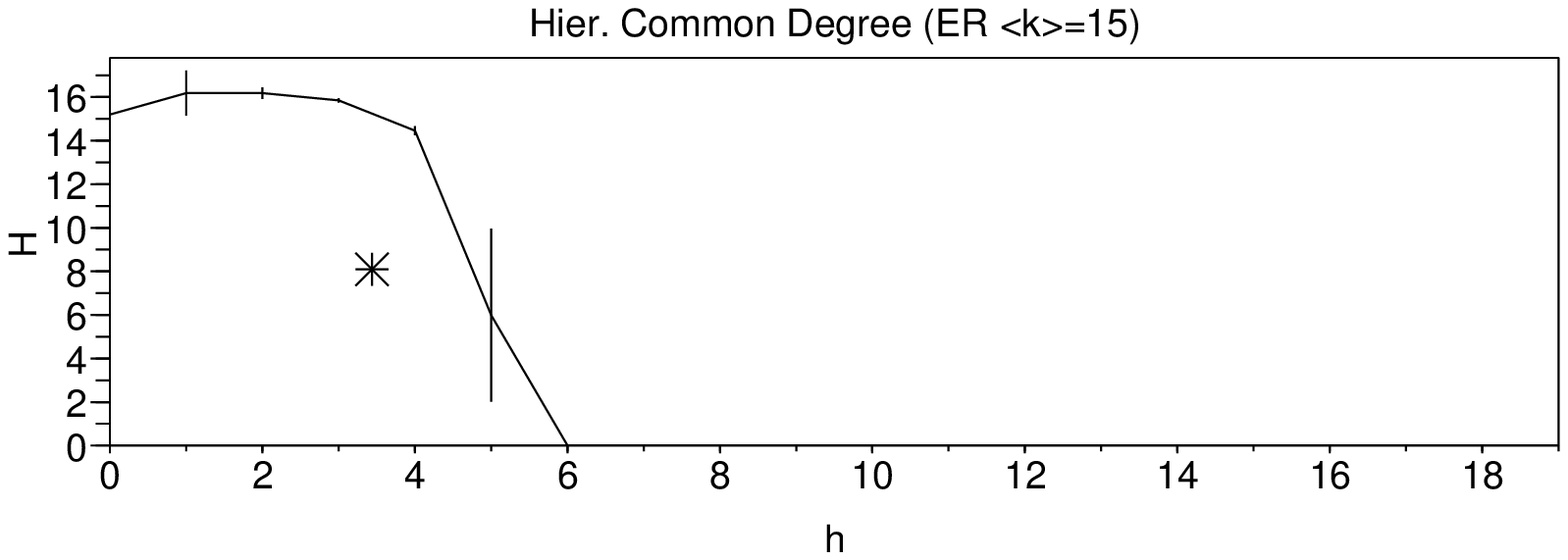}
\includegraphics[scale=0.4]{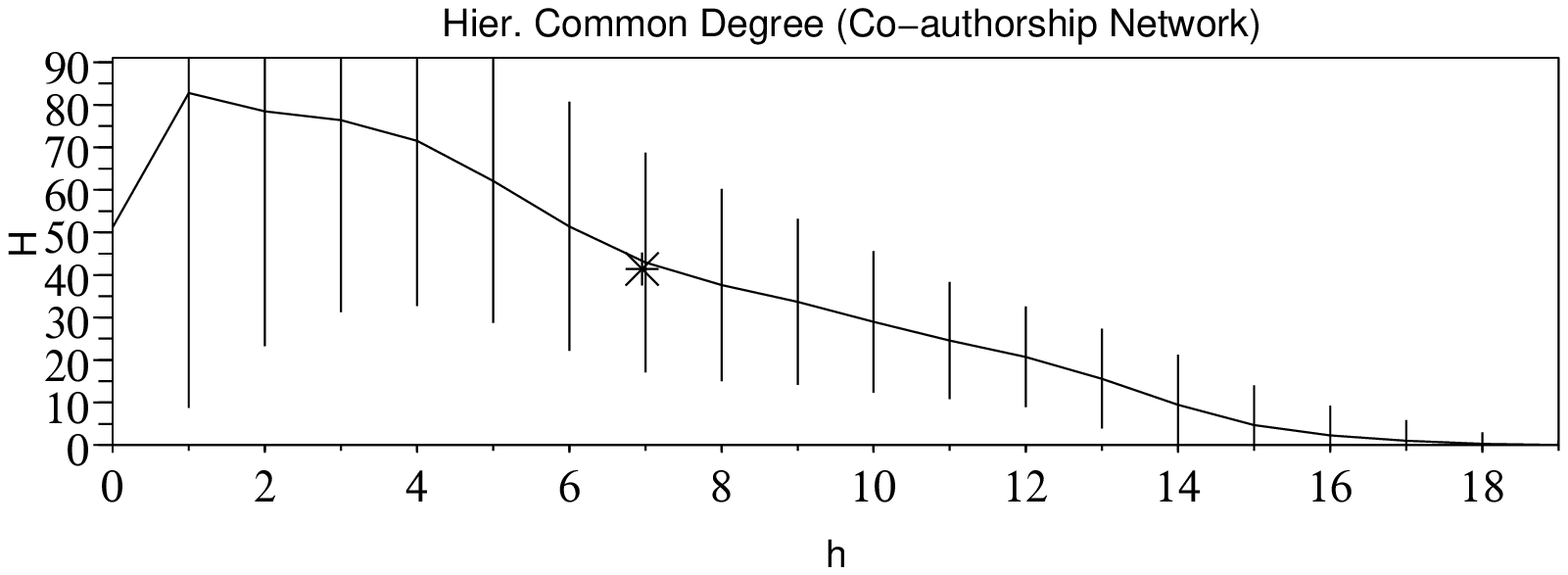}
\caption{Hierarchical common degree measurements with the respective
$\pm$ standard deviations obtained for the considered
models.~\label{fig:sim5}}
\end{figure}

\begin{figure}
\includegraphics[scale=0.4]{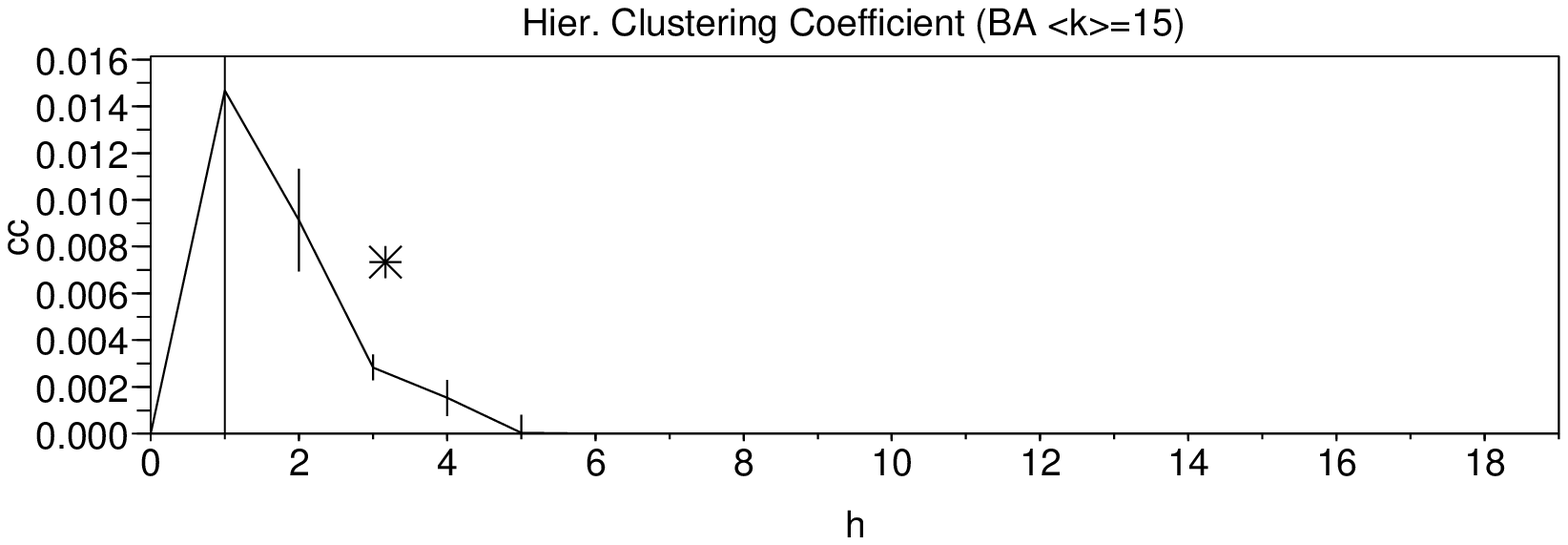}
\includegraphics[scale=0.4]{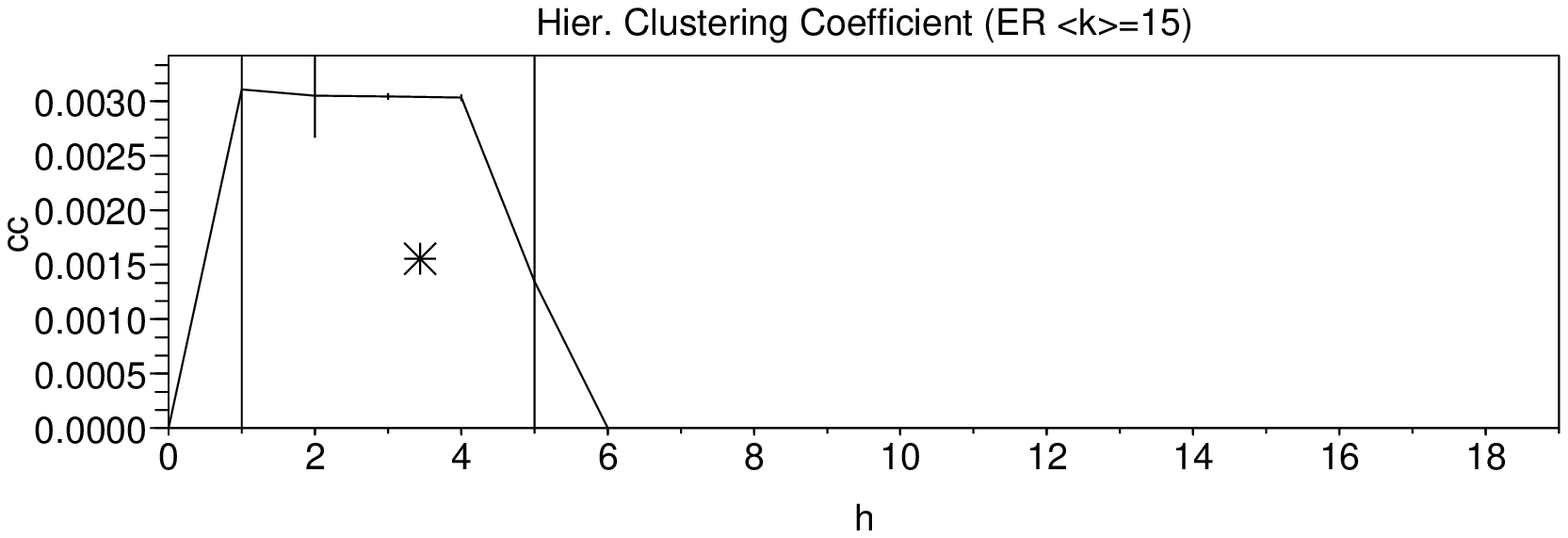}
\includegraphics[scale=0.4]{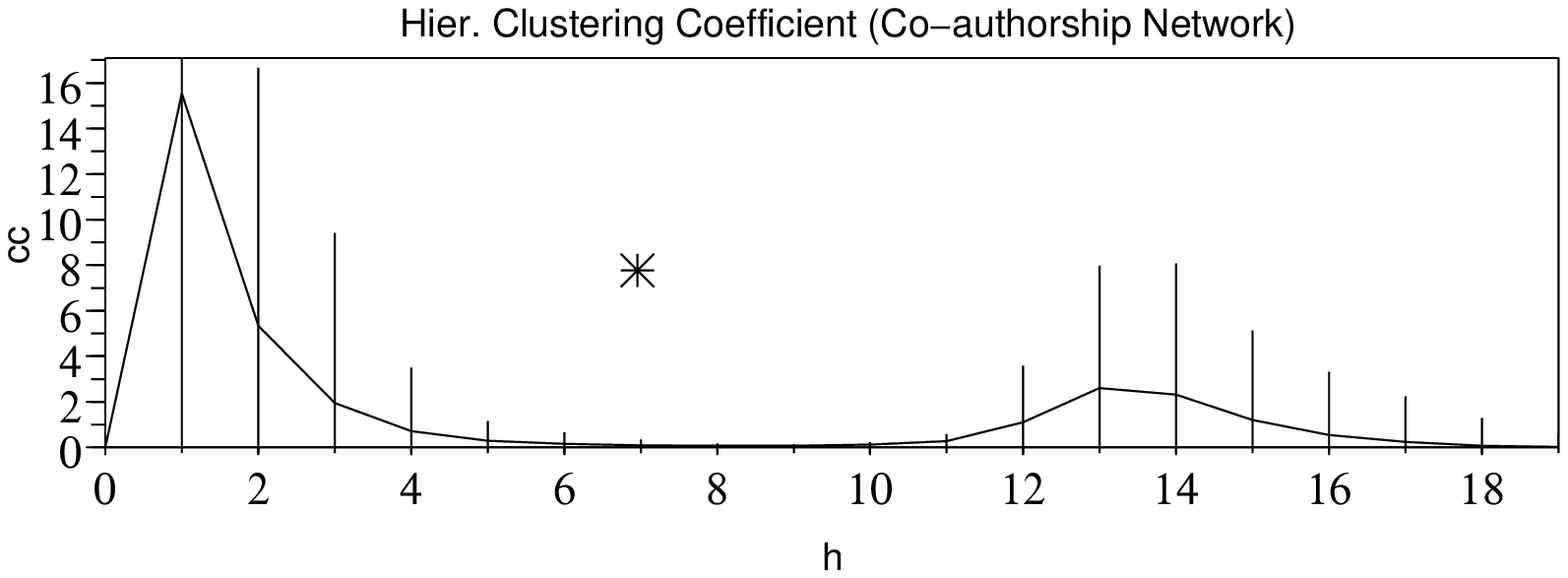}
\caption{Hierarchical clustering coefficient degree
measurements. Note the higher values of standard deviation
when compared to those in the other measurements.~\label{fig:sim6}}
\end{figure}

\begin{figure}
\includegraphics[scale=0.4]{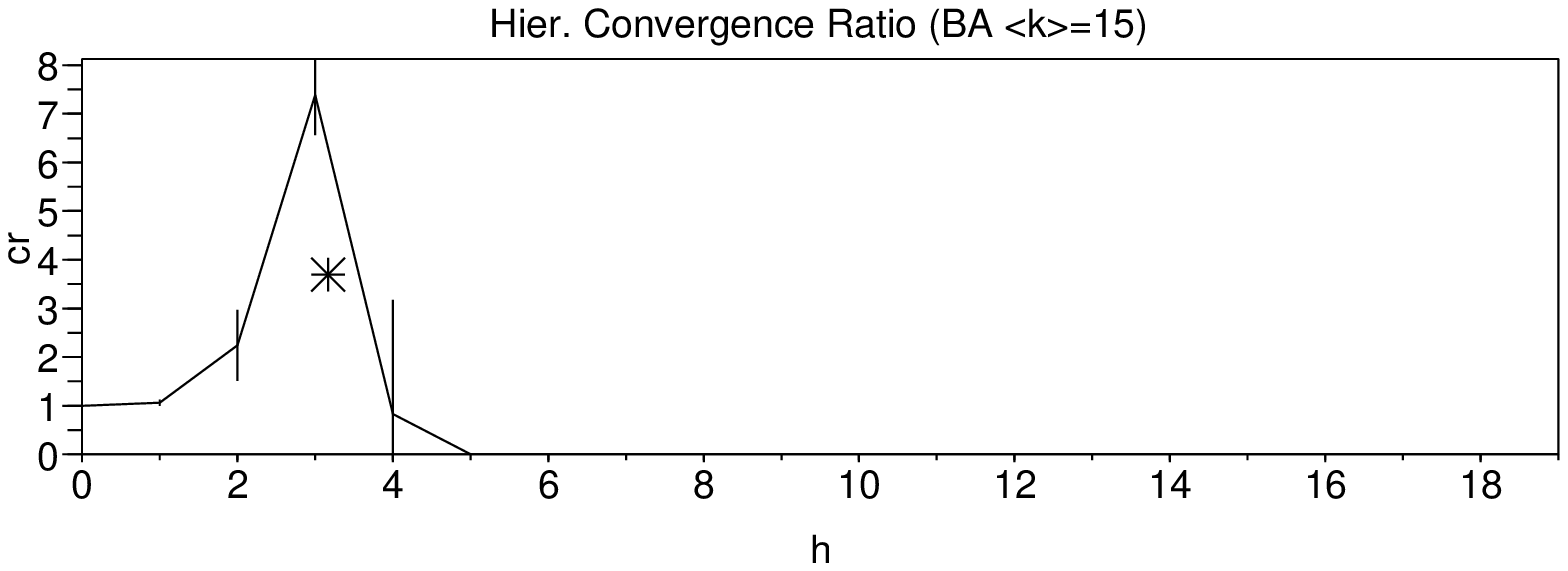}
\includegraphics[scale=0.4]{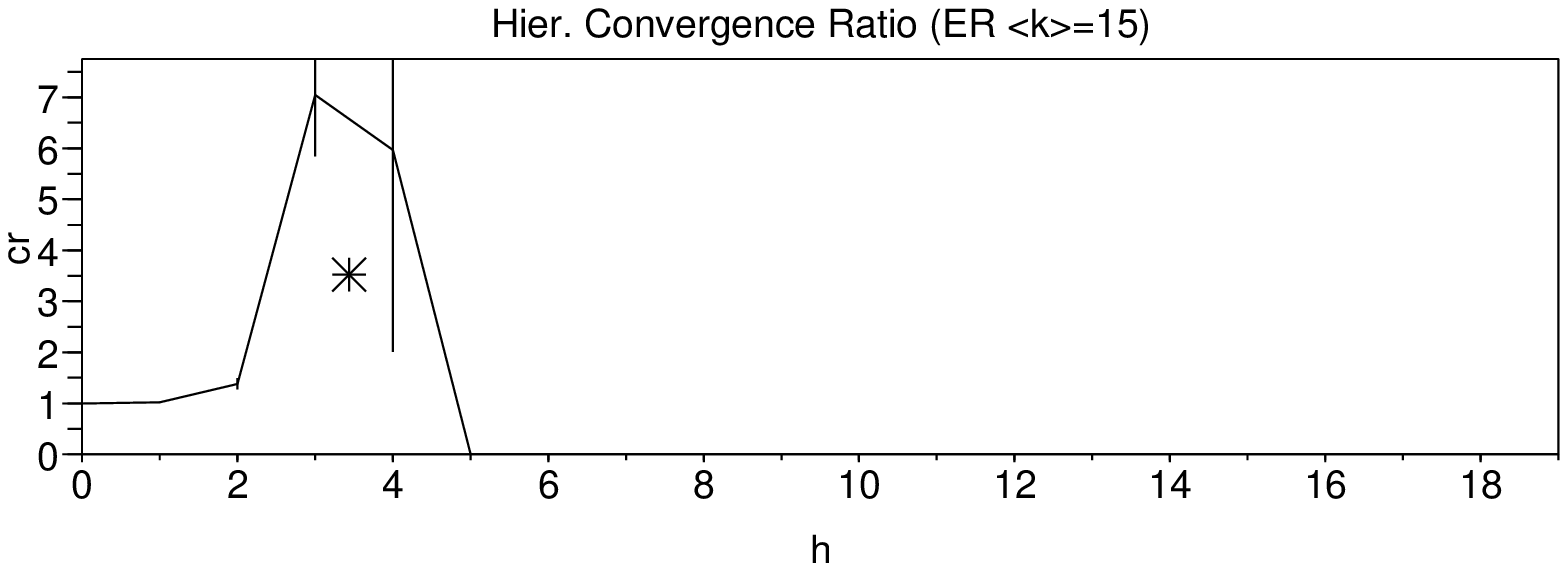}
\includegraphics[scale=0.4]{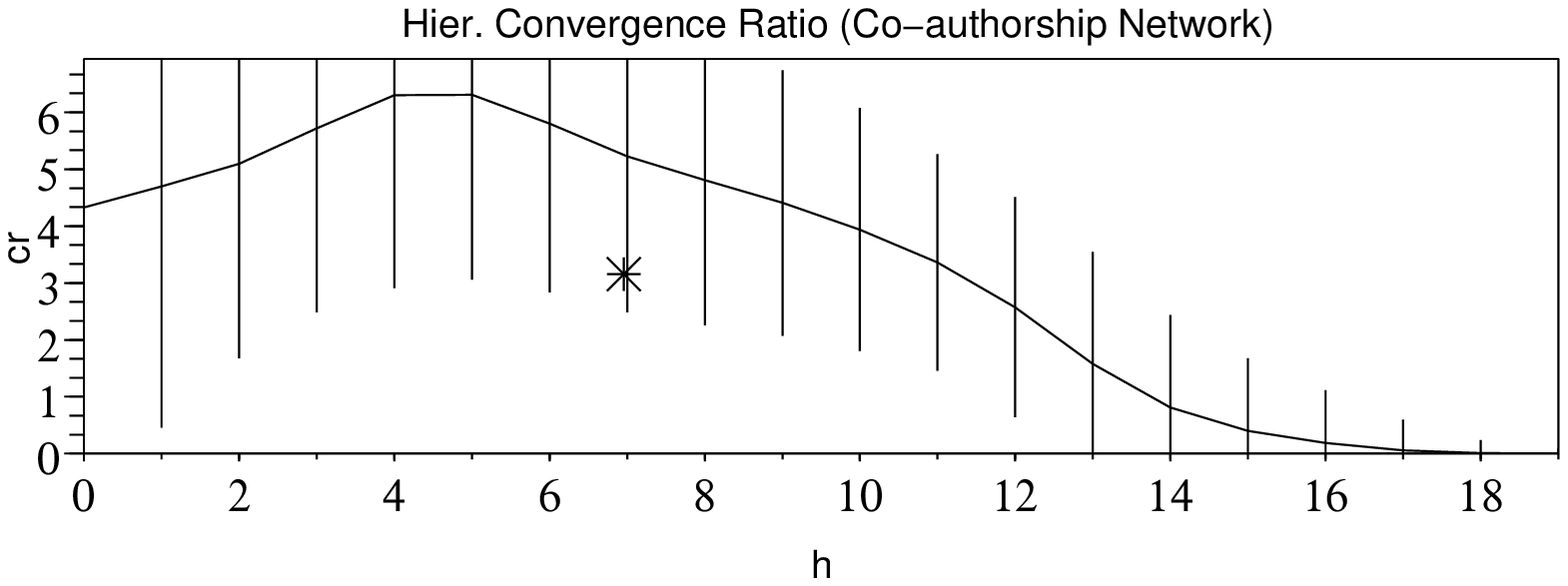}
\caption{Convergence Ratio measurements for the considered
networks.~\label{fig:sim7}}
\end{figure}

Figures \ref{fig:sim1} to \ref{fig:sim7} present the concentric
measurements distributions obtained for the three networks while
considering all the nodes. The asterisks indicate the position of the
average shortest path between any pair of nodes, which are included in
order to provide a reference for the hierarchical analysis.

Figure~\ref{fig:sim1} shows the concentric number of nodes (average
$\pm$ standard deviation) obtained for the considered networks. All
curves are characterized by a peak. Interestingly, the collaboration
network presents a considerably smoother curve and wide peak when
compared with the simulated models. In addition, its high values of
standard deviation suggest a wide variation of concentric features
among the nodes. The values of concentric node degrees, shown in
Figure~\ref{fig:sim2}, are similar to the respective measurements of
the concentric number of nodes.

The inter-ring degree curves, shown in Figure~\ref{fig:sim3}, are
monotonically decreasing after the first ring. While such curves for
the BA and ER model are clearly distinct, the curve for the
collaborative network shows a mix of both behavior. The collaborative
network curve begins with a constant value, like for the ER model, and
then decreases in a smooth fashion, like the curve for the BA
model. The curves obtained for the BA case show a peak at the first
ring, which is a consequence of the high chance of finding a hub at
that level. However, that characteristic seems not to be present on
the collaborative network. An explanation for this effect is that the
average topological distribution of hubs in the collaborative network
and BA model are very different.

The results for intra-ring degree, shown in Figure~\ref{fig:sim4}, are
very similar to the concentric number of nodes measurement,
characterized by a peak, except for the collaborative network, which
presents a wider peak centered at the left hand side of the graph.

Figure~\ref{fig:sim5} shows the values of concentric common degree for
the considered networks. These distributions are characterized by a
decreasing curve starting at the first level. Generally, these curves
are similar to those obtained for the inter-ring degrees, except that
the present curves are wider and the collaborative network has a well
defined peak at first level. Another observation is that the average
concentric common degree tends to be higher at the initial concentric
levels, which is a consequence of the fact that the largest hubs
present in the BA model tend to be reached sooner, providing bypasses
to the other nodes and therefore left-shifting the the peak and
reducing the number of concentric levels. This is the main reason why
the peak in the BA networks tends to be displaced to the lefthand side
than in the random network. As with the inter-ring degree, the
distribution of concentric degree for the collaborative network
results in a mix of the characteristics of the curves for both models,
supporting the evidence that the topological location of the hubs
tends to be more widespread than those in the BA model.

As shown in Figure ~\ref{fig:sim6}, the concentric clustering
coefficients curves are very distinct among both simulated models. The
curves for the ER model present a fast increase in value, followed by
a plateaux and then a rapid decrease. In fact, the nodes at each ring
of those networks are characterized by low interconnectivity. The
concentric clustering coefficient curves obtained for the BA and the
collaborative network, present much higher values and involve a
sharper peak. In addition, the curve for the co-authorship network
tends to present another peak along the last levels.

The convergence ratios obtained for each of the considered network
models, shown in Figure~\ref{fig:sim7}, yielded the most distinct
curves among the simulated models and collaborative network. The
curves for the BA and ER models are characterized by similar behavior
among themselves and a peak at the last levels (except for the regular
models), along which the concentric expansion tends to saturate,
i.e. after the peak is reached. Note also that sharper peaks tend to
be obtained for high values of $k$. The collaborative curve presents a
wider peak, with the center displaced to the lefthand side, far away
from the average shortest path. This is a consequence of the fact
that, differently of what is obtained for the BA, the hubs are reached
gradually along the concentric levels while starting from most nodes.

Indeed, as verified experimentally in~\cite{Costa:2006lr}, the
position and width of the peak of the convergence ratio is ultimately
defined by the distribution of hubs along the hierarchies.

The fact that the convergence ratios obtained for the co-authoship
networks, shown in Figure~\ref{fig:sim7}, tended to be relatively uniform
along the hierarchies indicates that the hubs are not highly
interconnected. In other words, the hubs tend to cover different
portions of the network. If we understand that the hubs are more
likely to correspond to leader scientists, it can be inferred from the
convergence ratios results that the multidisciplinarity would be
ultimately implemented by the respective co-authors connected to each
hub.

Among all the considered measurements, the concentric common degrees
and concentric clustering coefficients were found to provide the
most distinct curves for each network, revealing more information
about the distribution of hubs and the interconnectivity along the
concentric levels. Because the collaborative network curves have the
highest values of standard deviation, their individual nodes may have
distinct concentric features, and can be grouped into clusters of
similar features. The remainder of this section presents the results
obtained by application of an agglomerative hierarchical clustering
algorithm using the data obtained by the concentric clustering
coefficient and convergence ratios. The following graphs were
obtained, showing the average $\pm$ standard deviation of the
concentric measurements obtained at each respective level in the
dendrograms. Starting at the right hand side of the tree, the nodes
are progressively merged with basis on the similarity of their
concentric clustering coefficients, yielding the taxonomical
categorization of the nodes into meaningful clusters, identified by
each branching point in the tree.

\begin{figure*}
\begin{center}
\begin{picture}(410,560)(0,-20)
\put(0,0){ \includegraphics[scale=0.8]{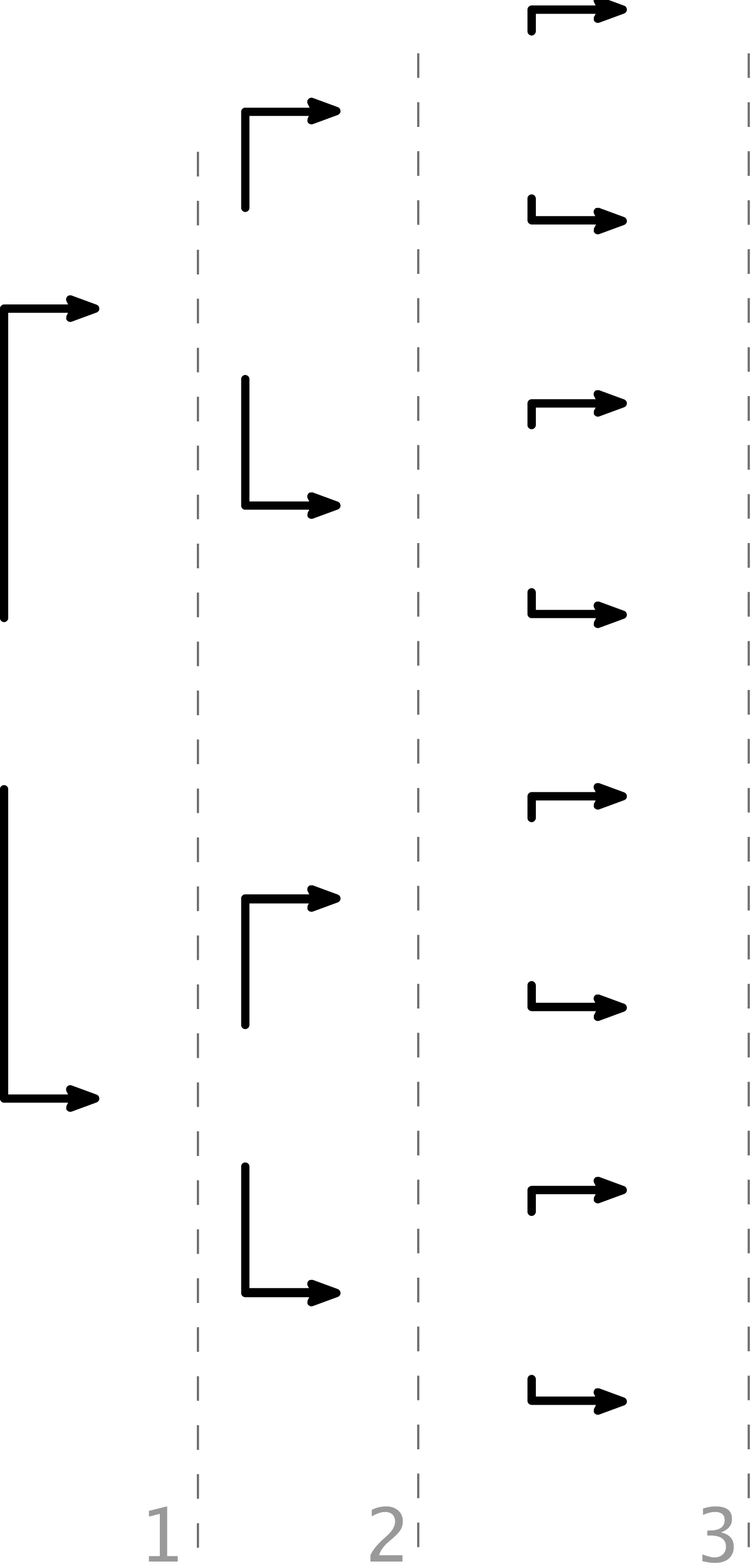}}
\begin{picture}(560,600)(20,-60)
\put(-40,255){A \includegraphics[scale=0.25]{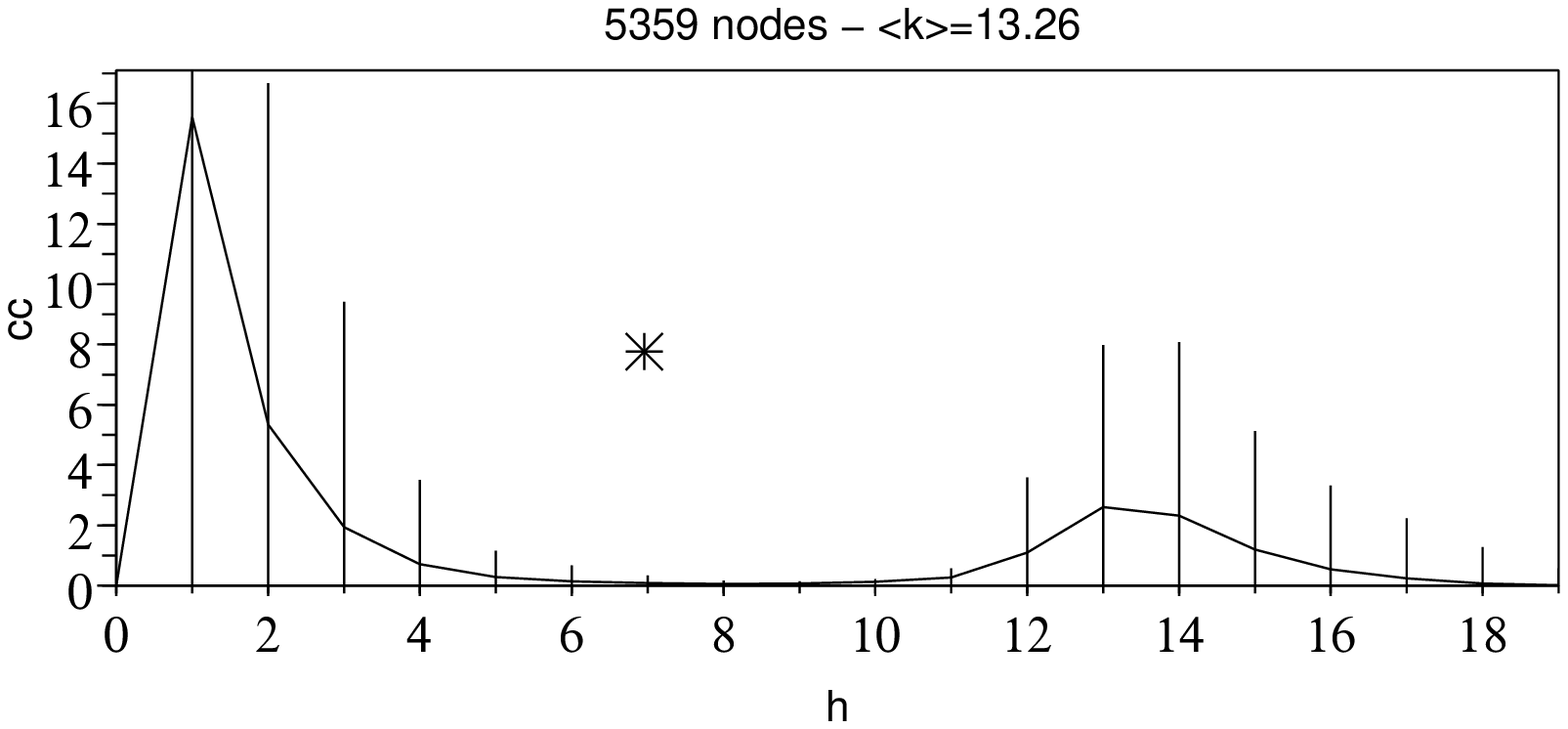}}
\put(55,415){B \includegraphics[scale=0.25]{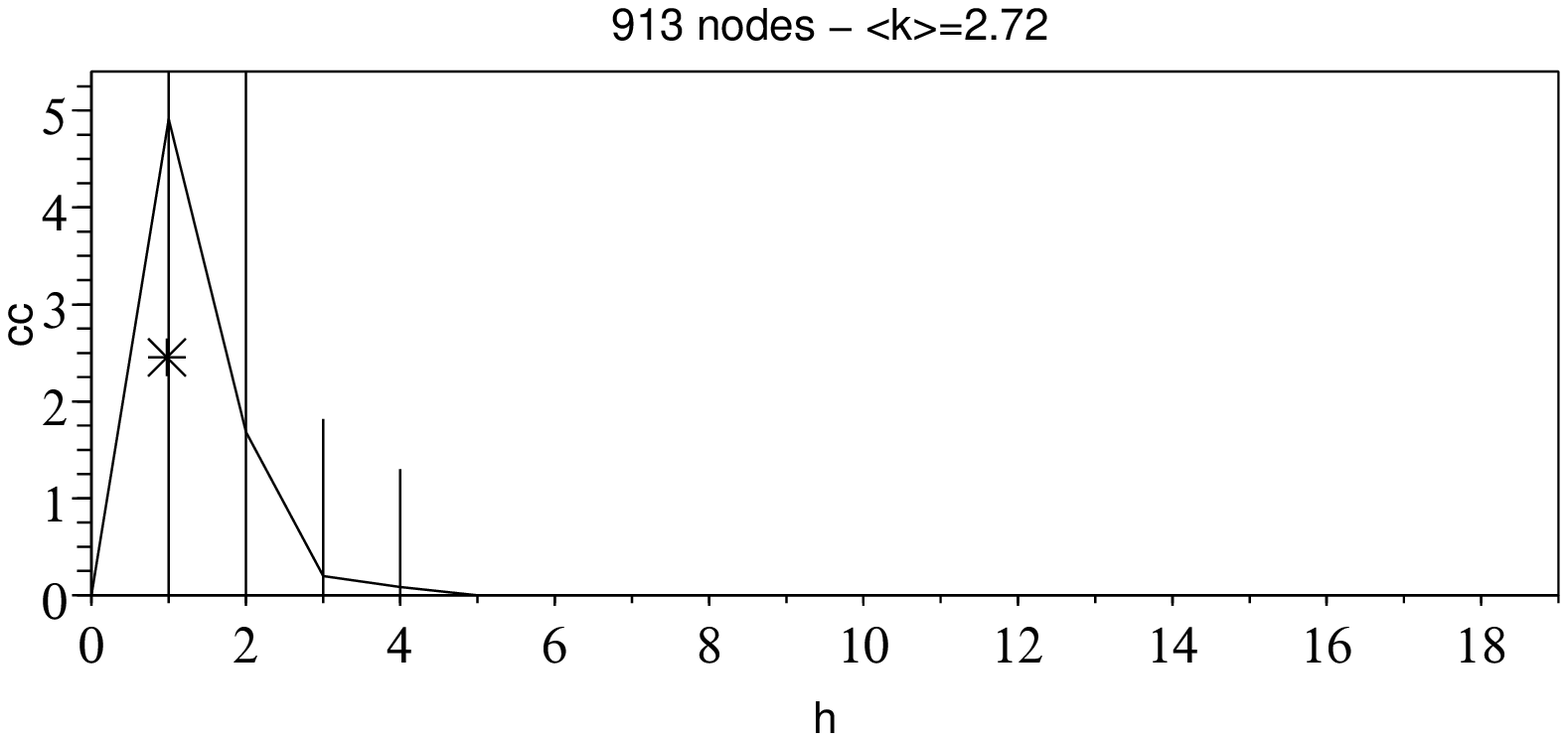}}
\put(55,100){C \includegraphics[scale=0.25]{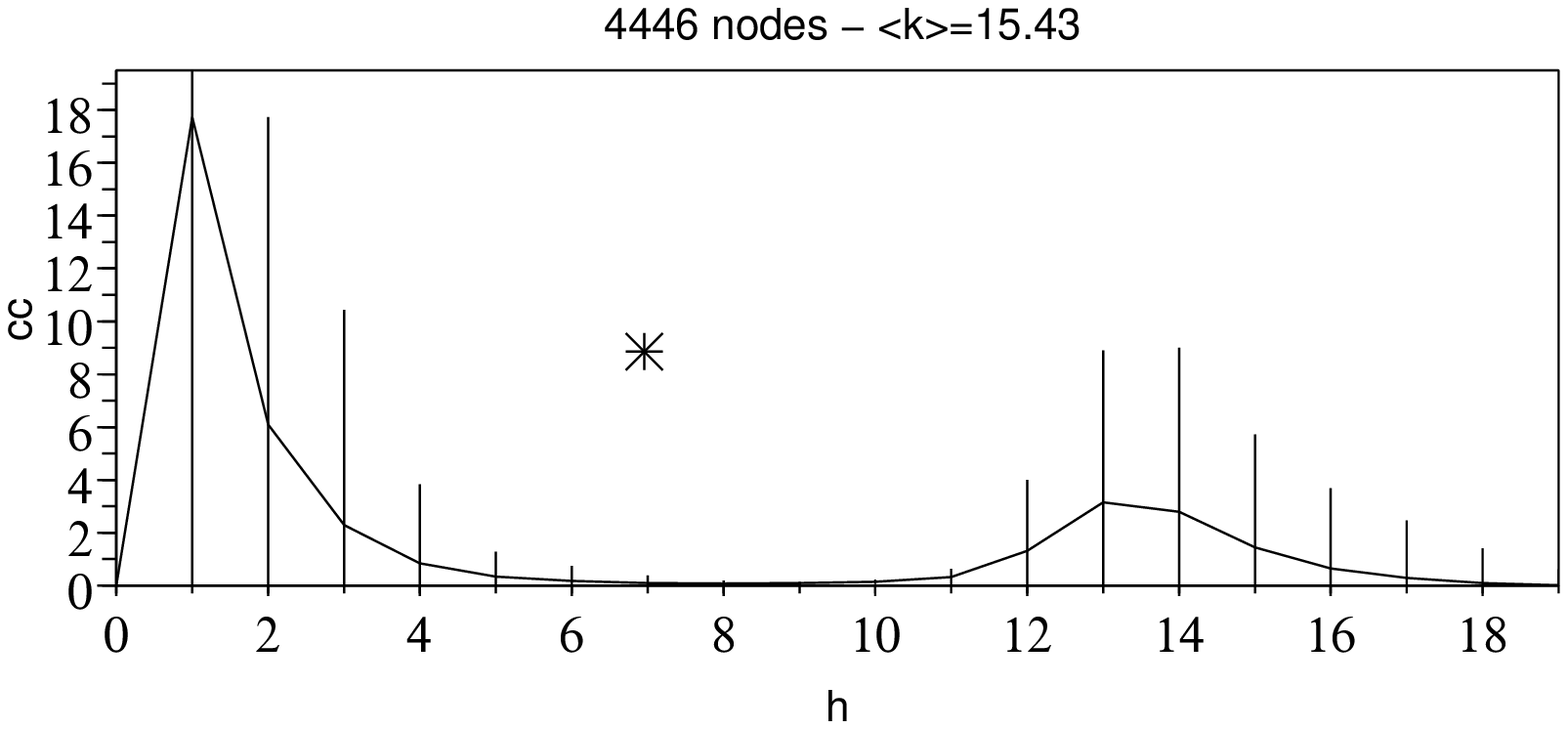}}
\put(150,485){D \includegraphics[scale=0.25]{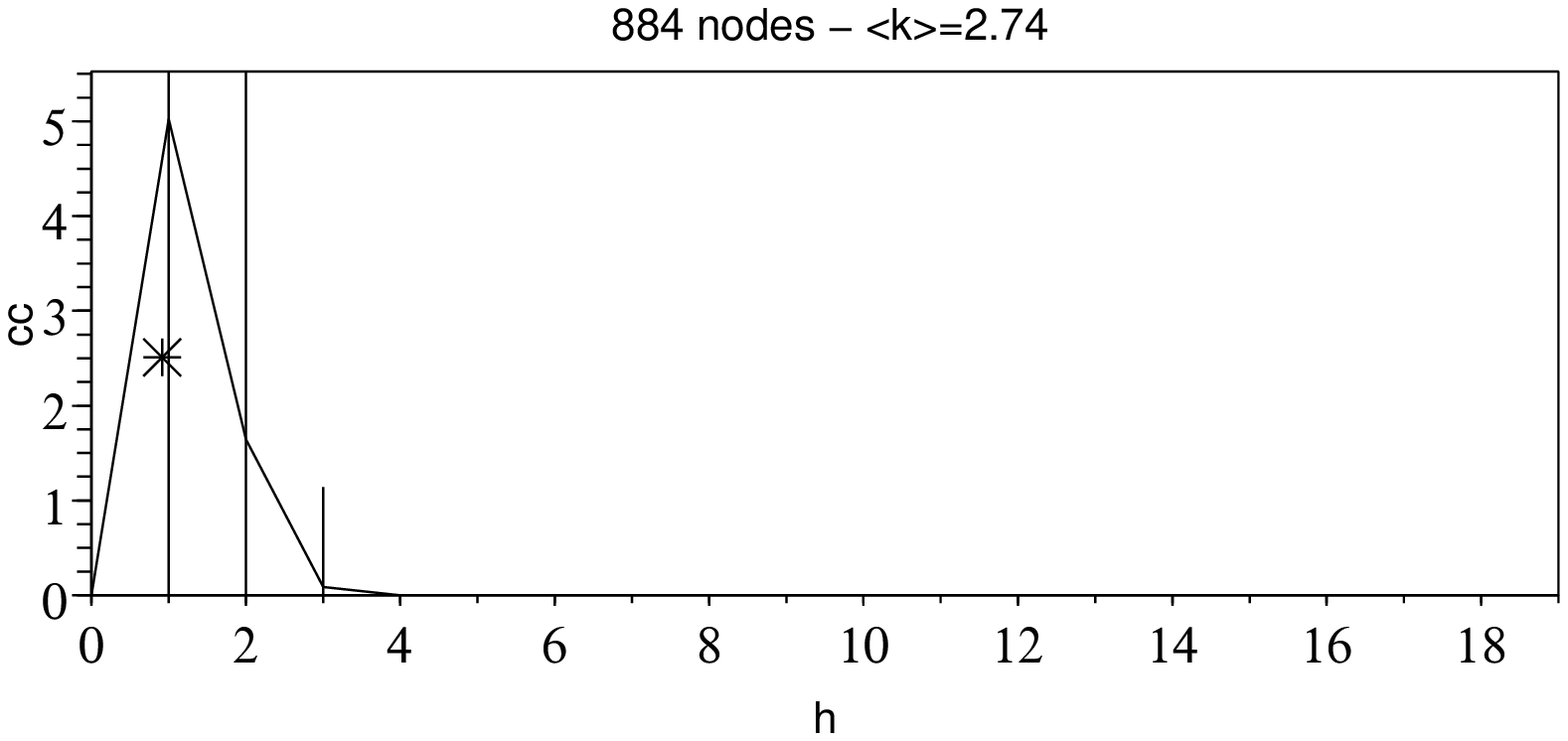}}
\put(150,330){E \includegraphics[scale=0.25]{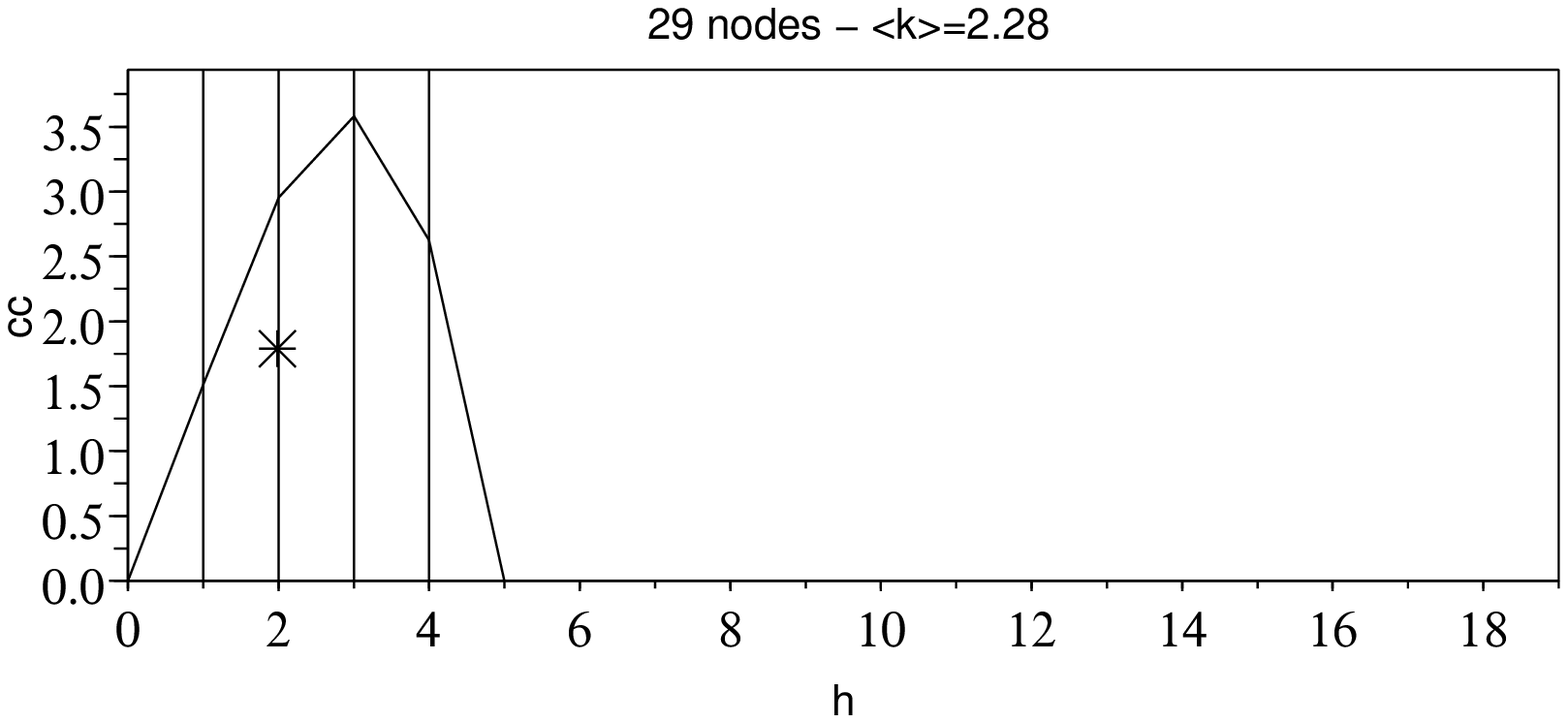}}
\put(150,175){F \includegraphics[scale=0.25]{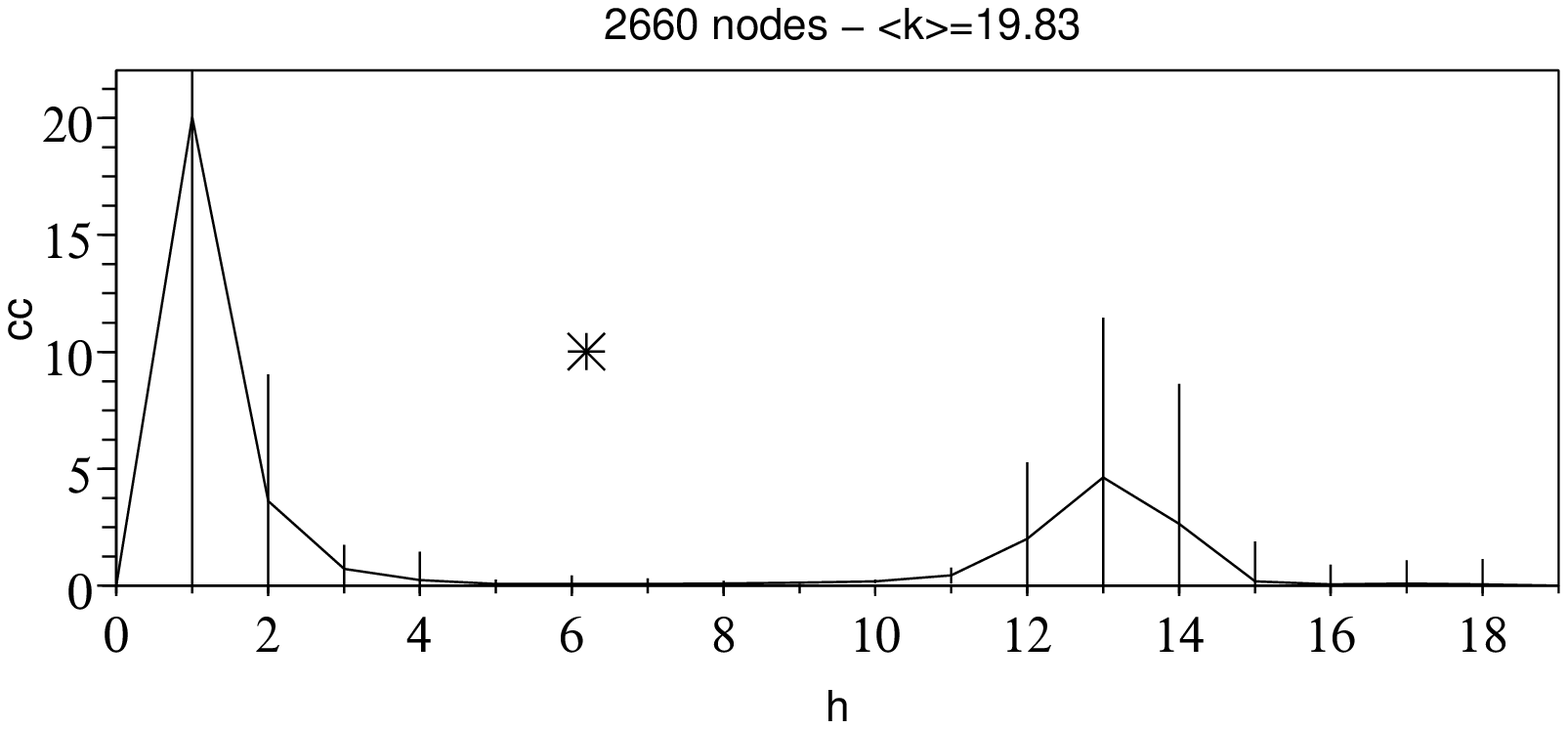}}
\put(150,20){G \includegraphics[scale=0.25]{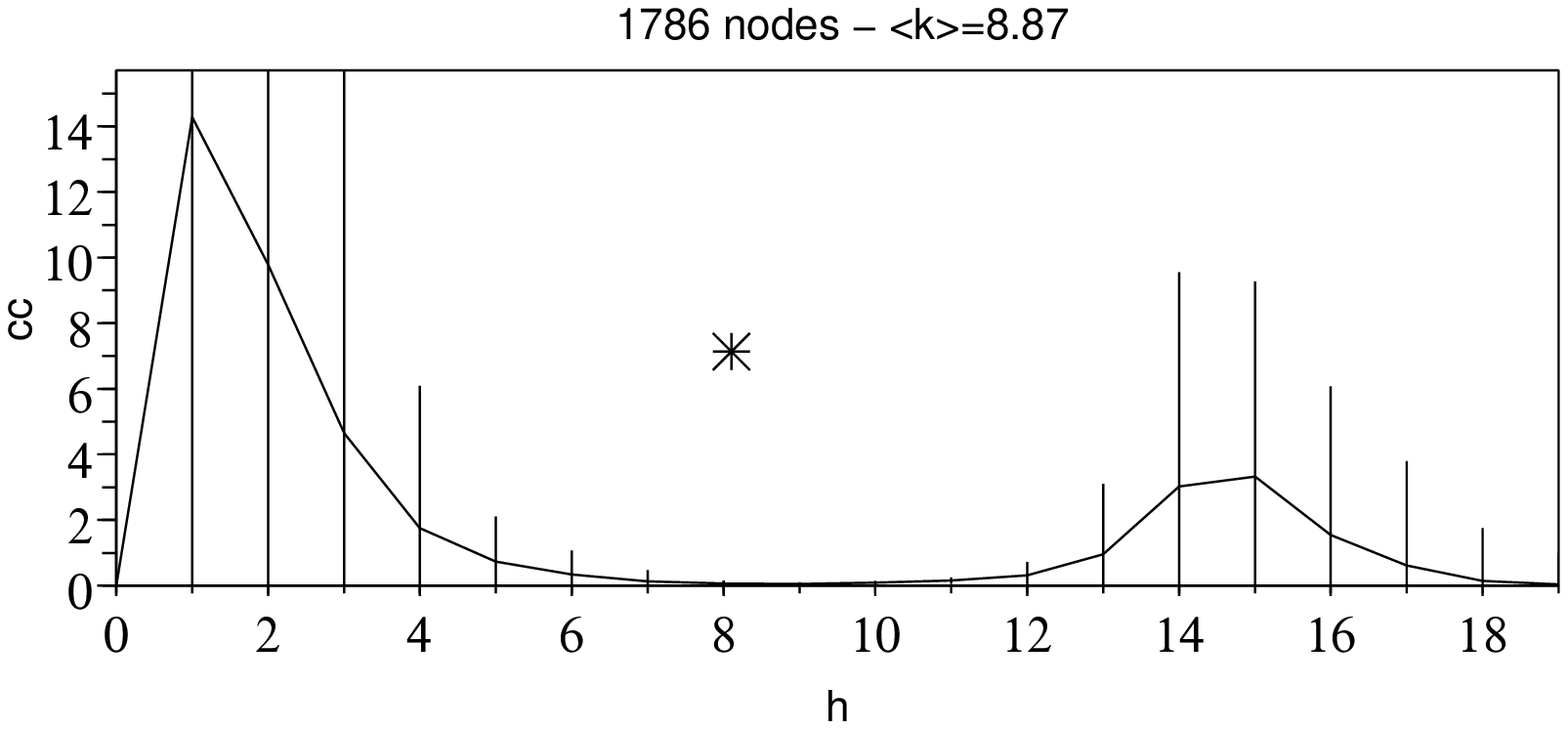}}
\put(300,525){H \includegraphics[scale=0.25]{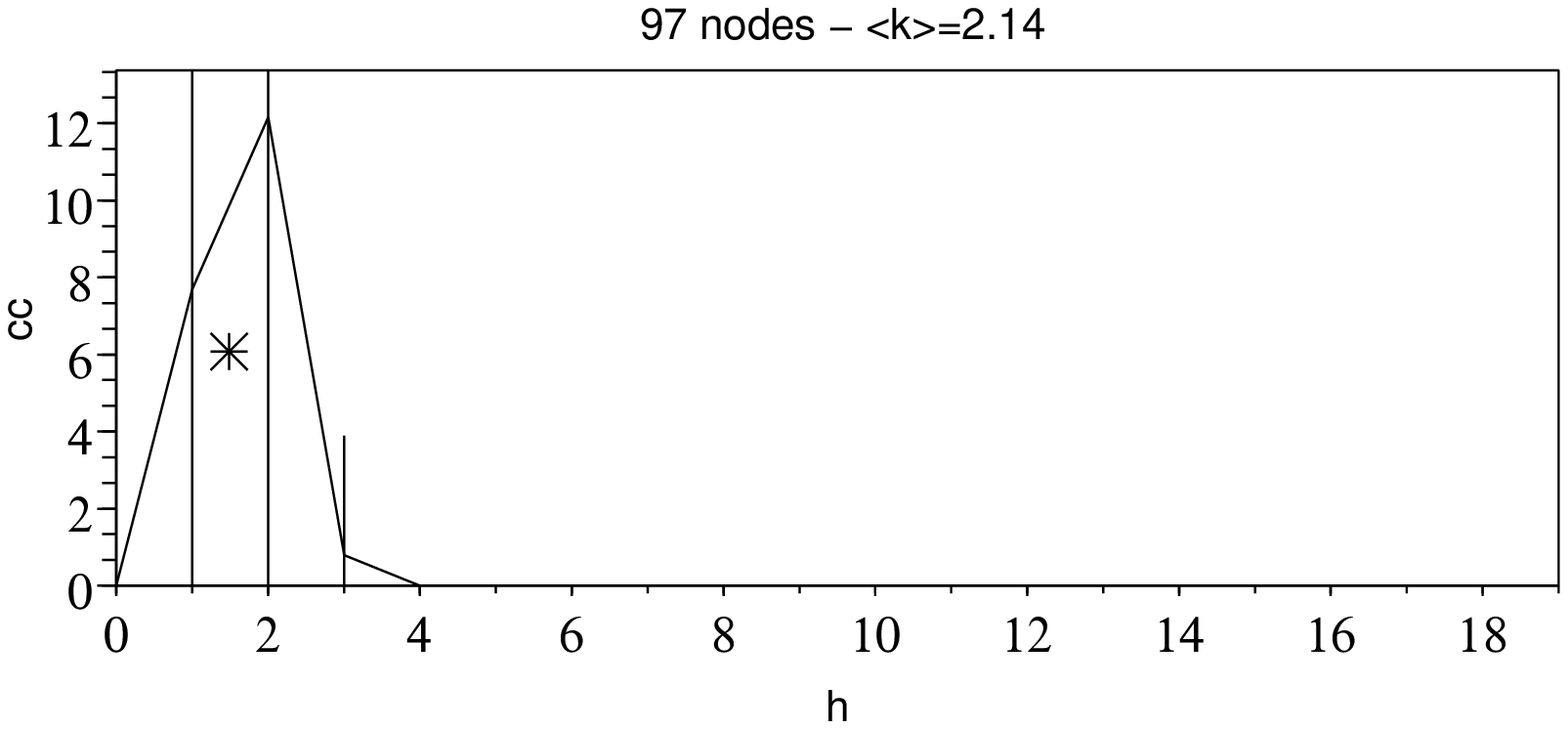}}
\put(300,445){I \includegraphics[scale=0.25]{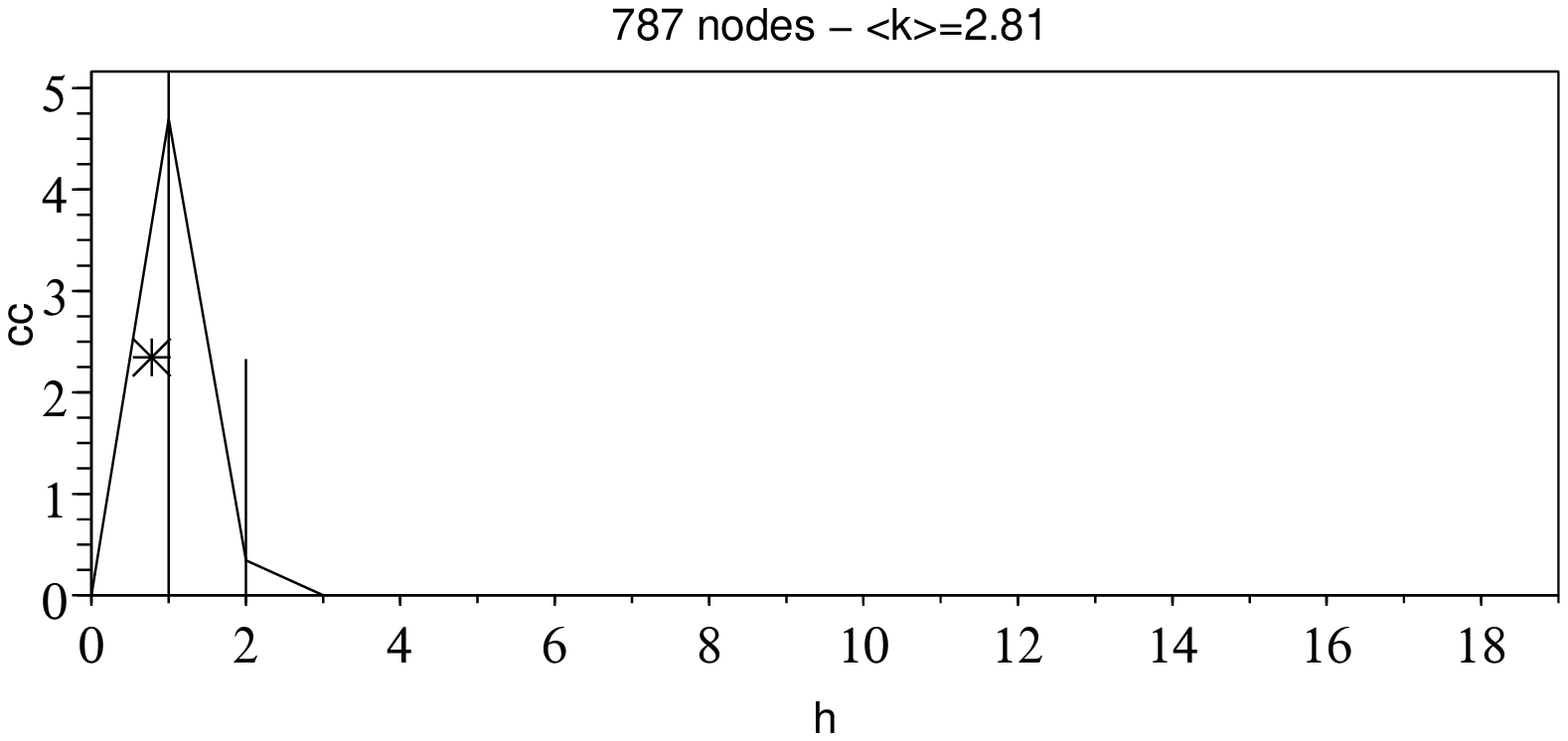}}
\put(300,370){J \includegraphics[scale=0.25]{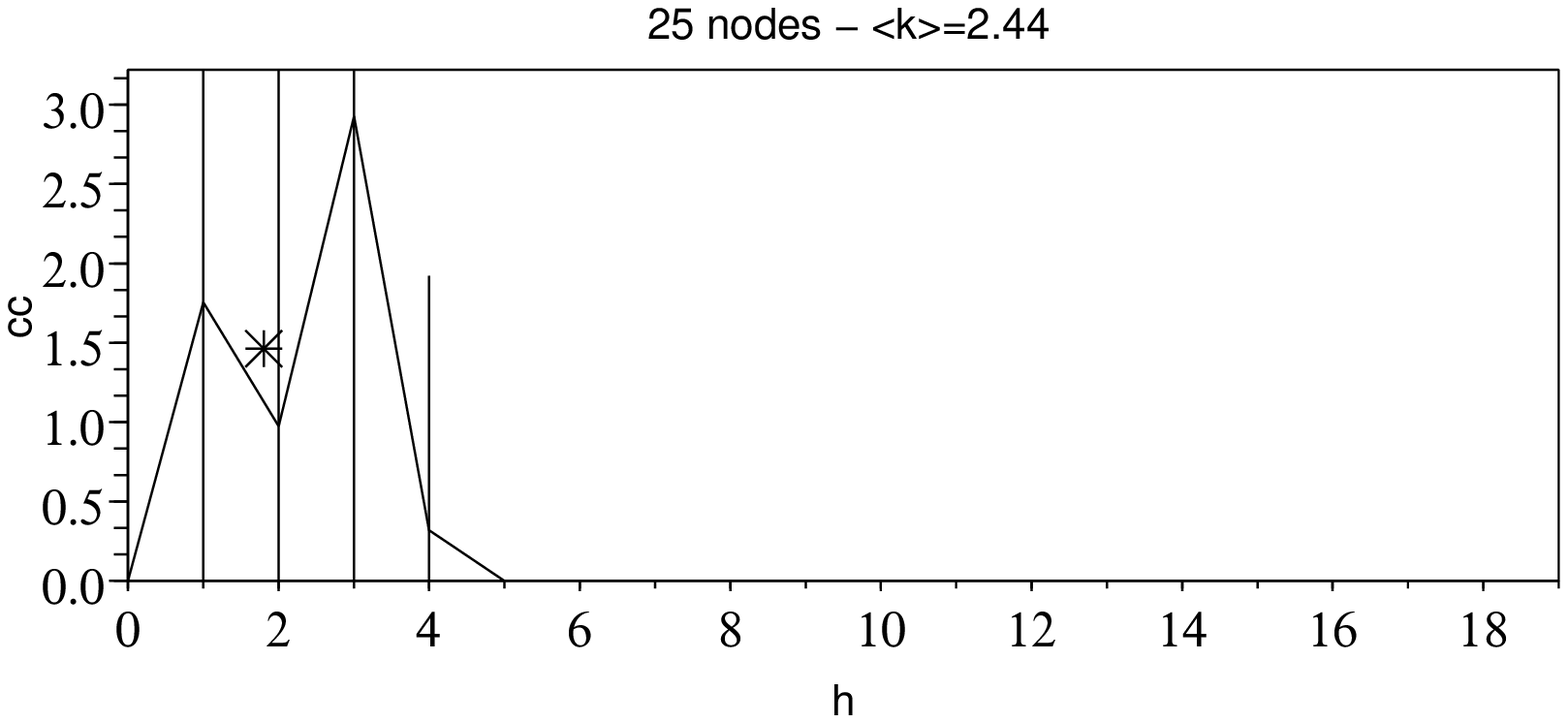}}
\put(300,290){K \includegraphics[scale=0.25]{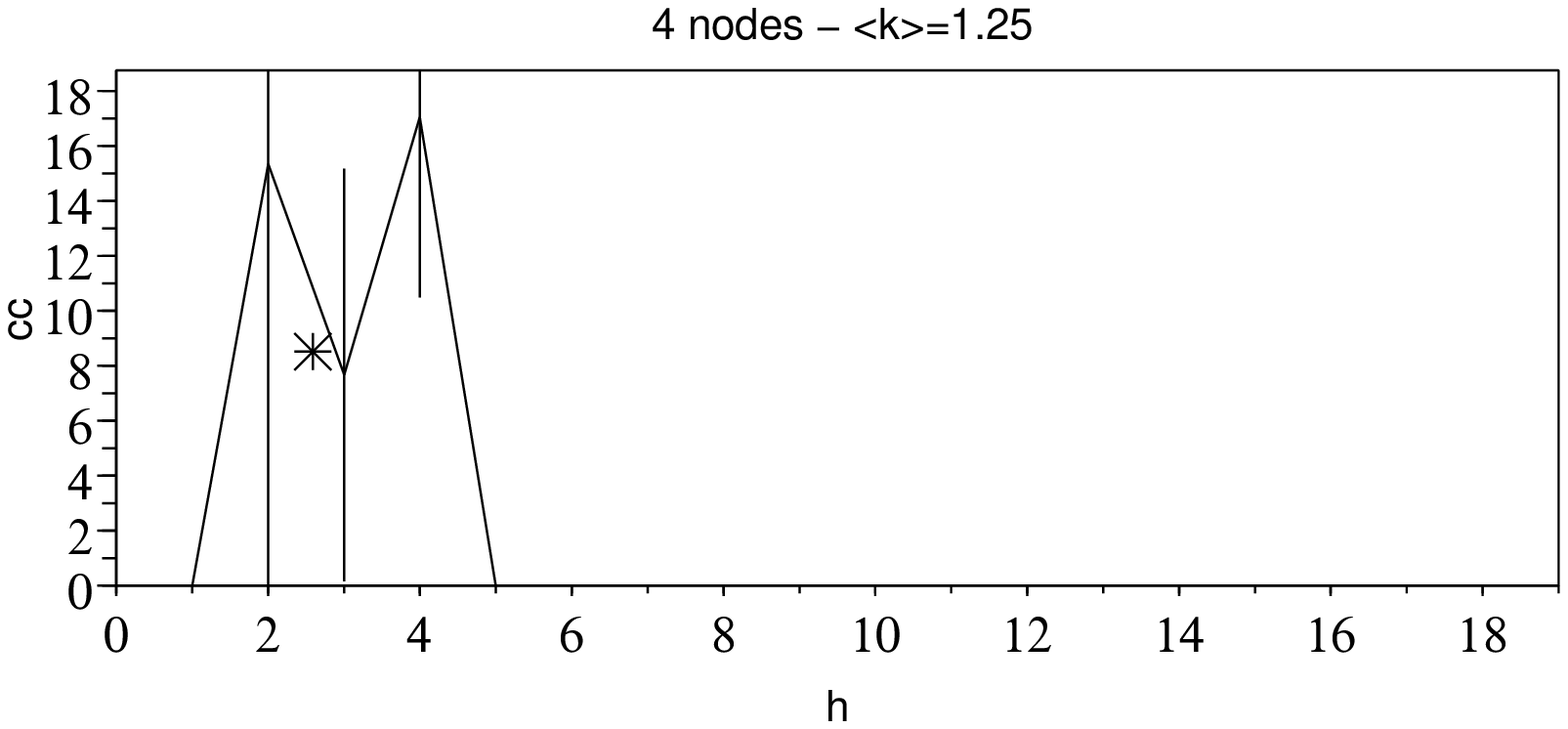}}
\put(300,215){L \includegraphics[scale=0.25]{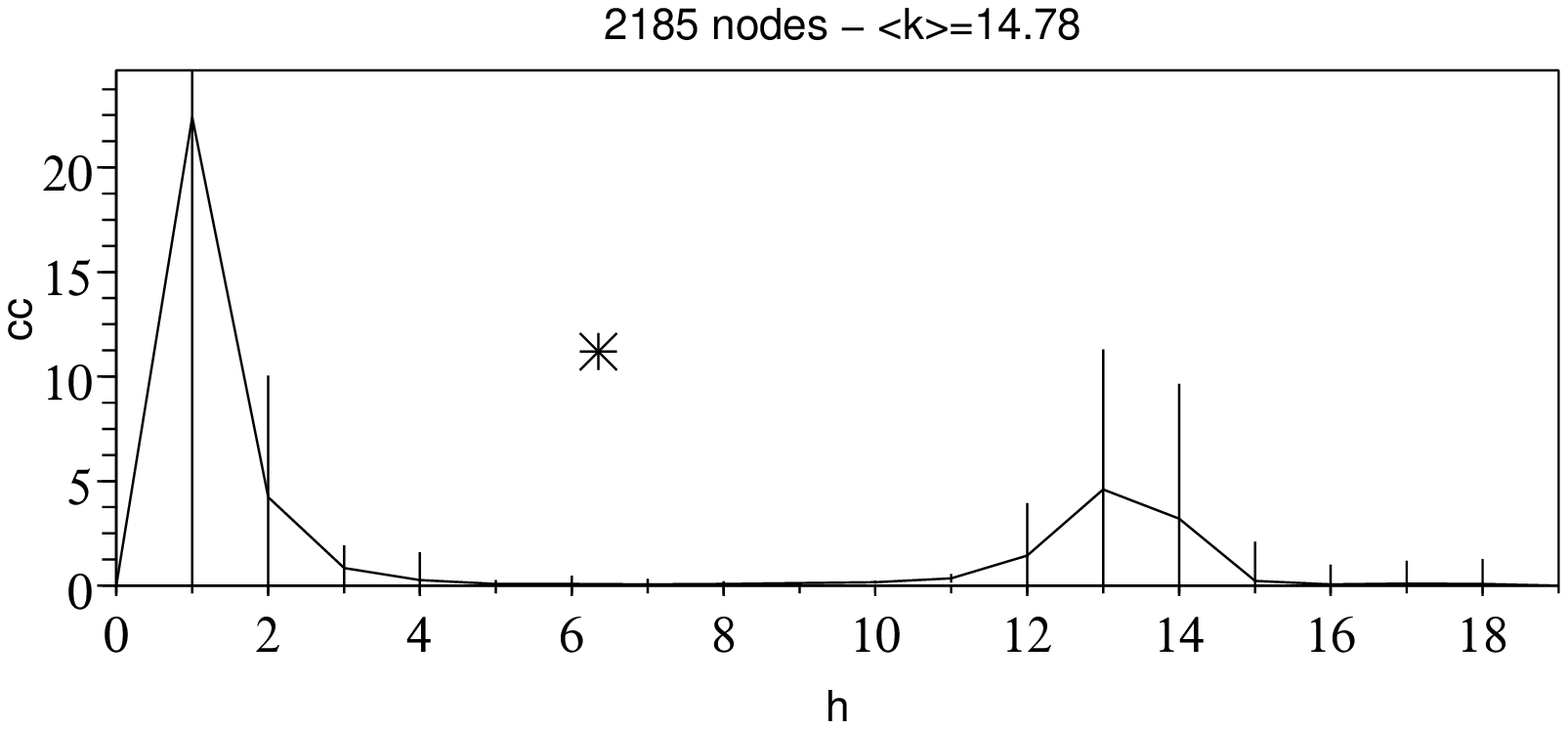}}
\put(300,135){M \includegraphics[scale=0.25]{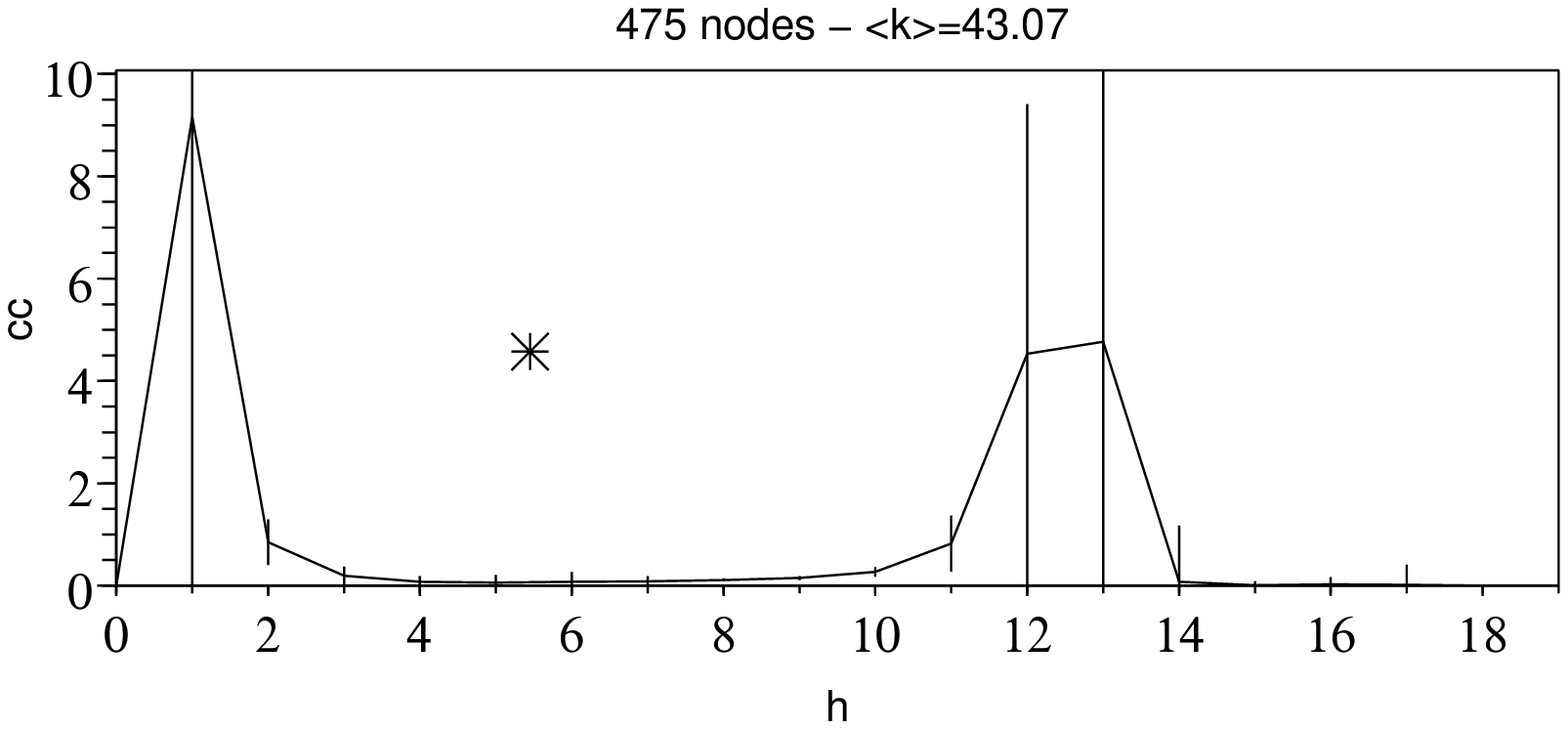}}
\put(300,060){N \includegraphics[scale=0.25]{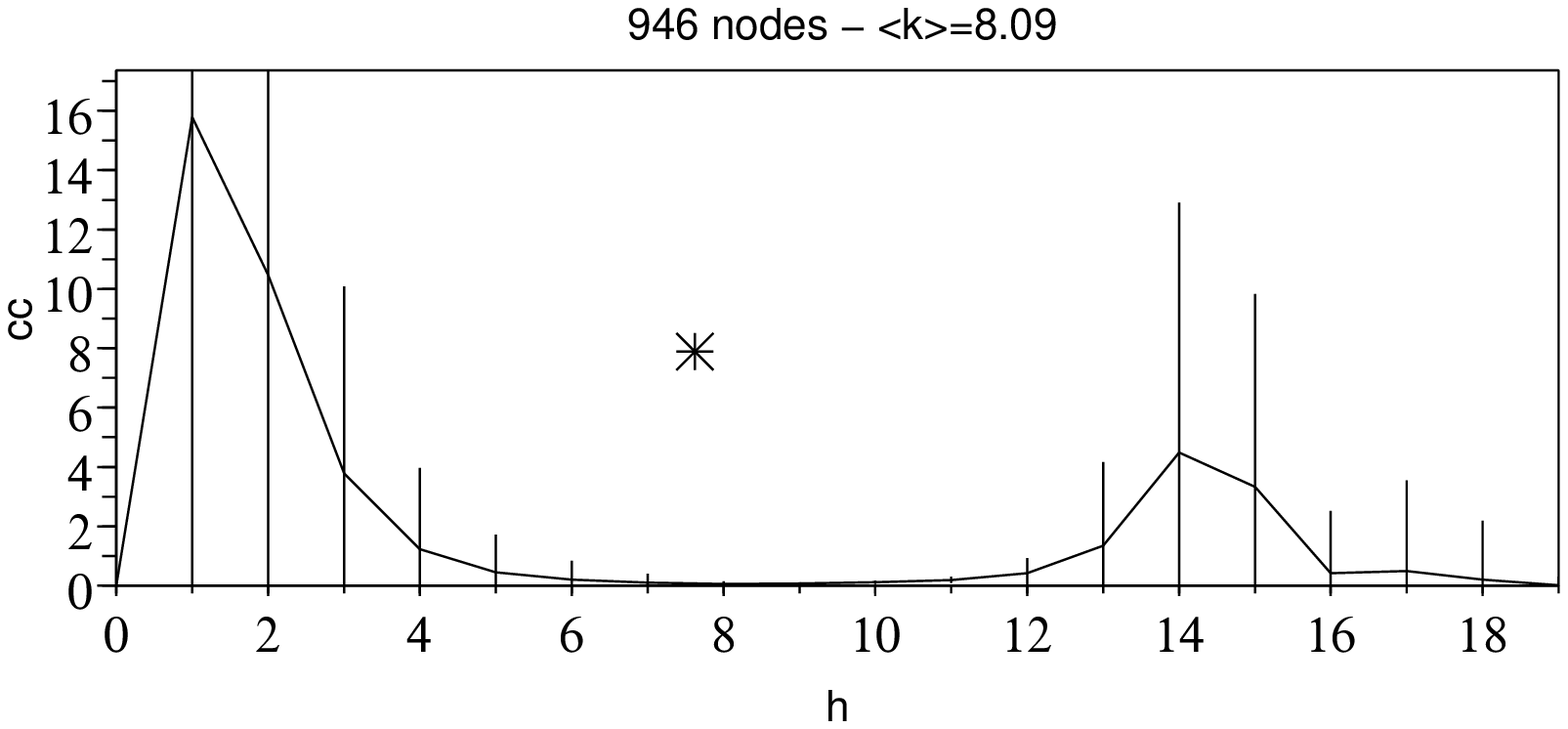}}
\put(300, -20){O\includegraphics[scale=0.25]{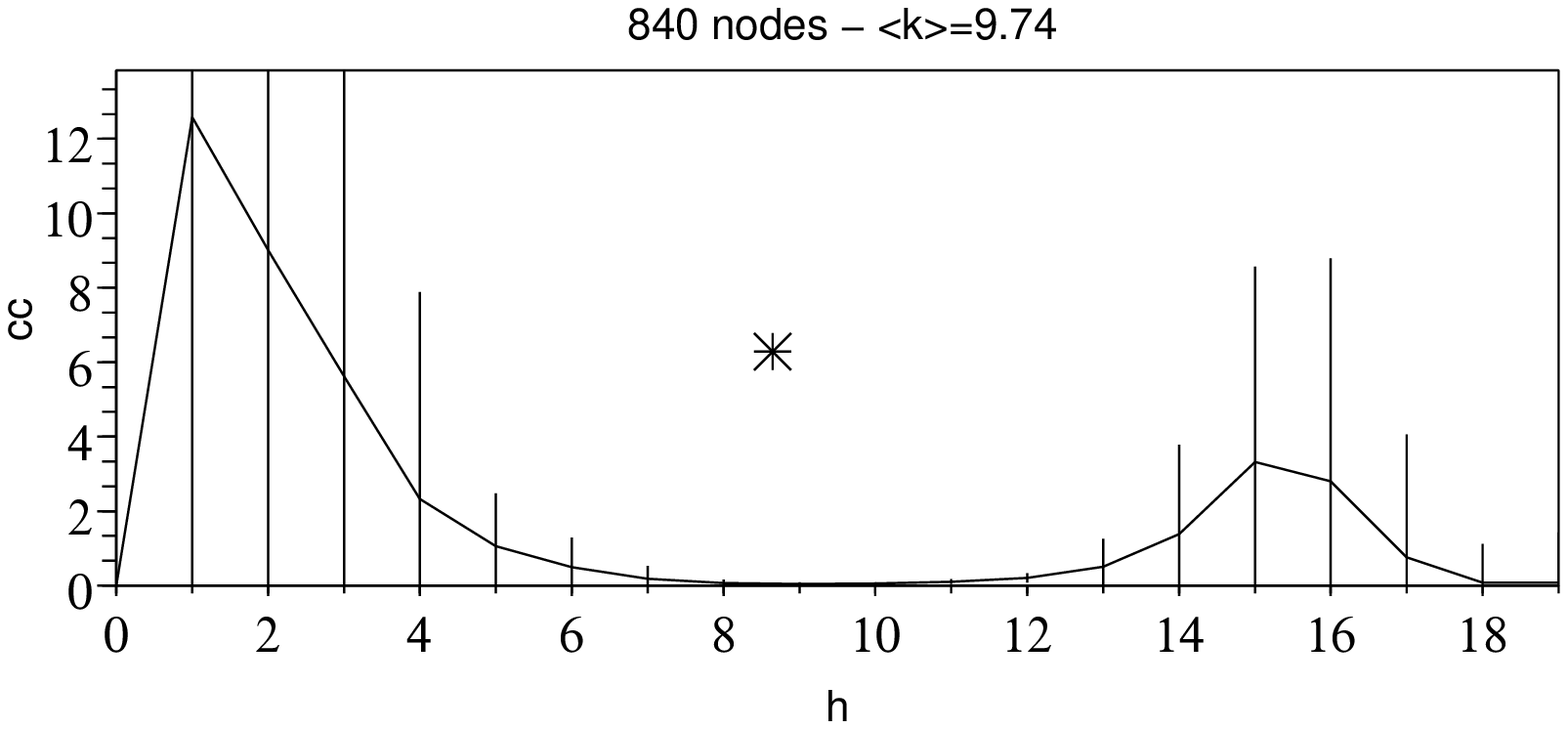}}
\end{picture}
\end{picture}
\end{center}
\caption{Graphs of the average $\pm$ standard deviation of the
concentric clustering coefficient obtained for the co-authorship
network. Only four levels of the dendogram obtained by the
agglomerative hierarchical clustering are shown.~\label{fig:tree1}}

\end{figure*}

Figure ~\ref{fig:tree1} shows the graph for concentric clustering
coefficients and the respectively obtained clusters. The mean degree
and number of nodes of each cluster are given above each graphic. As
can be seen, in the first branch point leading to the first clusters
(i.e. B and C), about $17\%$ of nodes of the network are in cluster B
and, differently from the curve for all nodes(i.e. A), the clustering
coefficient distribution shows only a peak, reveling that those nodes
have distinct hierarchical behavior when compared with those from
cluster C. Note that cluster B leads to a great variety of types of
curves, while the curve for cluster H has a peak centered at the
concentric level 2, those for cluster I are centered at level
1. Cluster E shows that about 25 nodes have very specific distribution
including two peaks centered at levels 1 and 2 (as in J) or two peaks
centered at levels 2 and 3 (as in K). The branch corresponding to
cluster C shows two basic structures (G and F), both including the
second peak, but the first one is wider for the cluster G when
compared with F.

\begin{figure*}
\begin{center}
\begin{picture}(410,560)(0,-20)
\put(0,0){ \includegraphics[scale=0.8]{Images/base.eps}}
\begin{picture}(560,600)(20,-60)
\put(-40,255){A \includegraphics[scale=0.25]{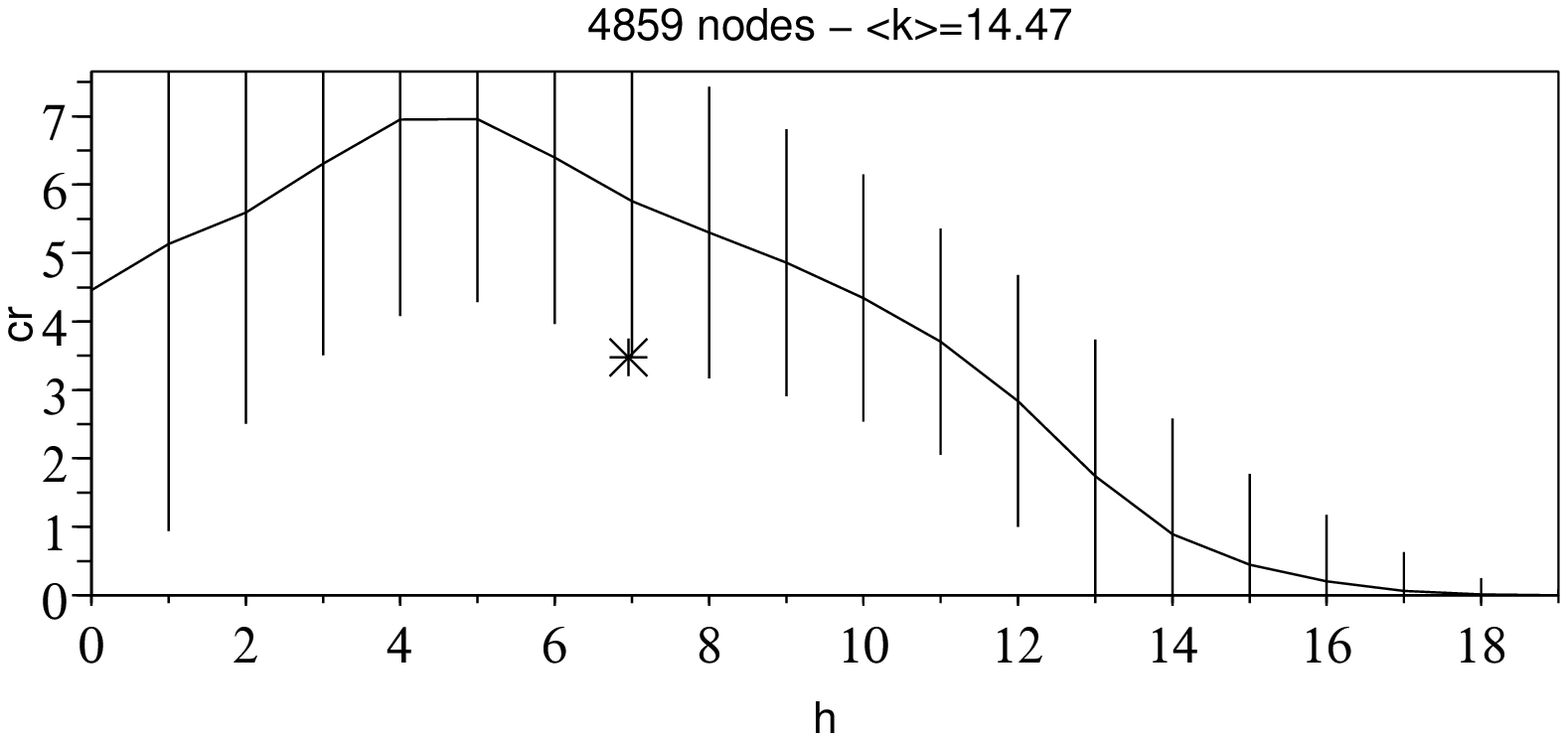}}
\put(55,415){B \includegraphics[scale=0.25]{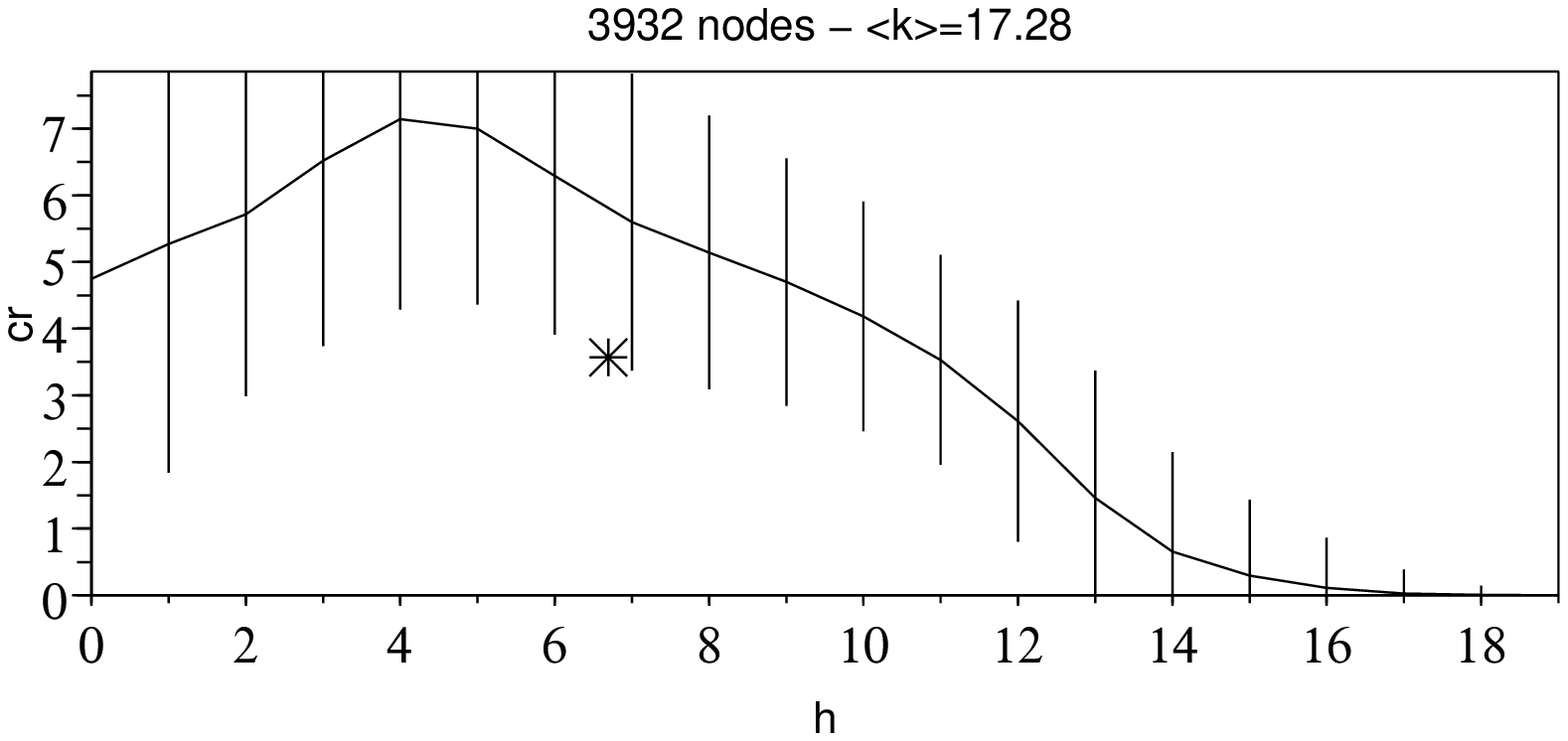}}
\put(55,100){C \includegraphics[scale=0.25]{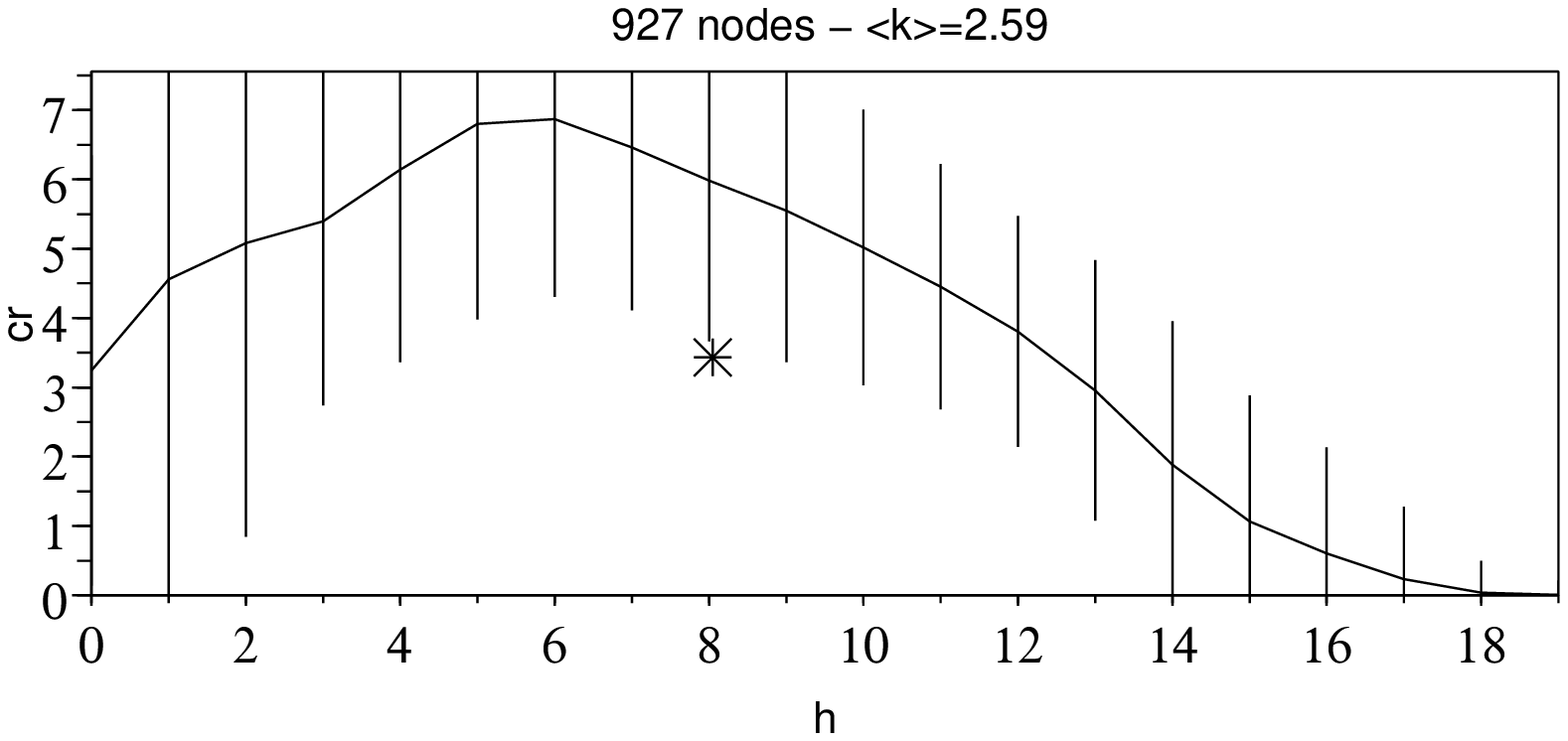}}
\put(150,485){D \includegraphics[scale=0.25]{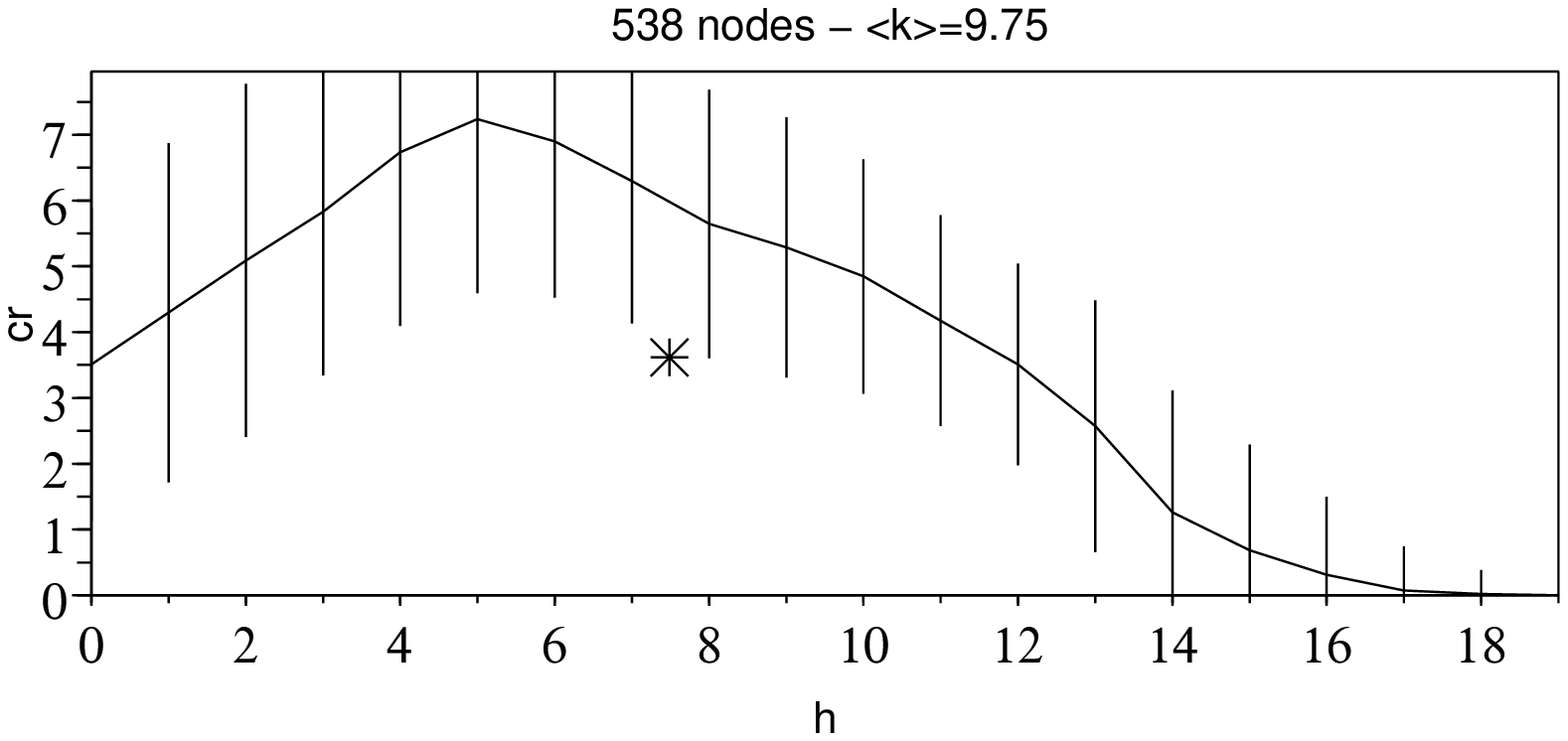}}
\put(150,330){E \includegraphics[scale=0.25]{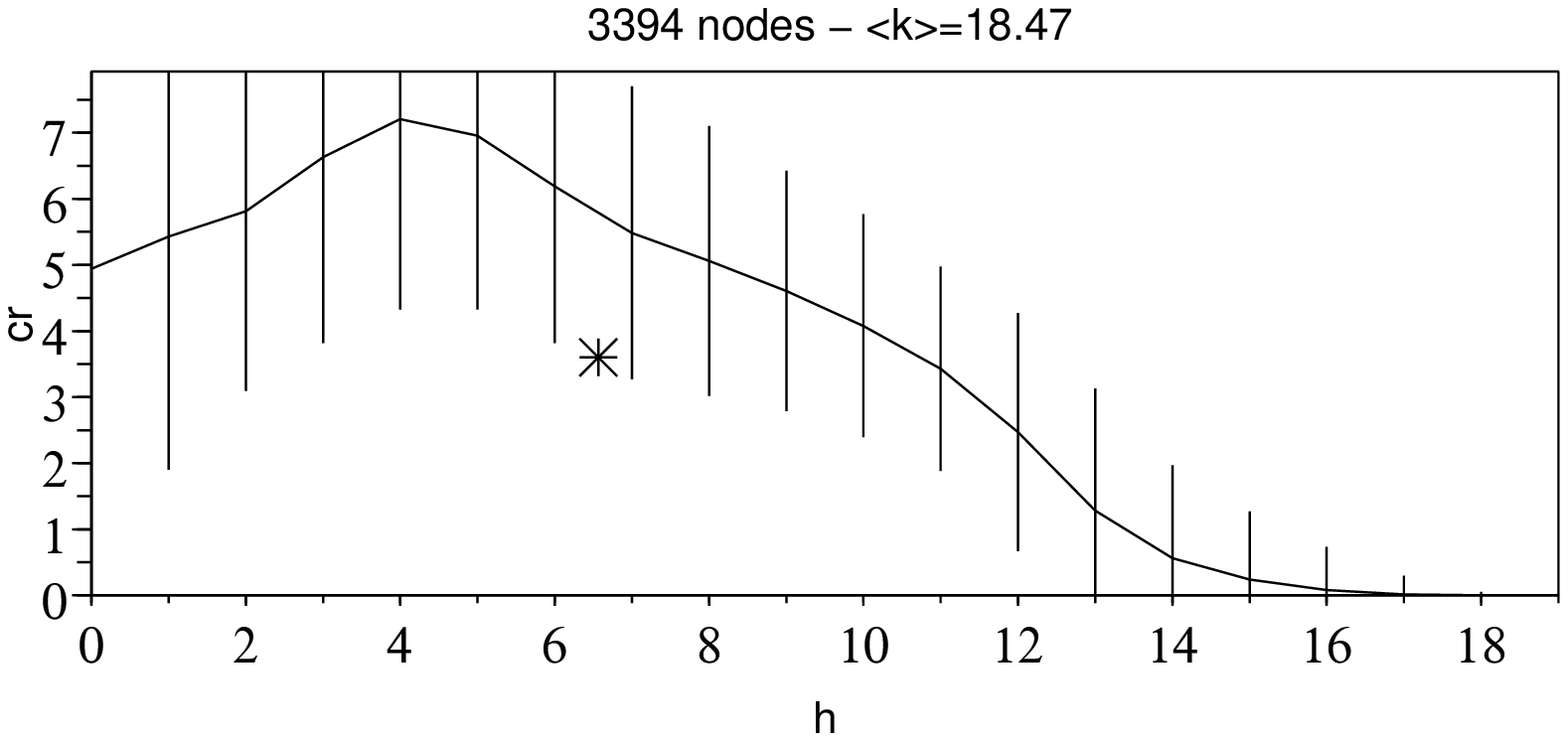}}
\put(150,175){F \includegraphics[scale=0.25]{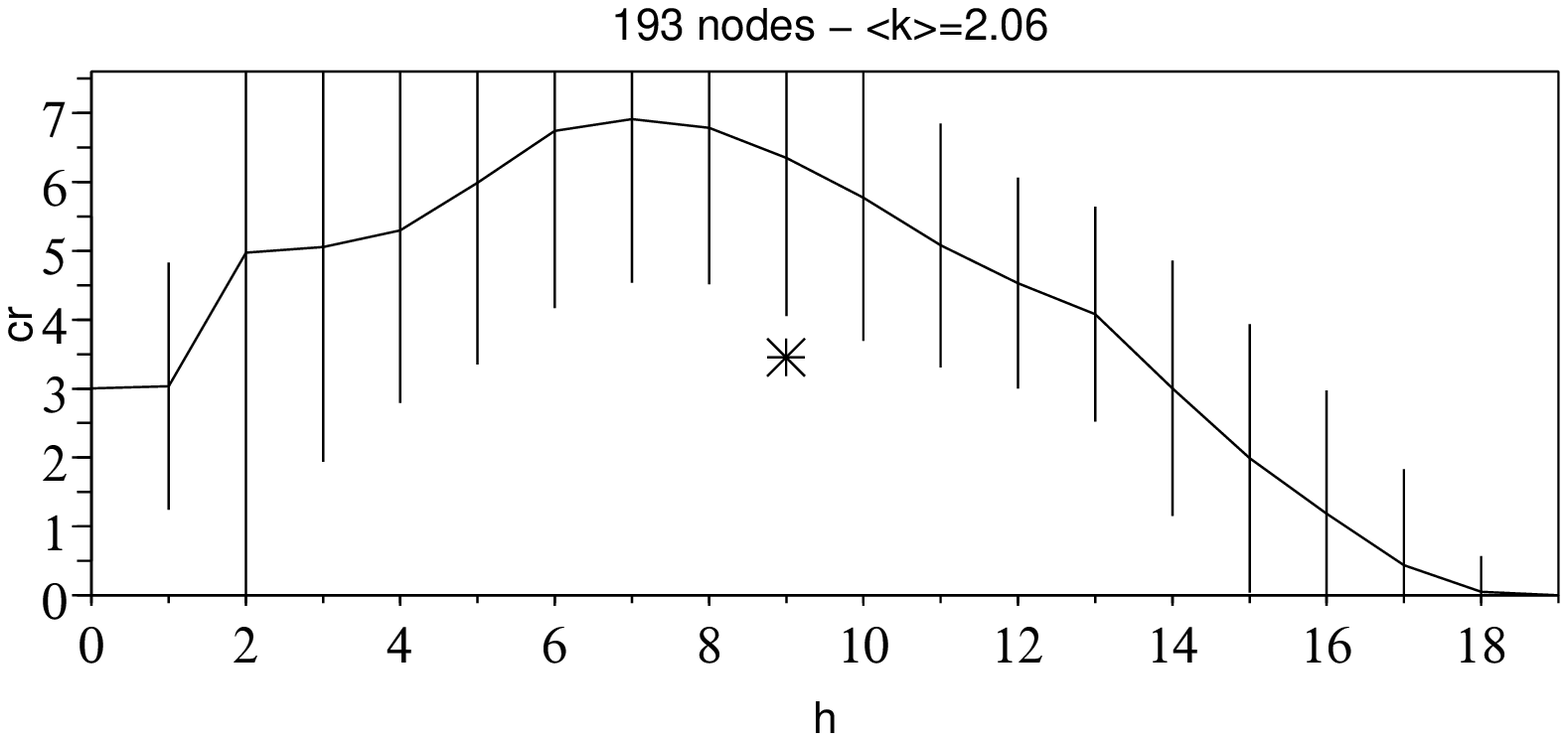}}
\put(150,20){G \includegraphics[scale=0.25]{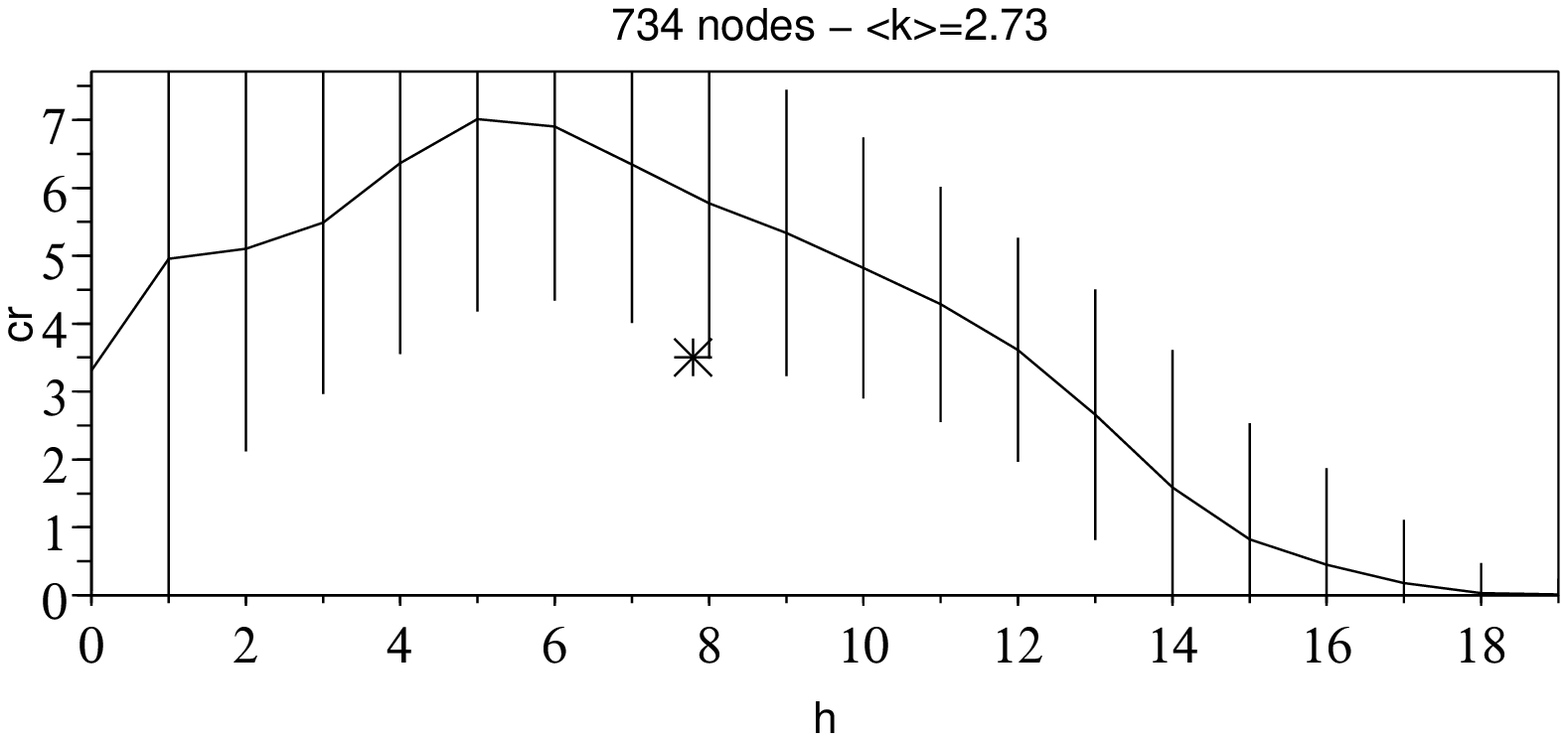}}
\put(300,525){H \includegraphics[scale=0.25]{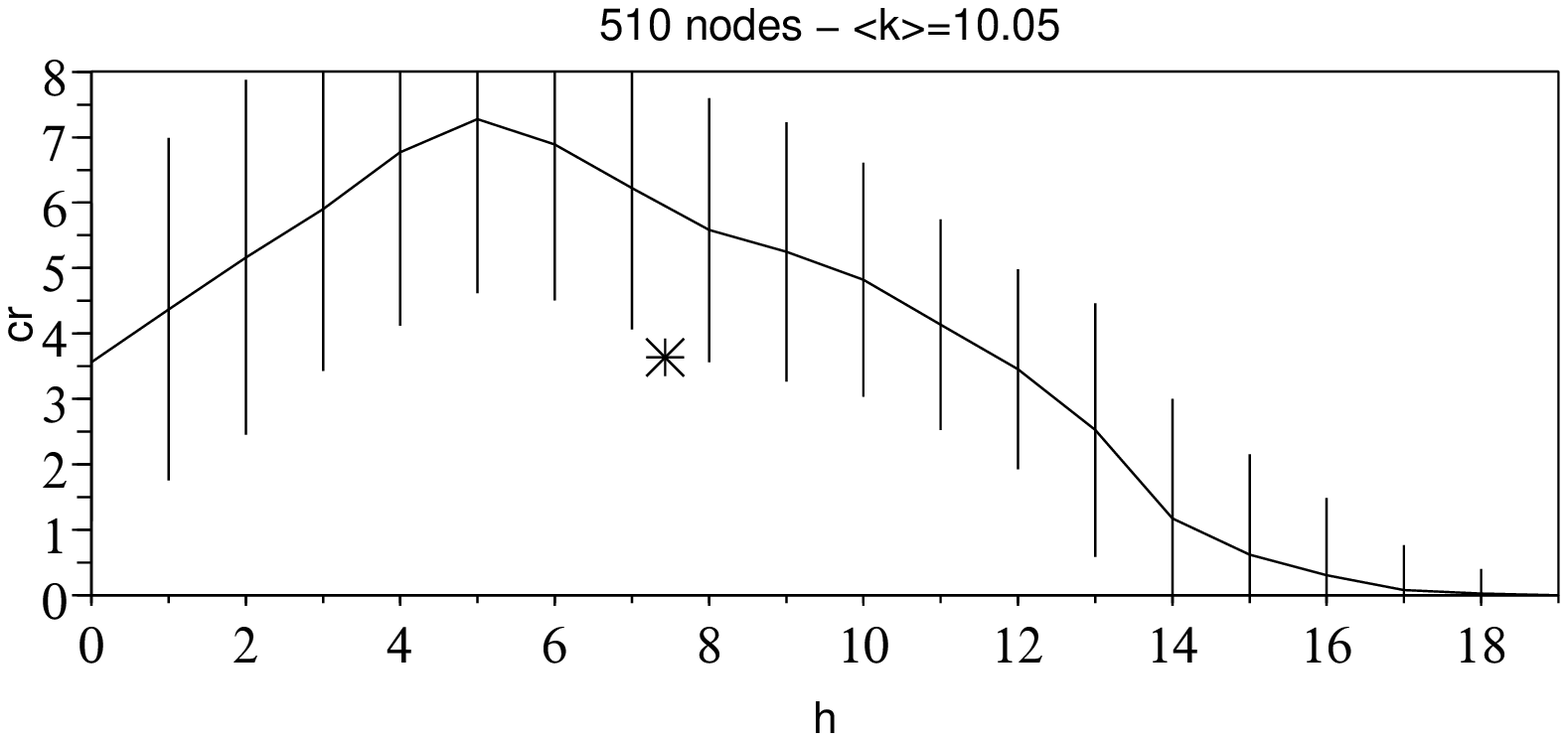}}
\put(300,445){I \includegraphics[scale=0.25]{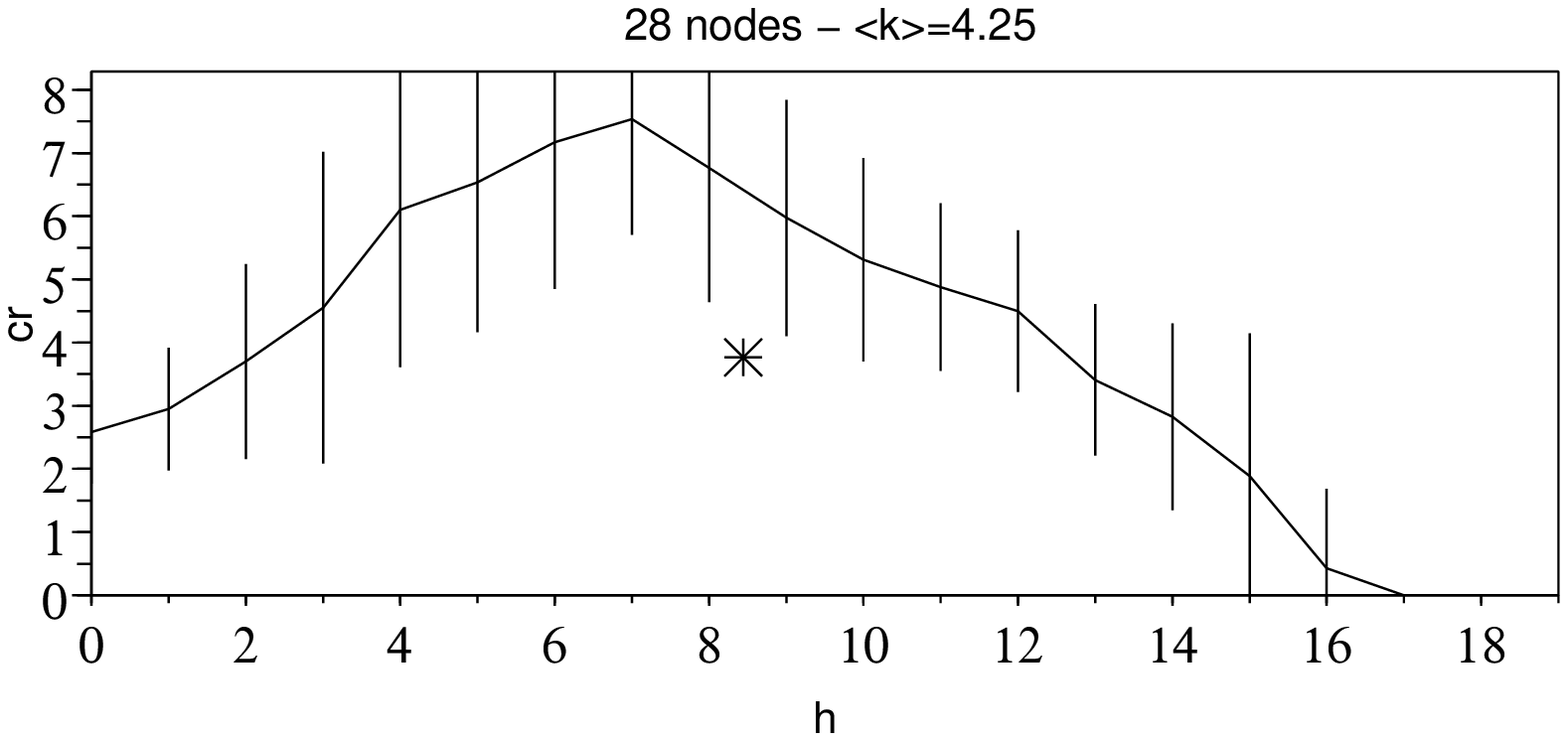}}
\put(300,370){J \includegraphics[scale=0.25]{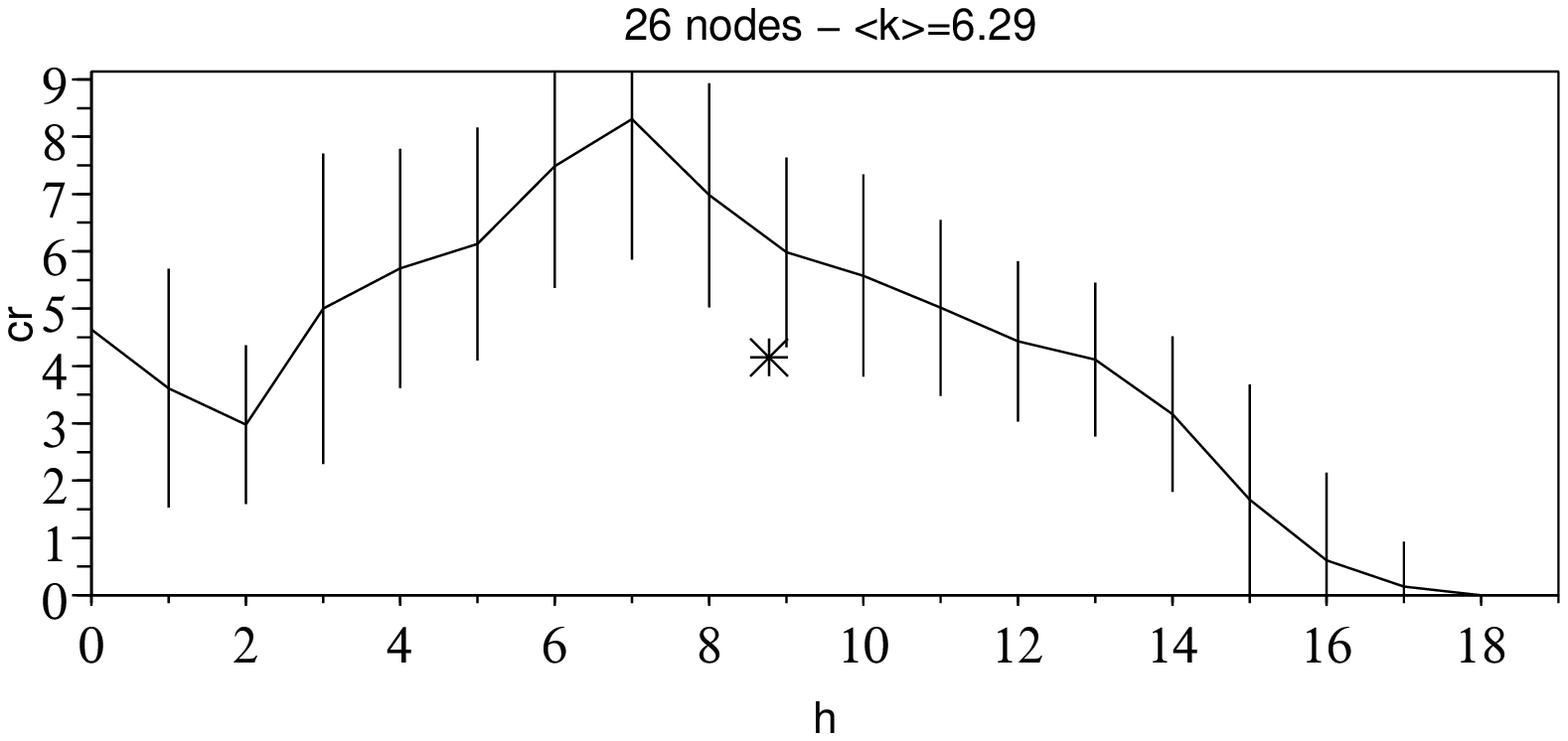}}
\put(300,290){K \includegraphics[scale=0.25]{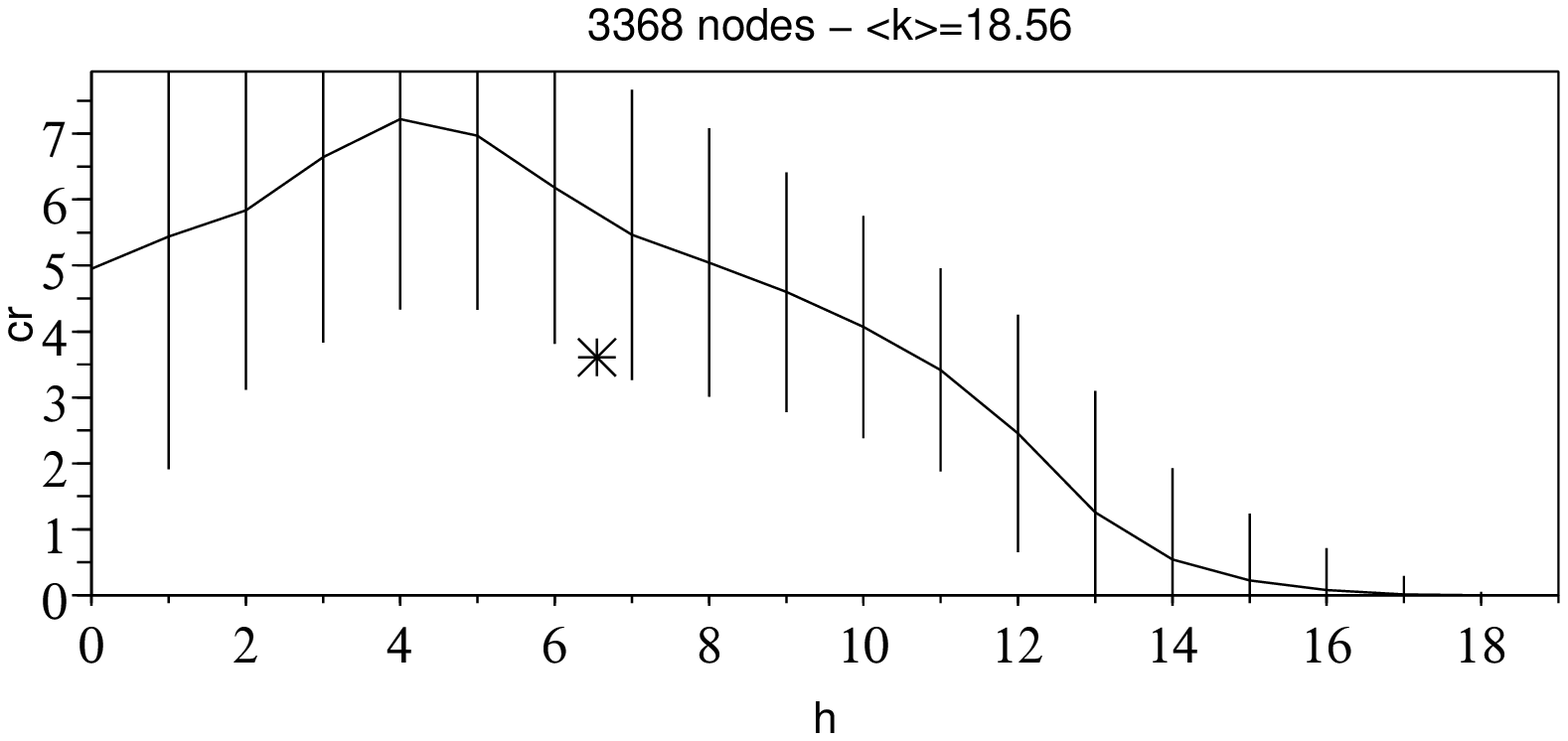}}
\put(300,215){L \includegraphics[scale=0.25]{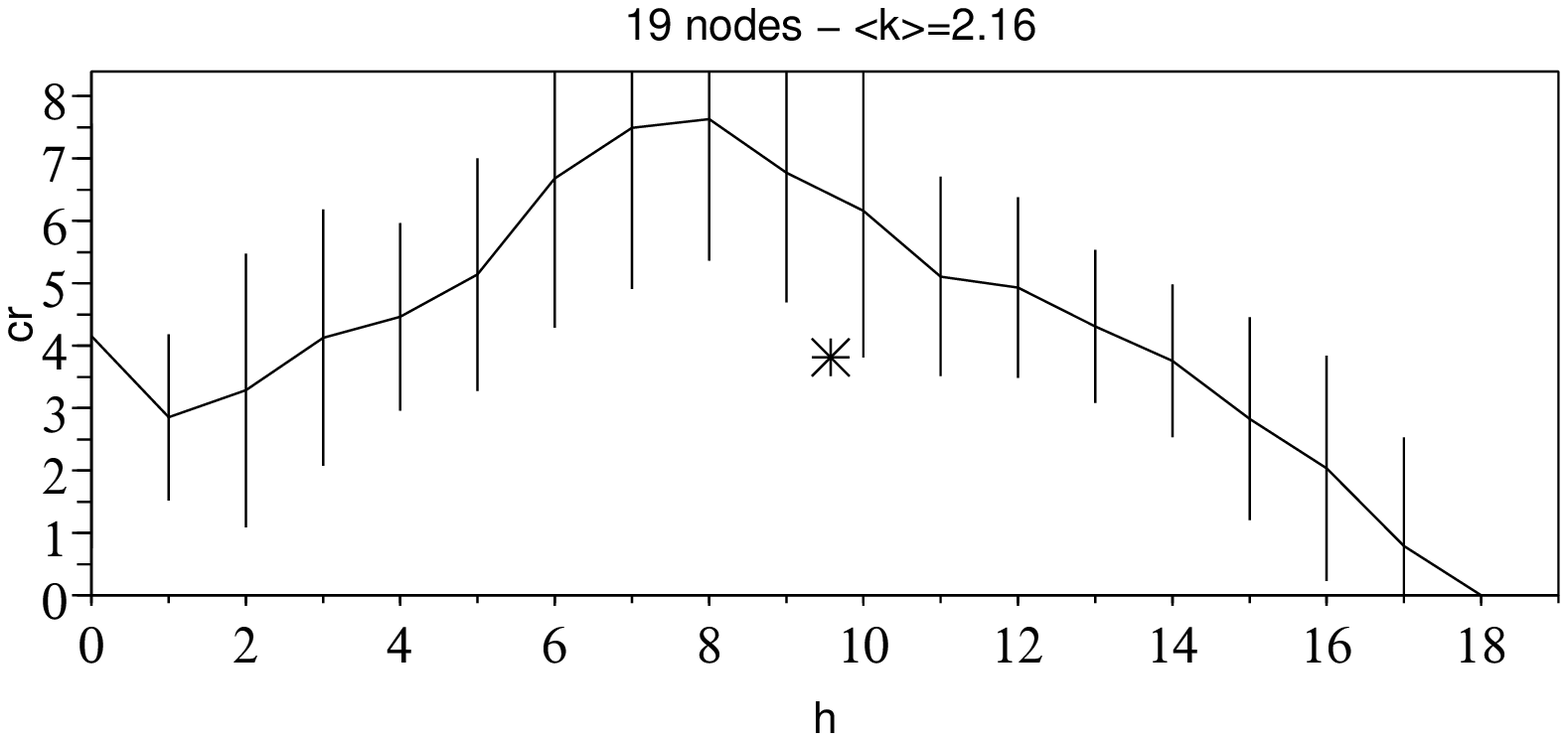}}
\put(300,135){M \includegraphics[scale=0.25]{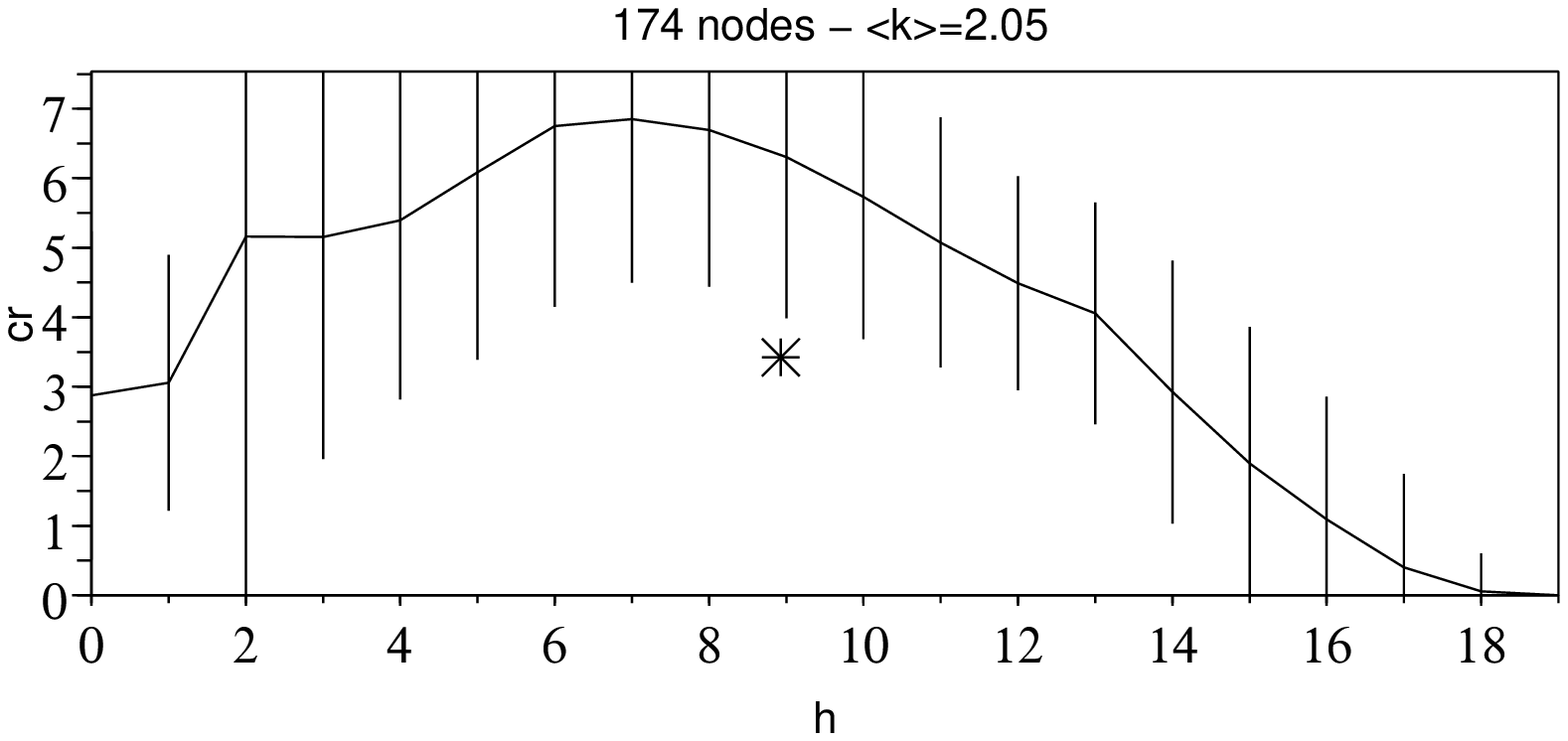}}
\put(300,060){N \includegraphics[scale=0.25]{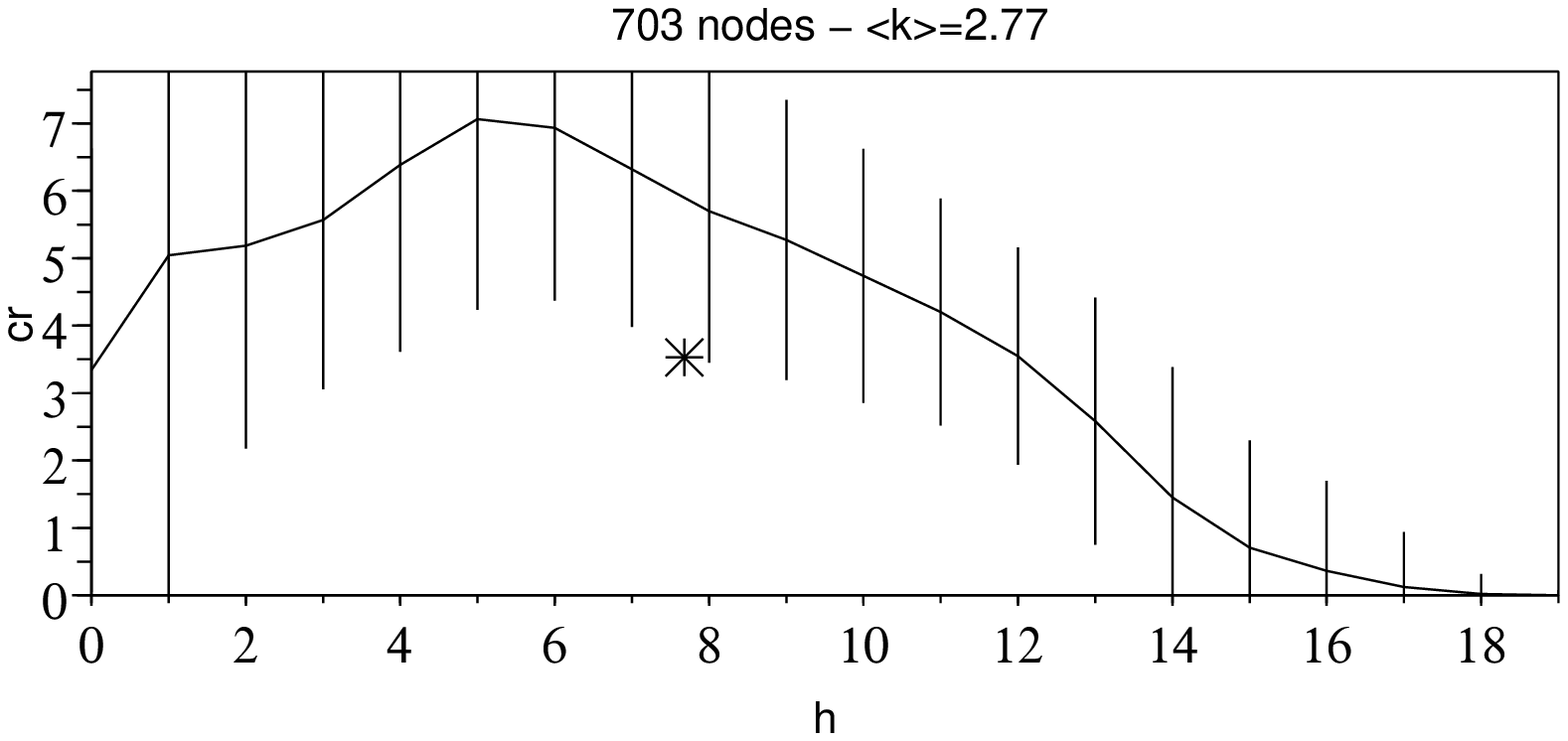}}
\put(300, -20){O\includegraphics[scale=0.25]{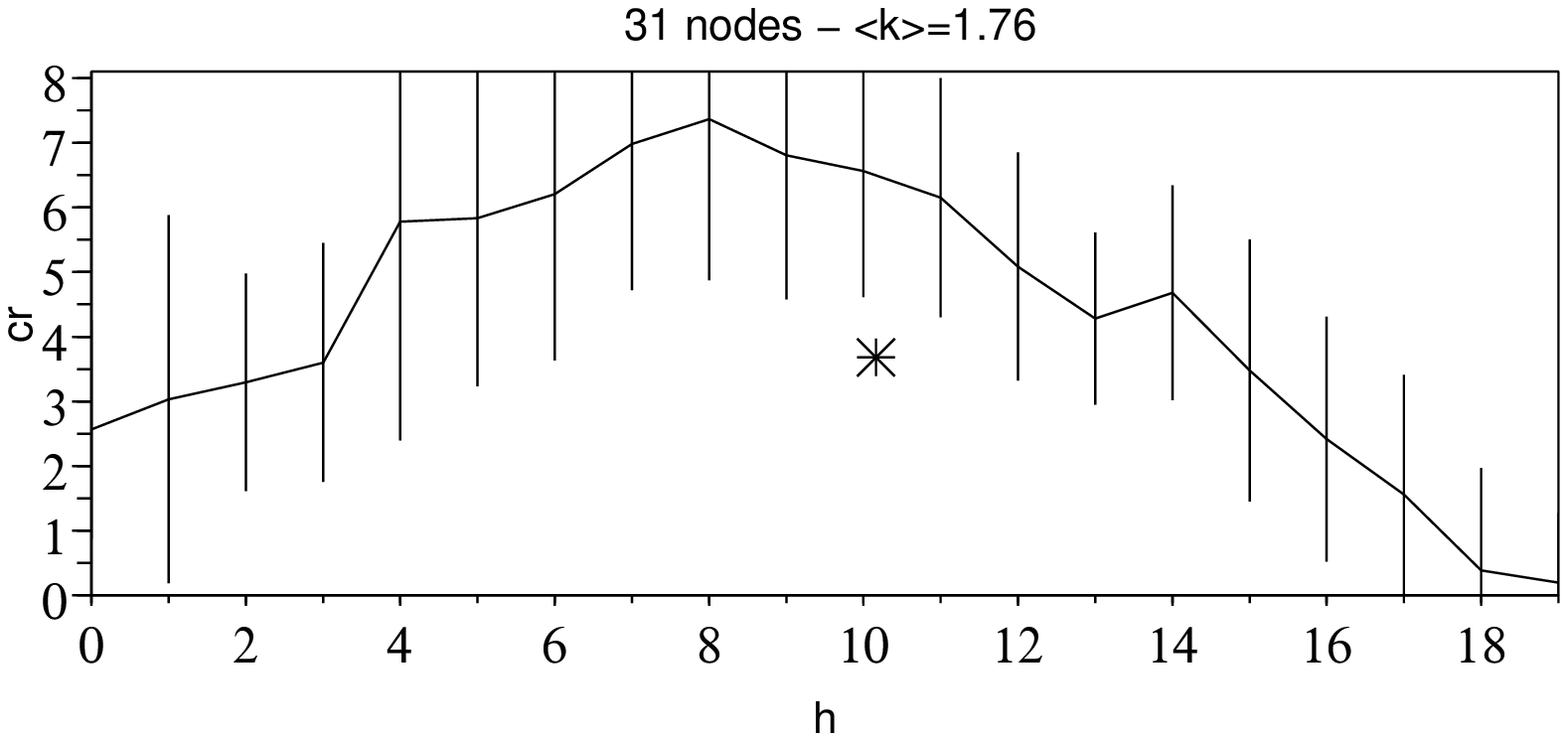}}
\end{picture}
\end{picture}
\end{center}
\caption{Graphs of the average $\pm$ standard deviation of the
concentric convergence ratio obtained for the co-authorship
network. Only four levels of the dendogram obtained by the
agglomerative hierarchical clustering are shown.~\label{fig:tree2}}

\end{figure*}

The results obtained for the convergence ratio can be seen in
Figure~\ref{fig:tree2}. The main distinction between the final groups
are the positions of the center of the peak for each curve, varying
from peaks centered at the concentric level 4 --- as in cluster K,
to around level 8 --- as in M. In the case of the convergence ratio,
the displacement indicates how close the groups of nodes are to the
hubs.

\begin{figure*}
\begin{center}
\begin{picture}(560,560)(0,-20)
\put(30,0){ \includegraphics[scale=0.8]{Images/base.eps}}
\begin{picture}(560,500)(-2,-55)
\put(-010,255){A\includegraphics[scale=0.1]{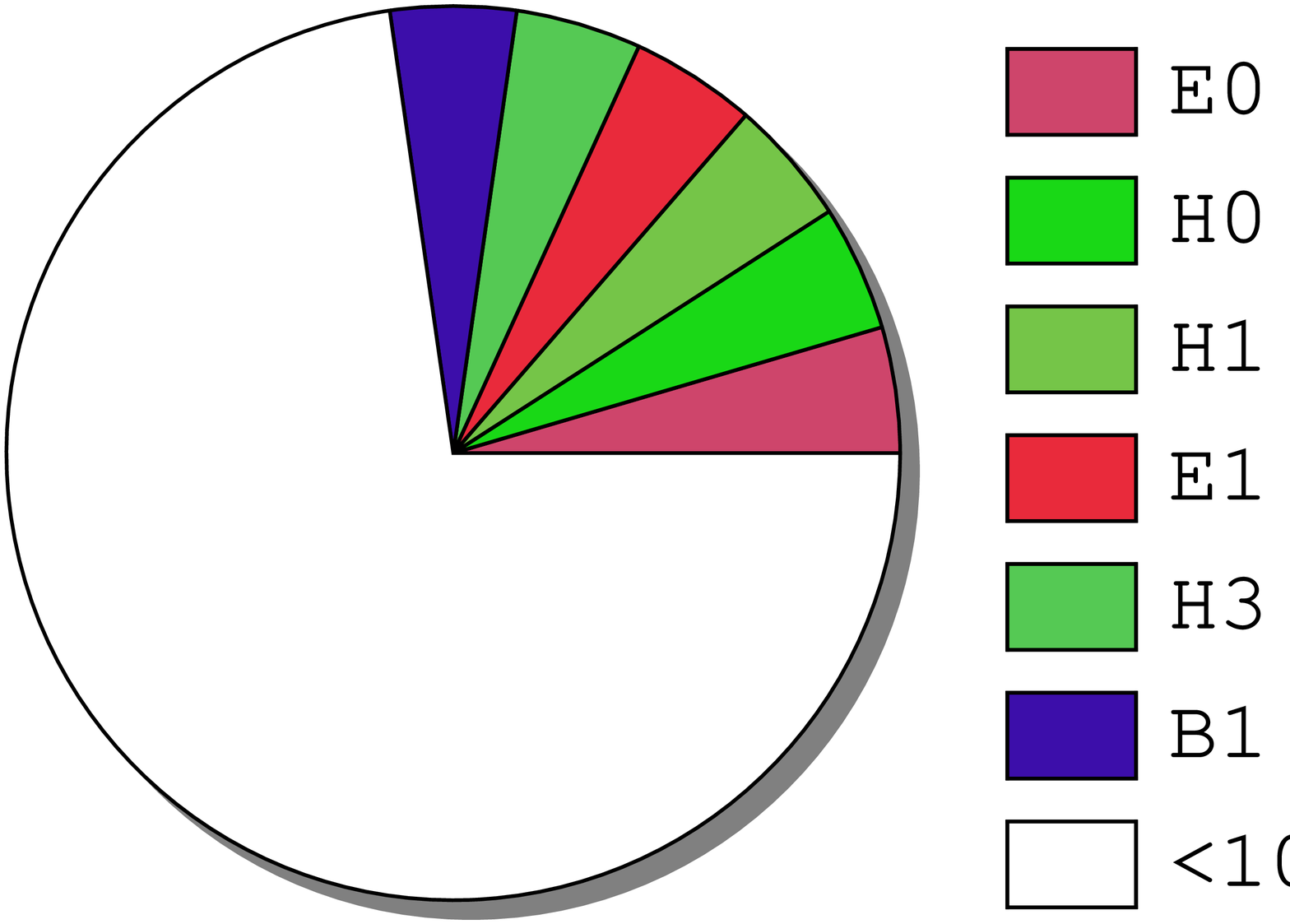}}
\put(75,415){B\includegraphics[scale=0.1]{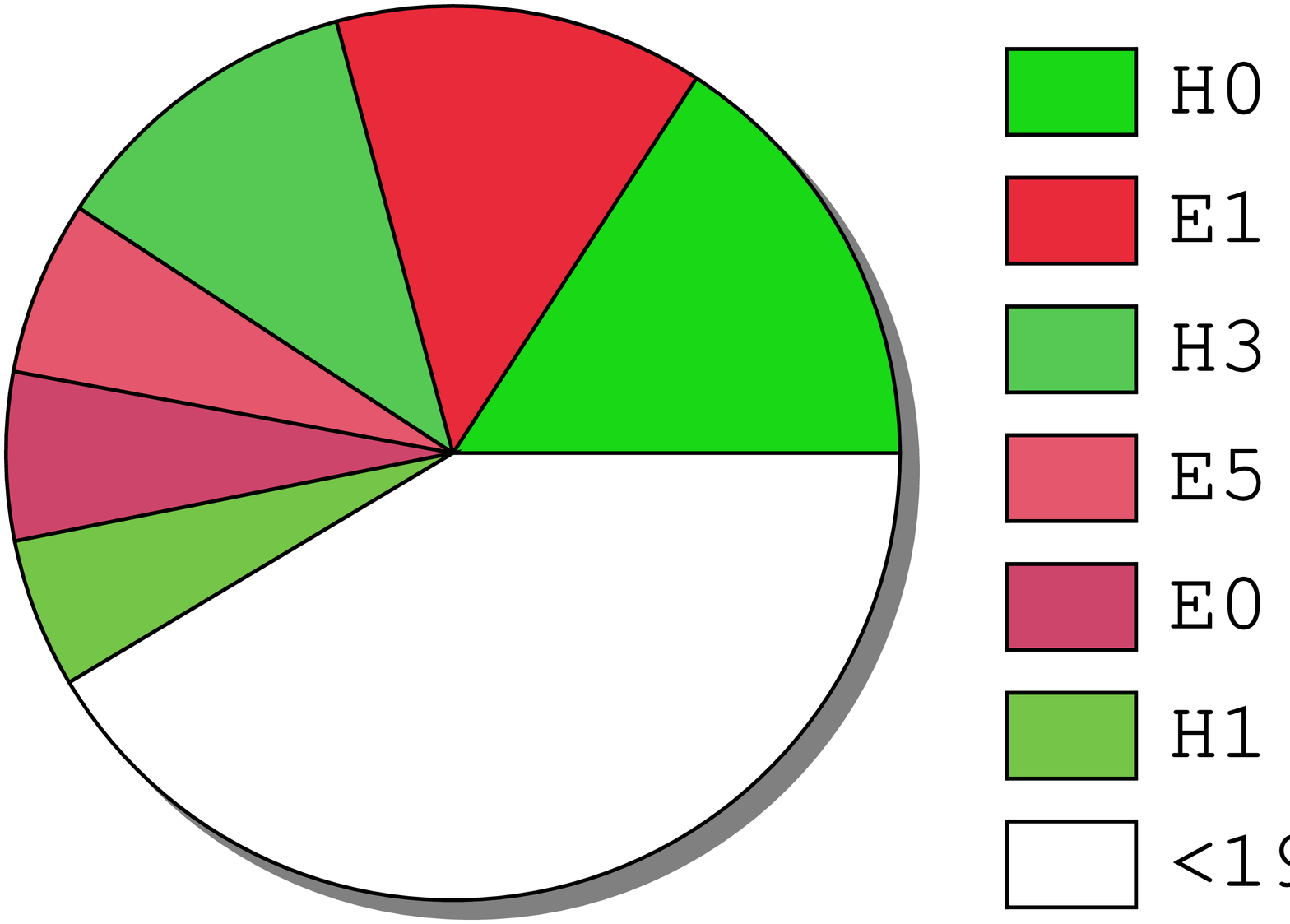}}
\put(75,100){C\includegraphics[scale=0.1]{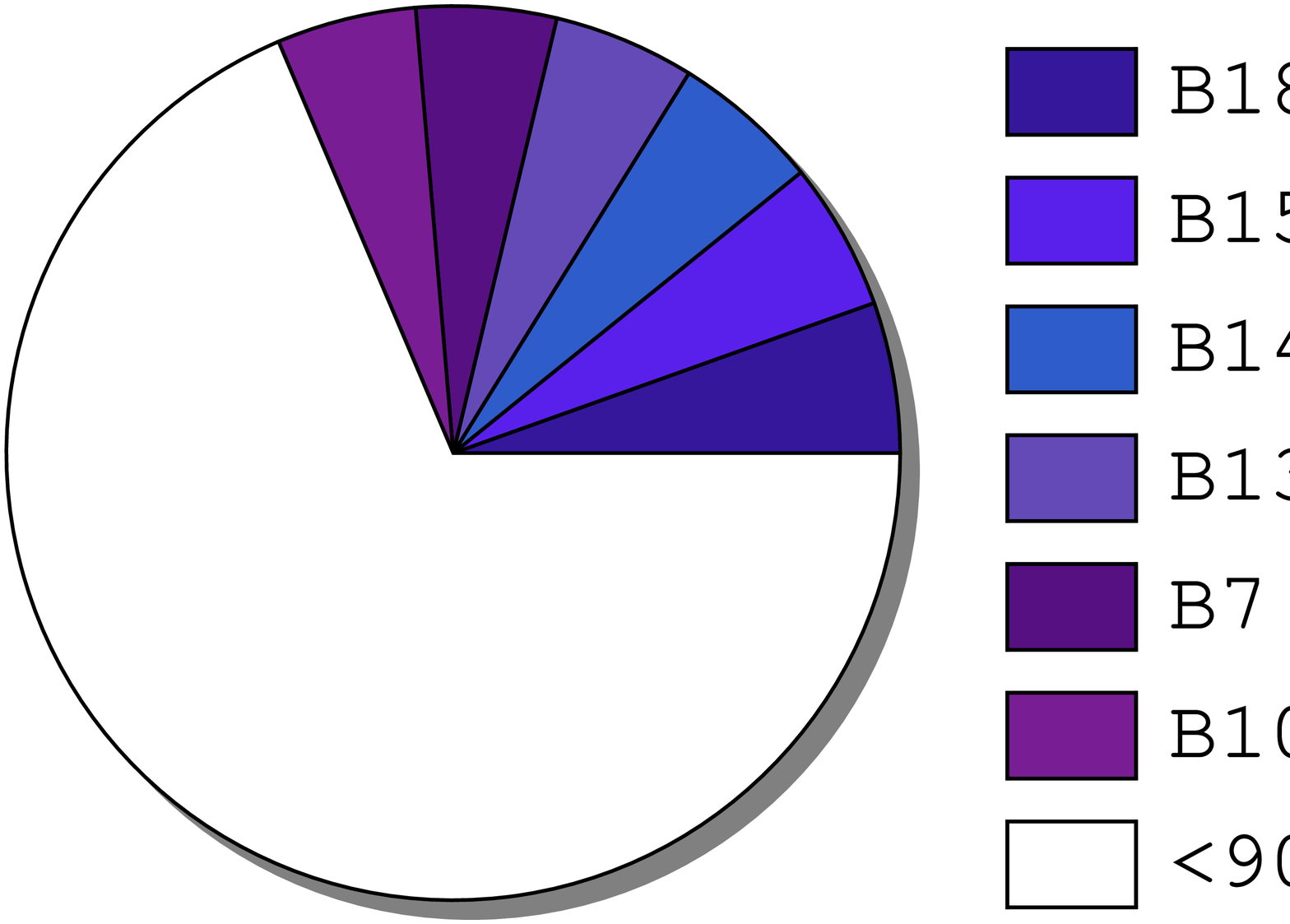}}
\put(160,485){D\includegraphics[scale=0.1]{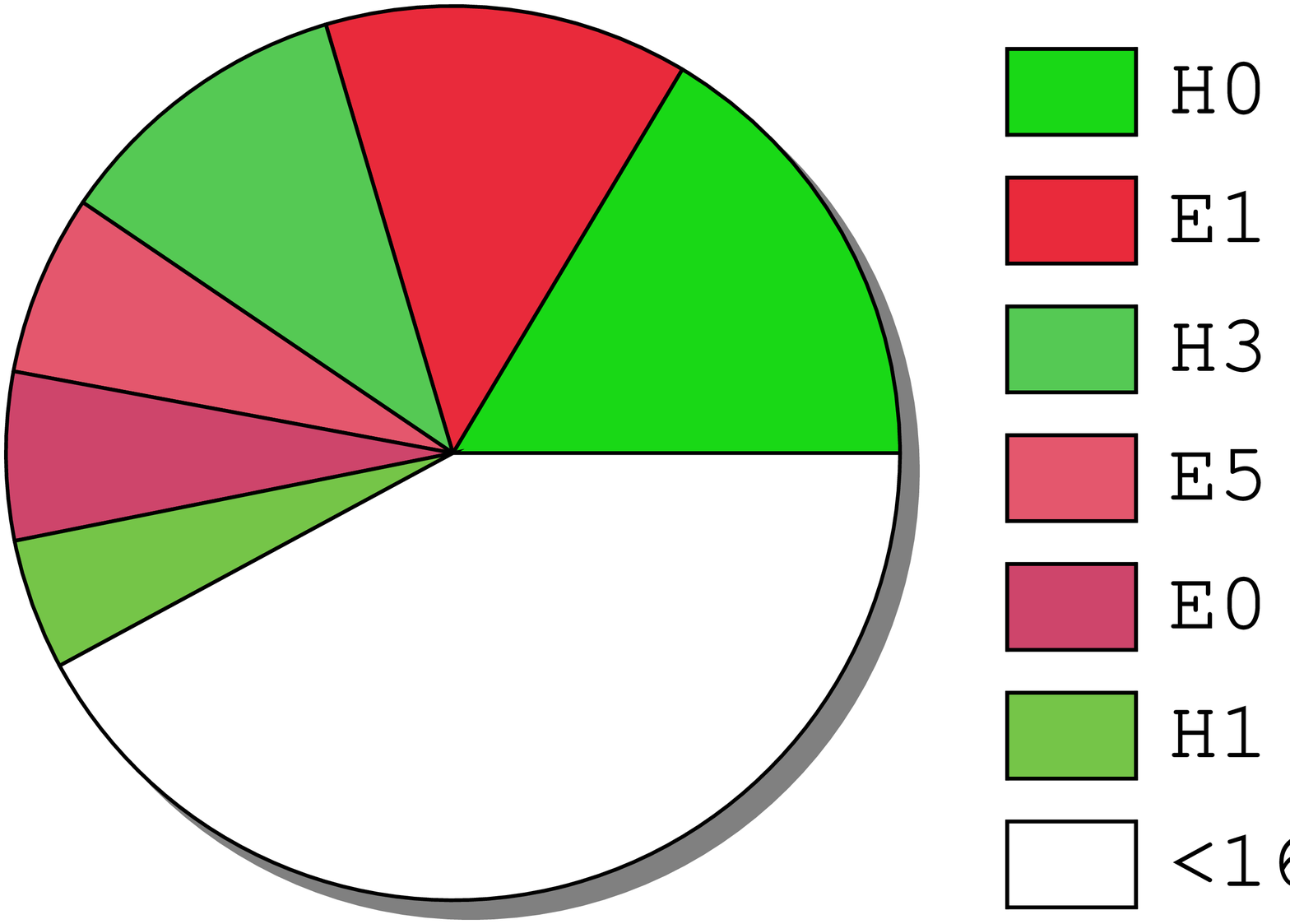}}
\put(160,335){E\includegraphics[scale=0.1]{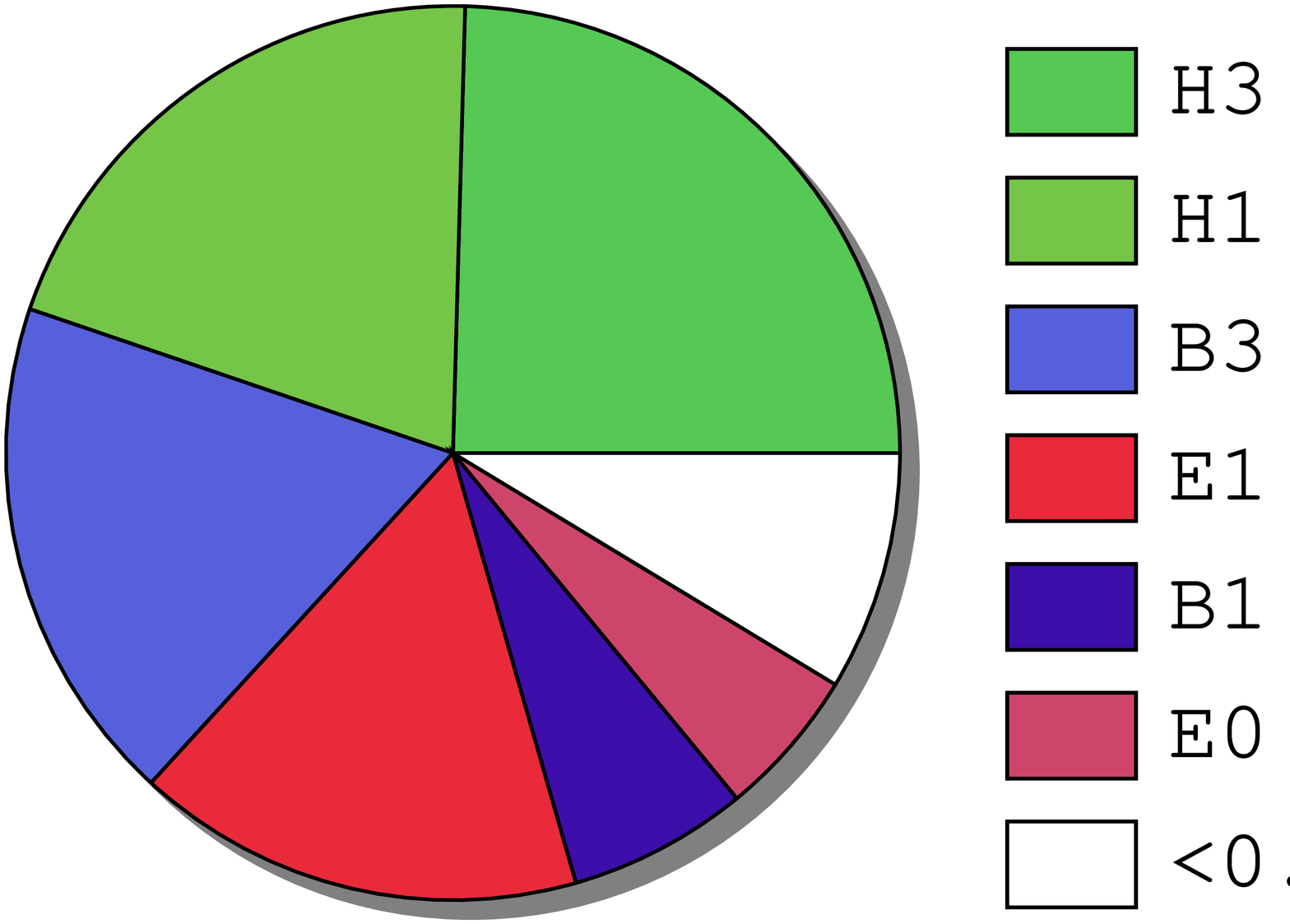}}
\put(160,175){F\includegraphics[scale=0.1]{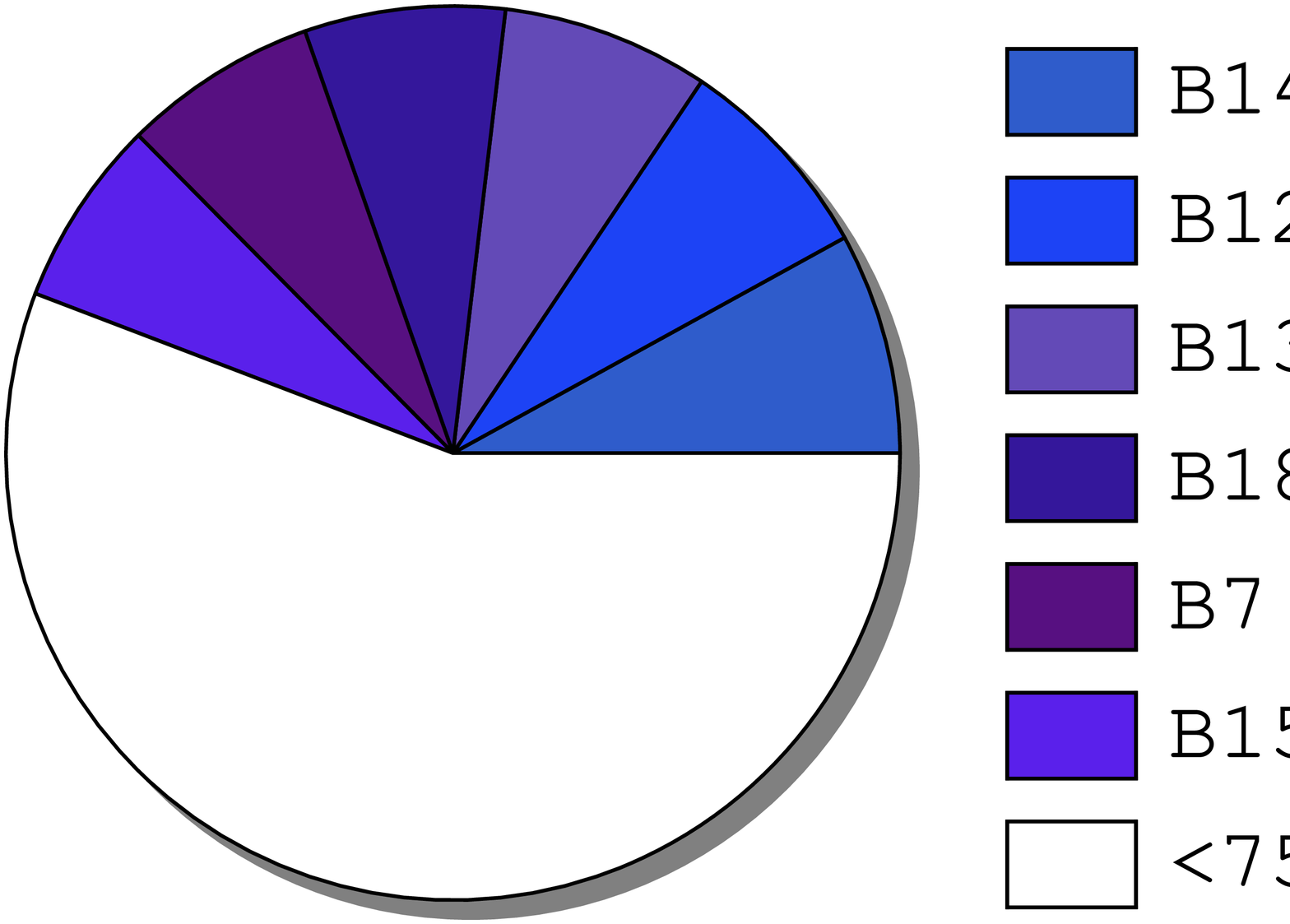}}
\put(160,20){G\includegraphics[scale=0.1]{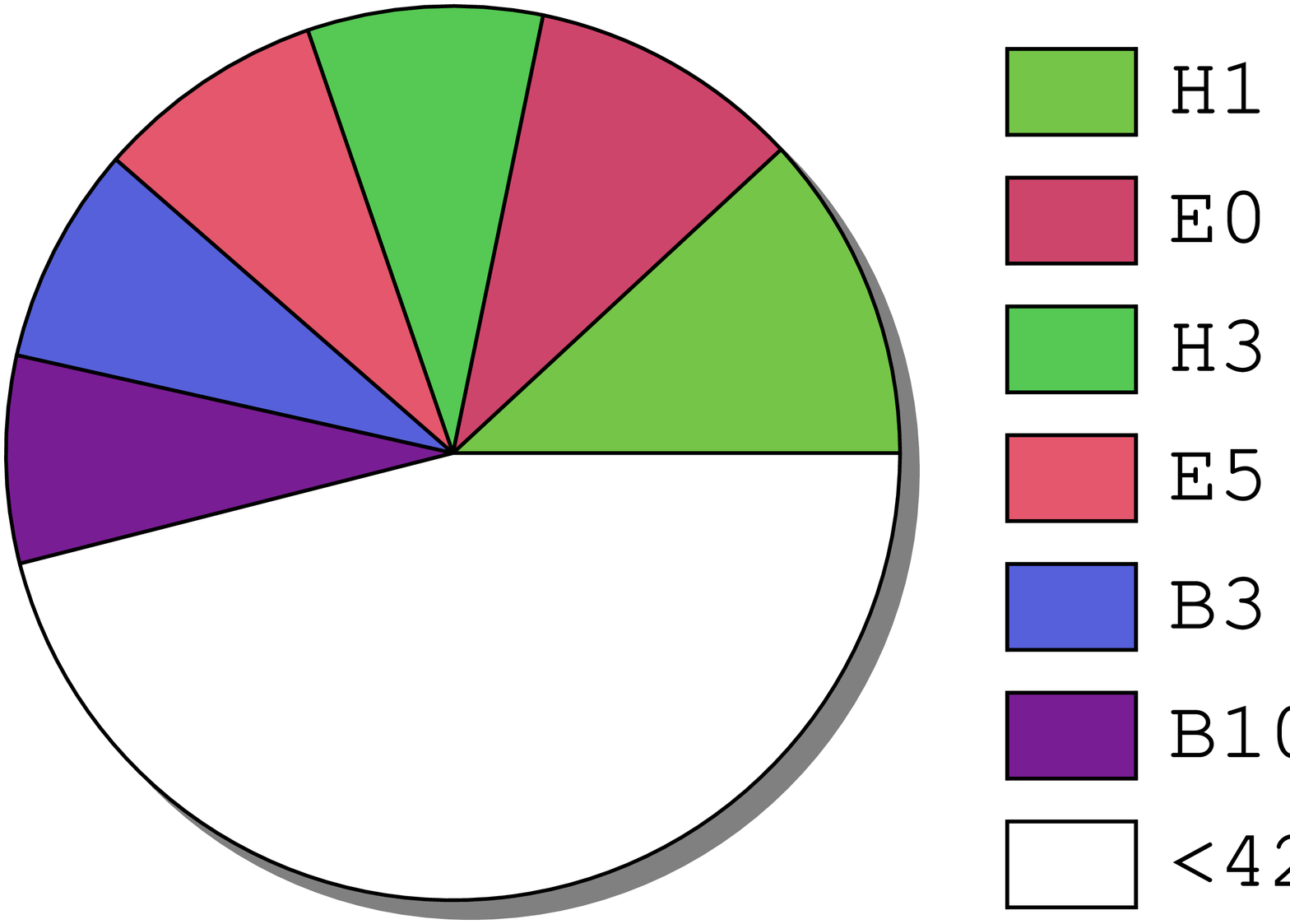}}
\put(290,525){H\includegraphics[scale=0.1]{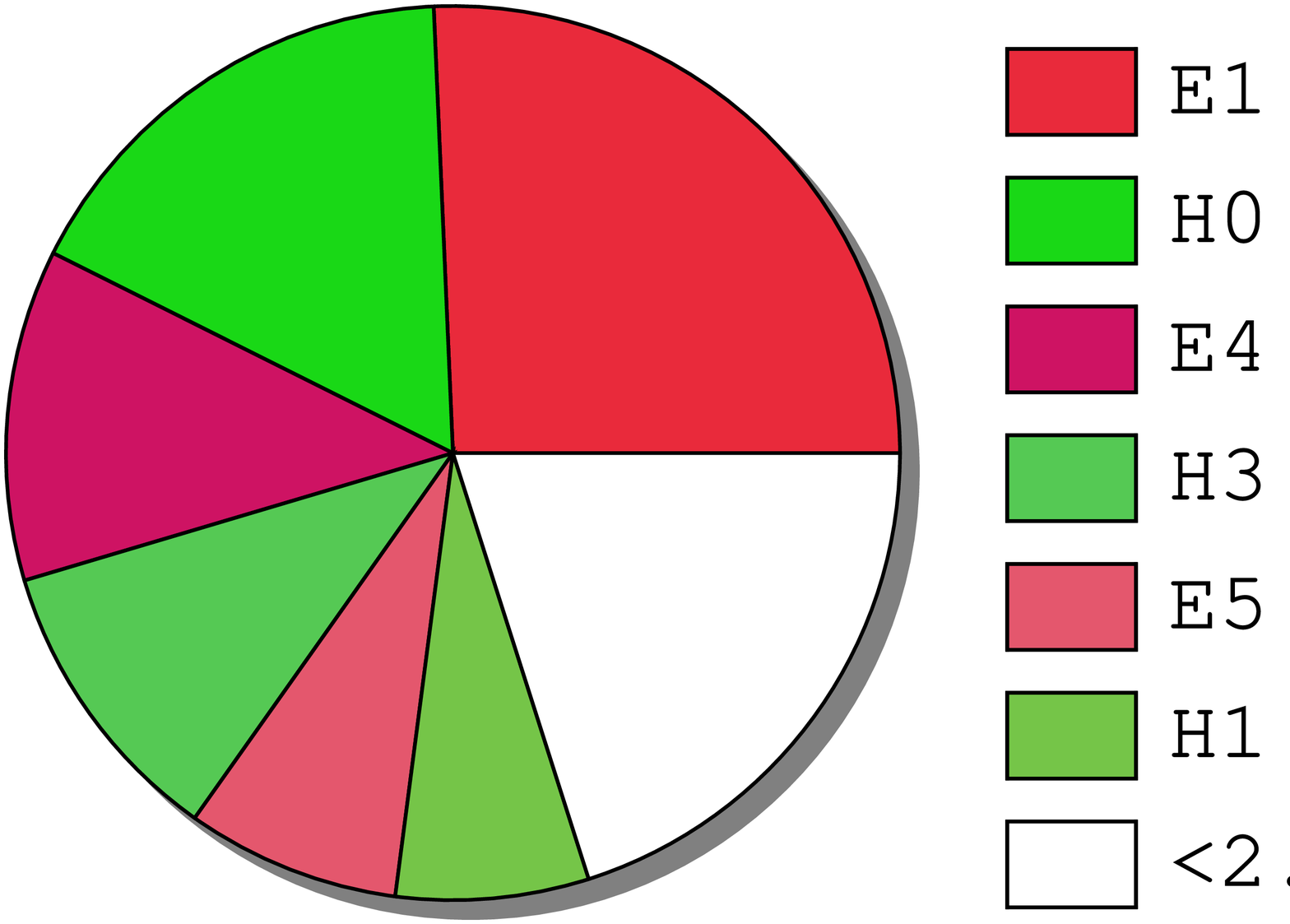}}
\put(290,445){I \includegraphics[scale=0.1]{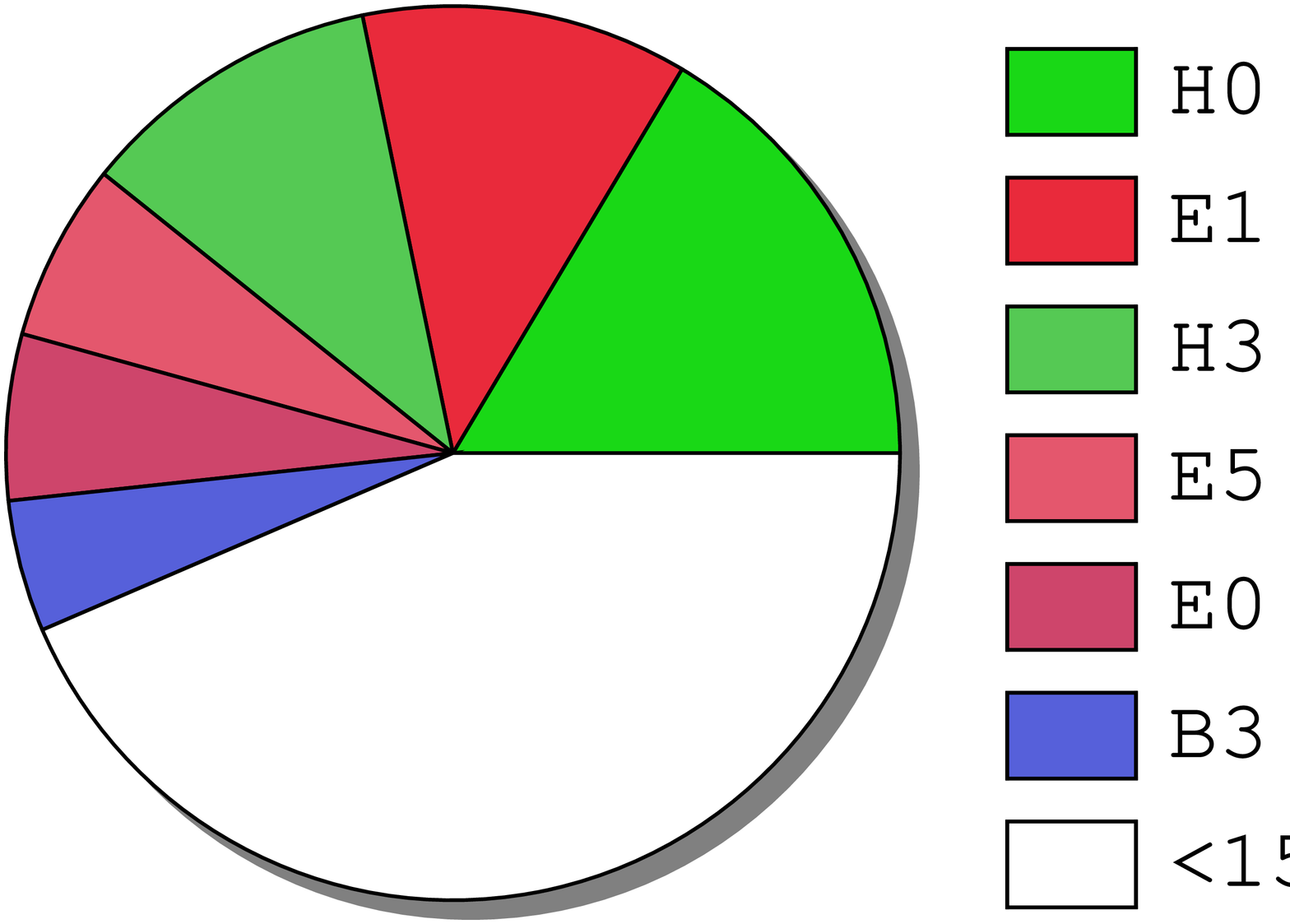}}
\put(290,370){J \includegraphics[scale=0.1]{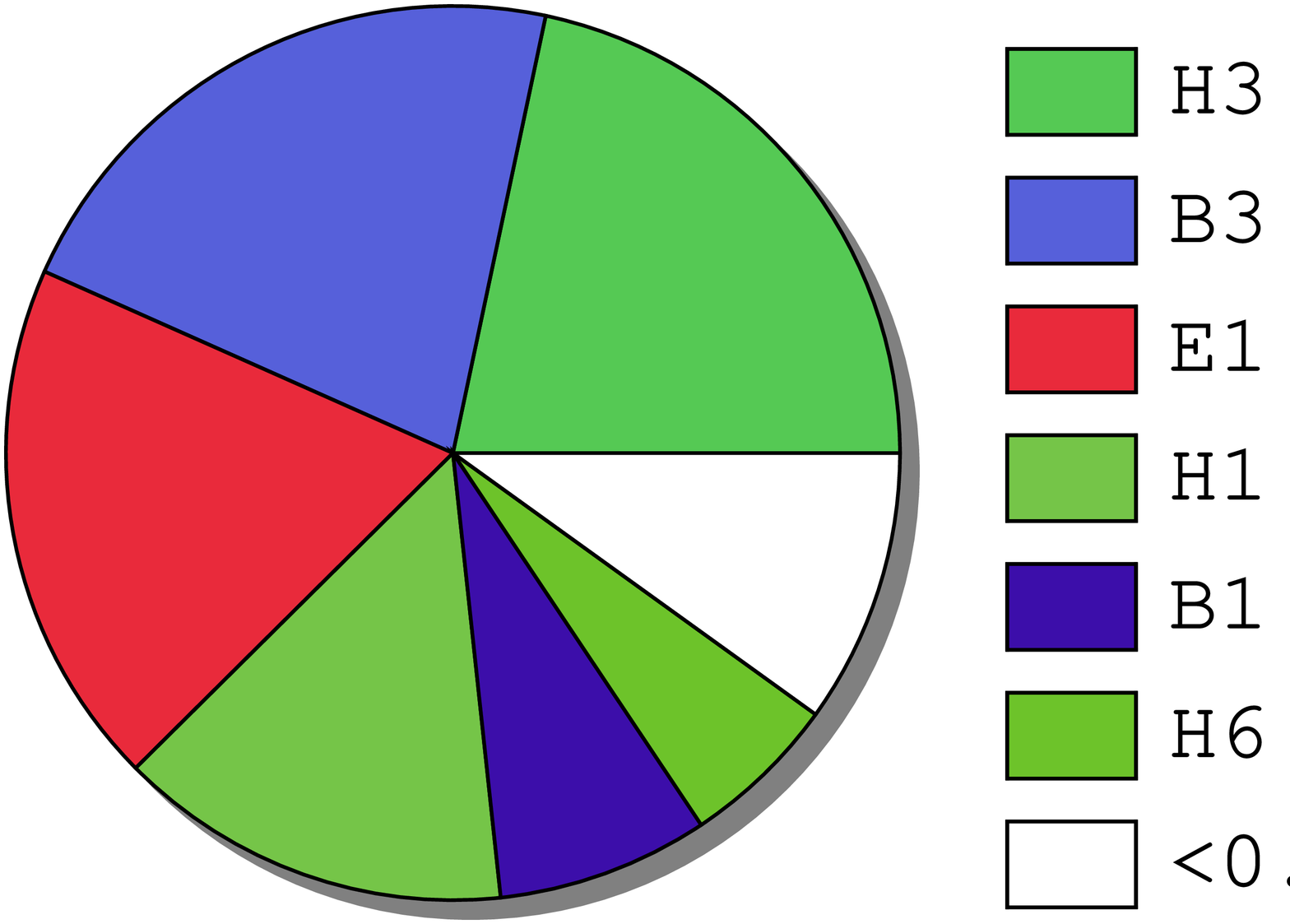}}
\put(290,290){K\includegraphics[scale=0.1]{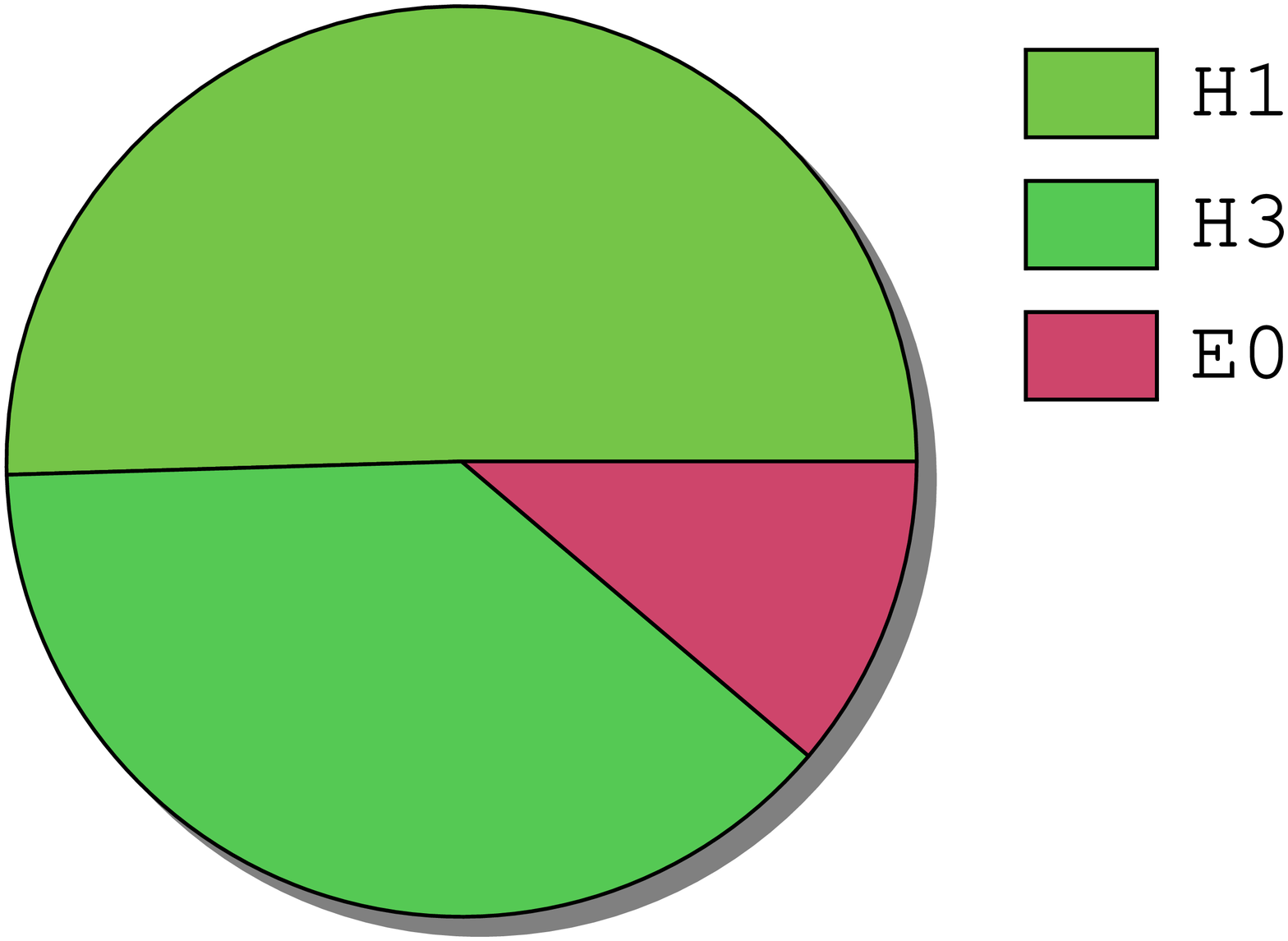}}
\put(290,215){L \includegraphics[scale=0.1]{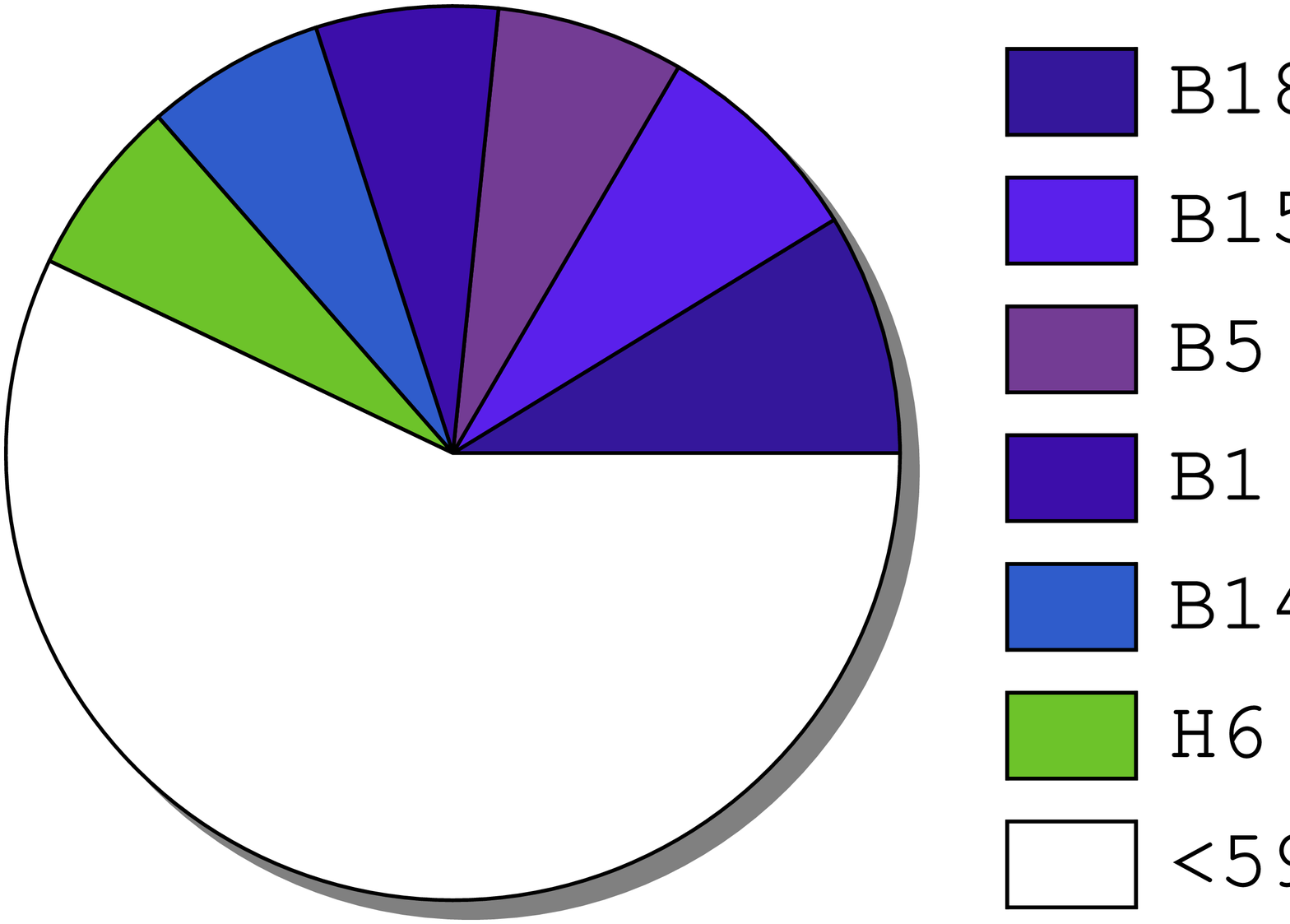}}
\put(290,135){M\includegraphics[scale=0.1]{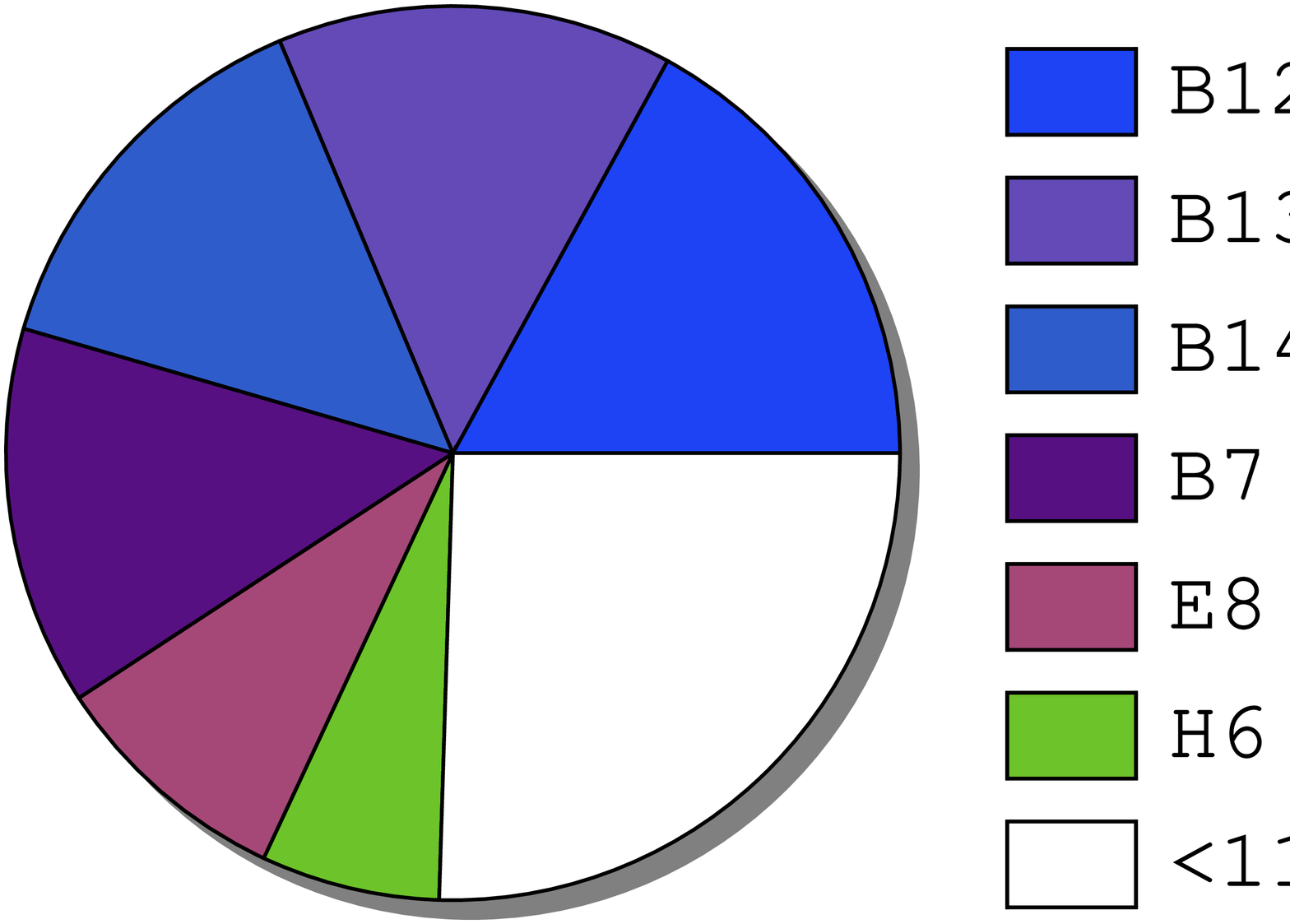}}
\put(290,060){N\includegraphics[scale=0.1]{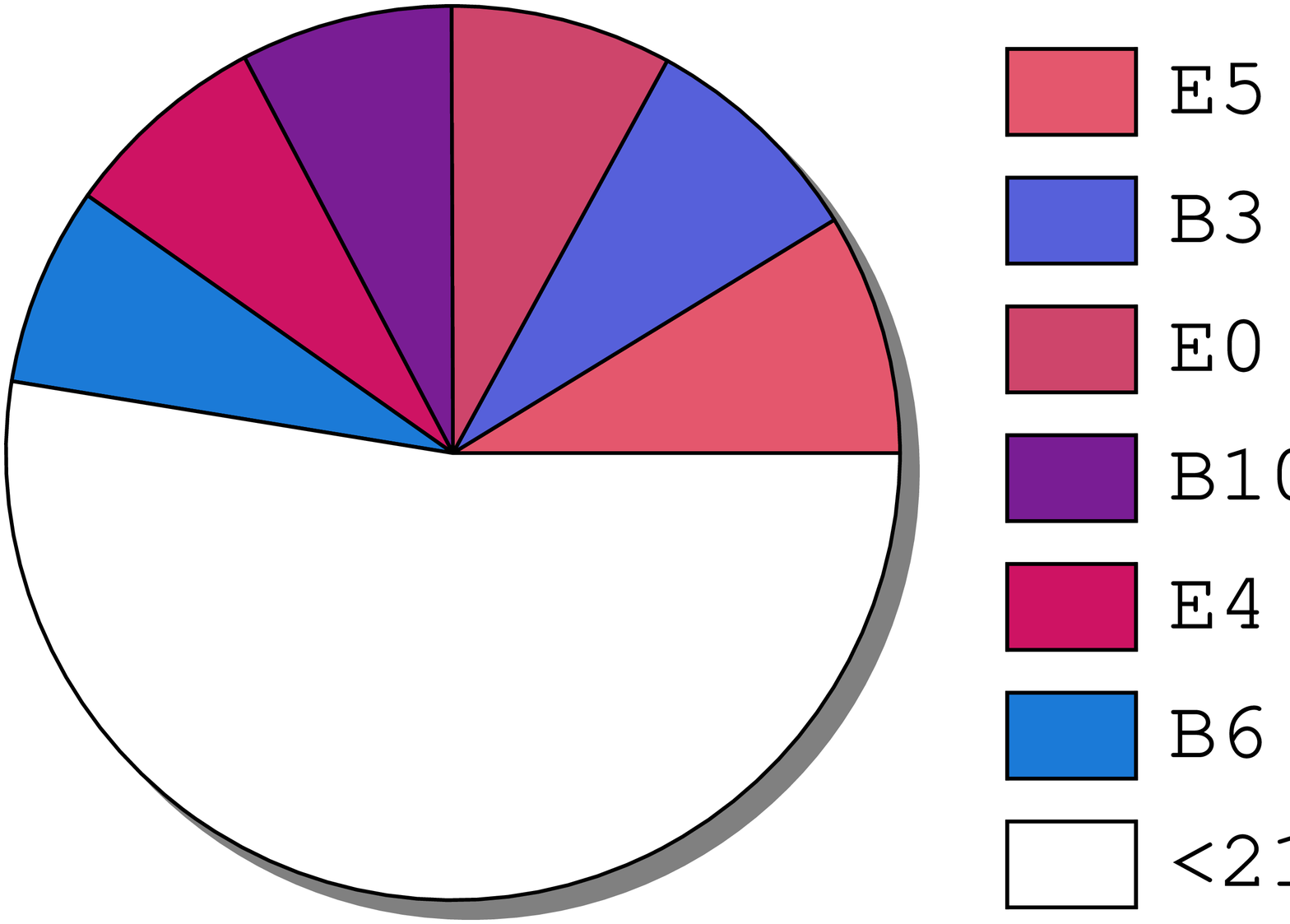}}
\put(290, -20){O\includegraphics[scale=0.1]{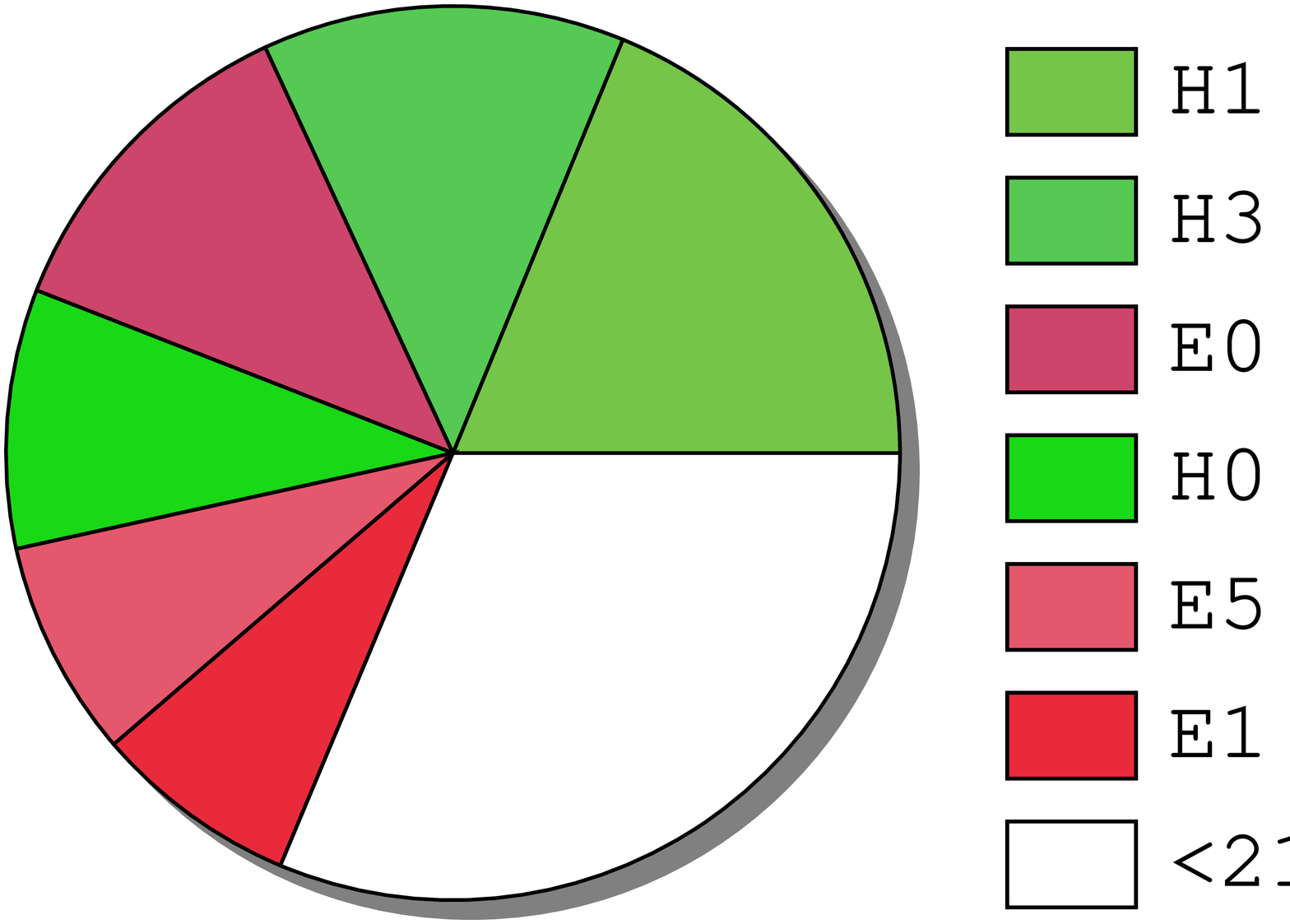}}
\end{picture}
\end{picture}
\end{center}
\caption{Pie charts with the percentage of nodes in each department for each
cluster obtained by using the Clustering Coefficient.}
\label{fig:tree3}
\end{figure*}

\begin{figure*}
\begin{center}
\begin{picture}(560,560)(0,-20)
\put(30,0){ \includegraphics[scale=0.8]{Images/base.eps}}
\begin{picture}(560,500)(-2,-55)
\put(-010,255){A\includegraphics[scale=0.1]{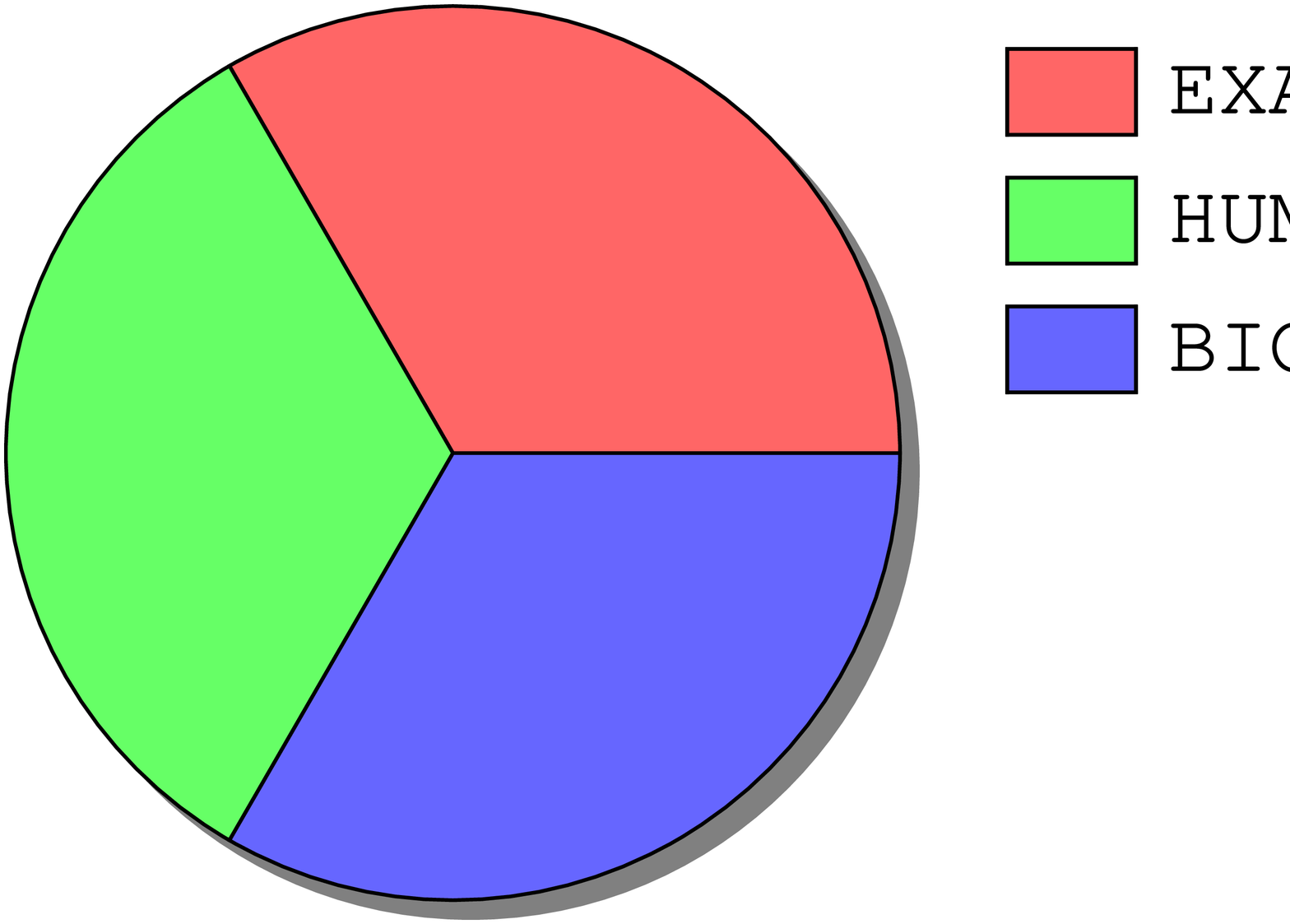}}
\put(75,415){B\includegraphics[scale=0.1]{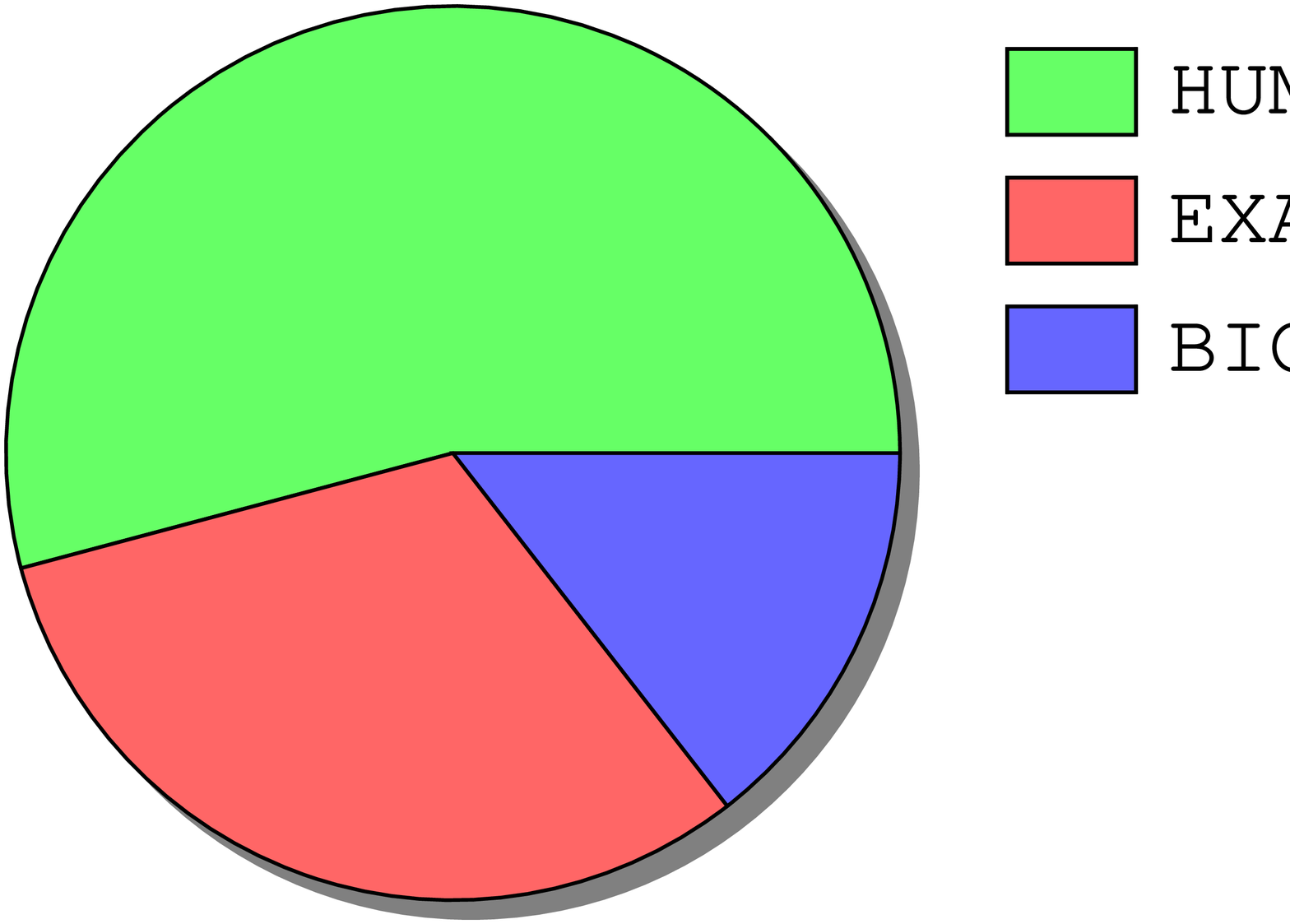}}
\put(75,100){C\includegraphics[scale=0.1]{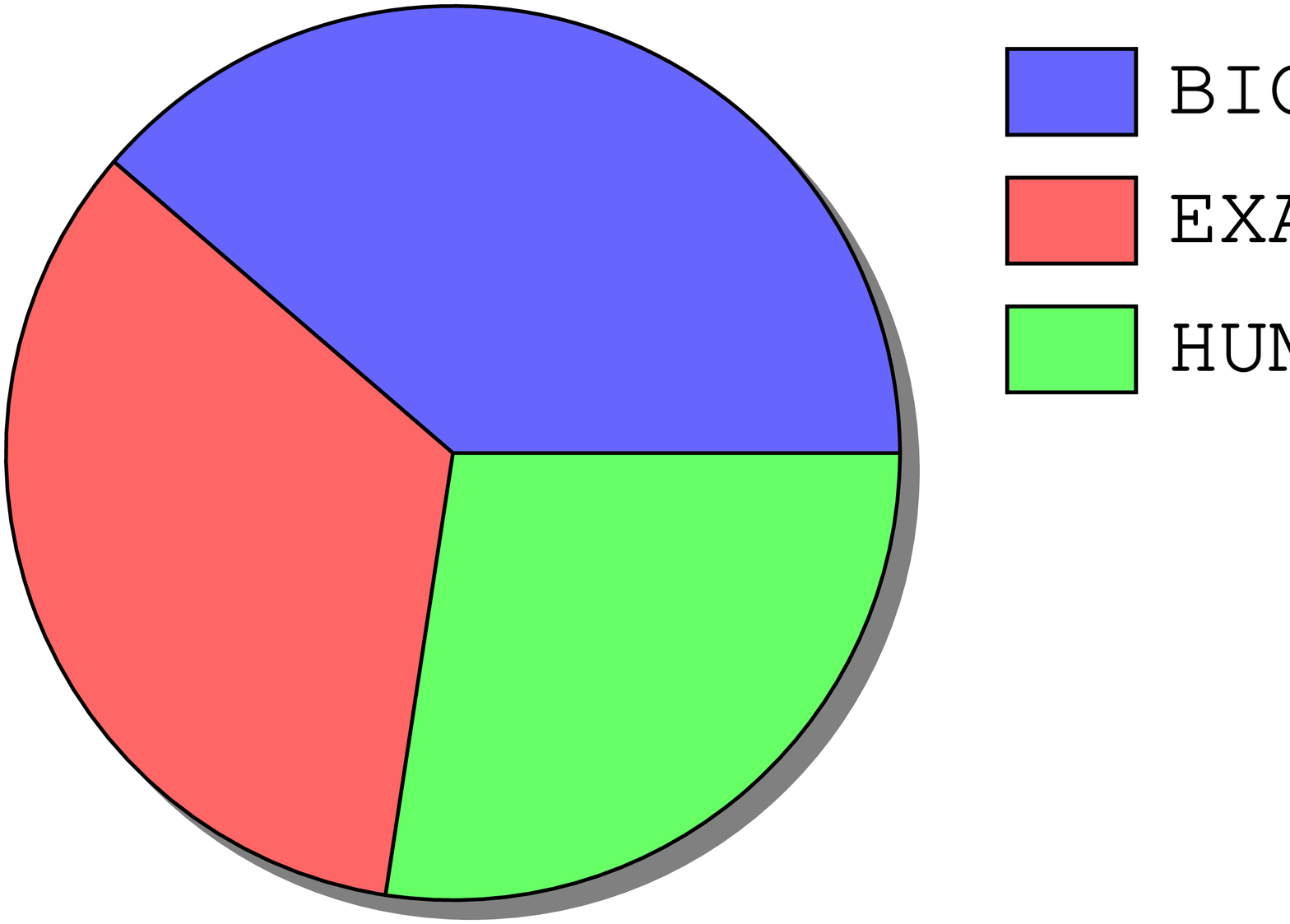}}
\put(160,485){D\includegraphics[scale=0.1]{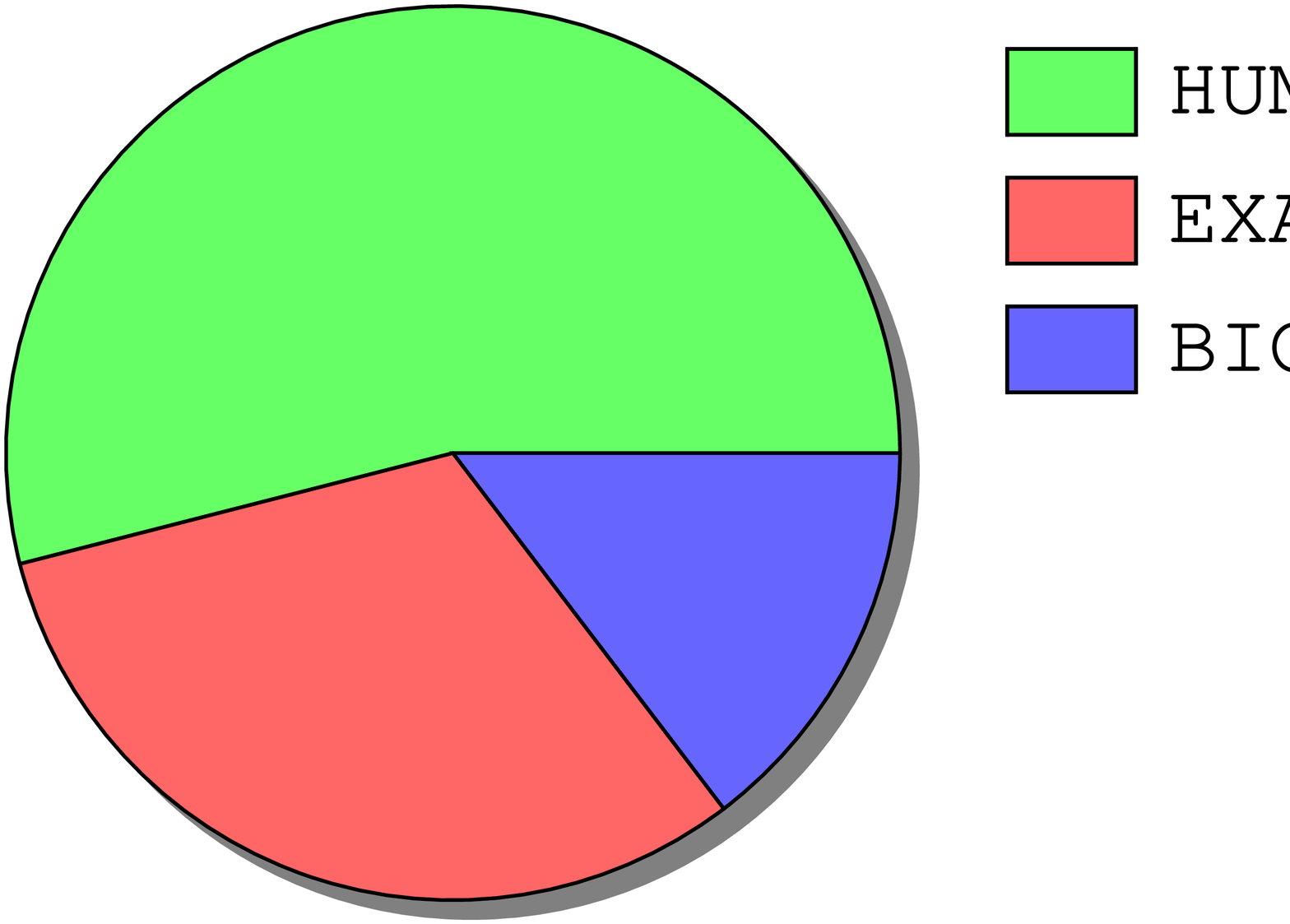}}
\put(160,335){E\includegraphics[scale=0.1]{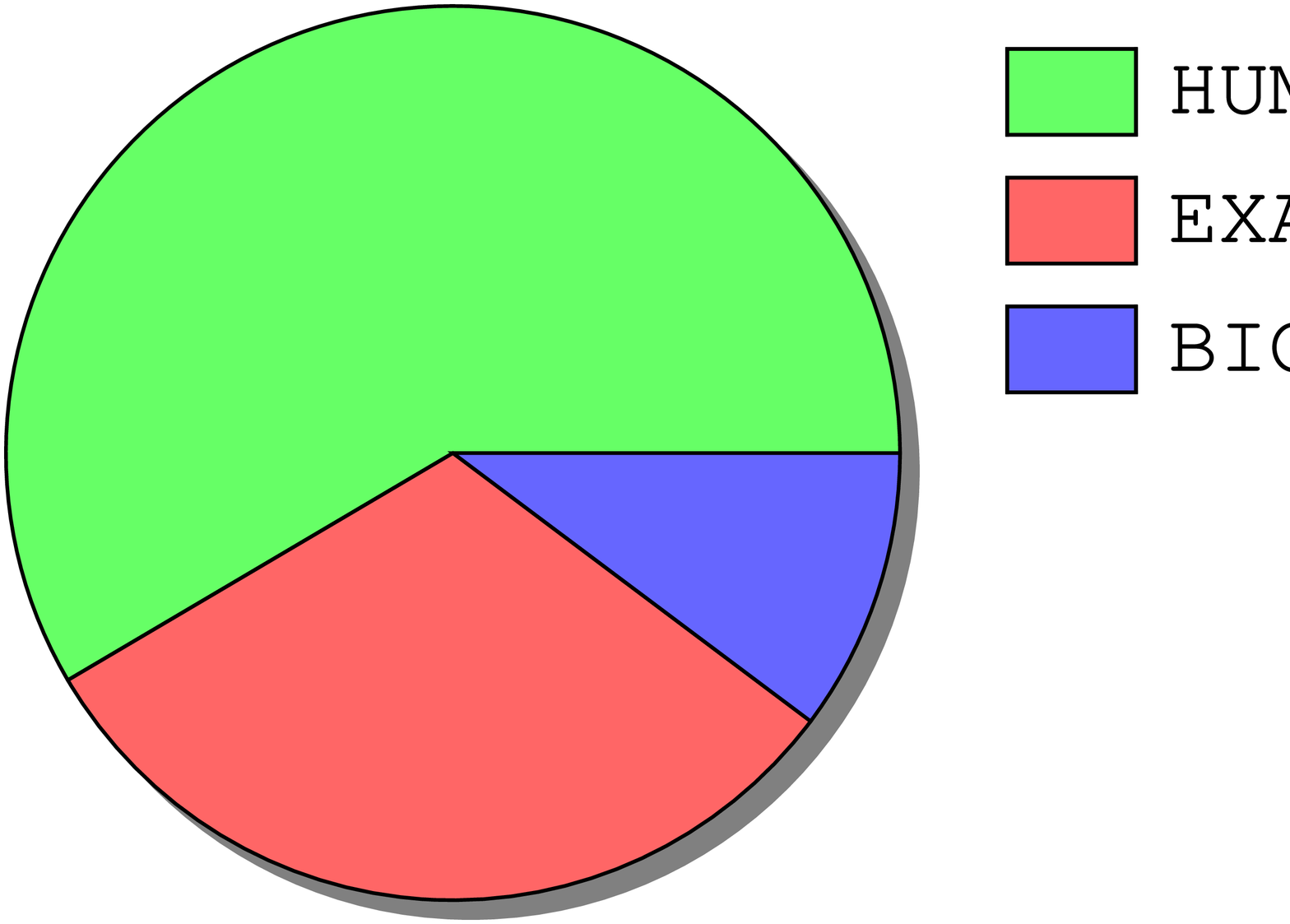}}
\put(160,175){F\includegraphics[scale=0.1]{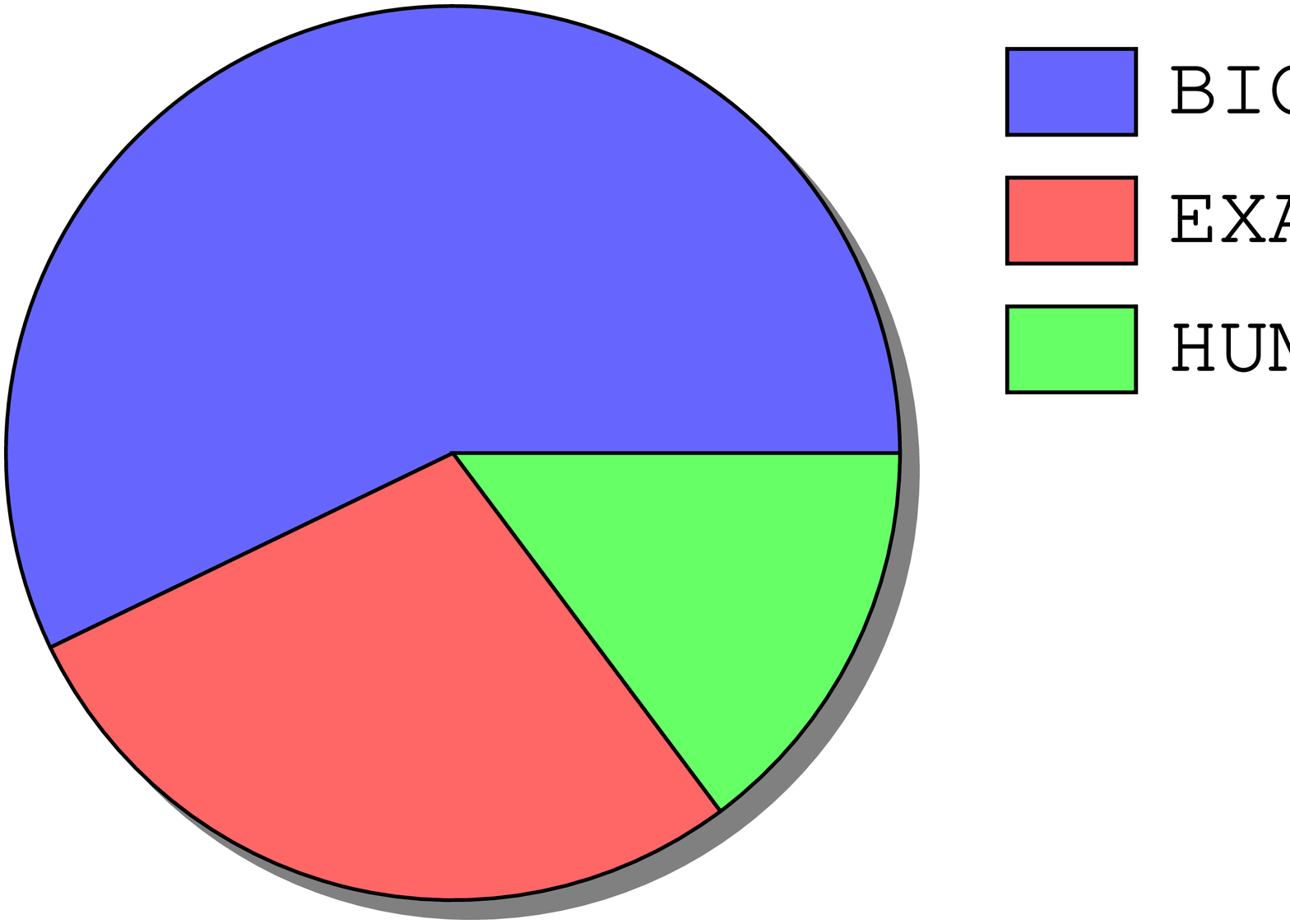}}
\put(160,20){G\includegraphics[scale=0.1]{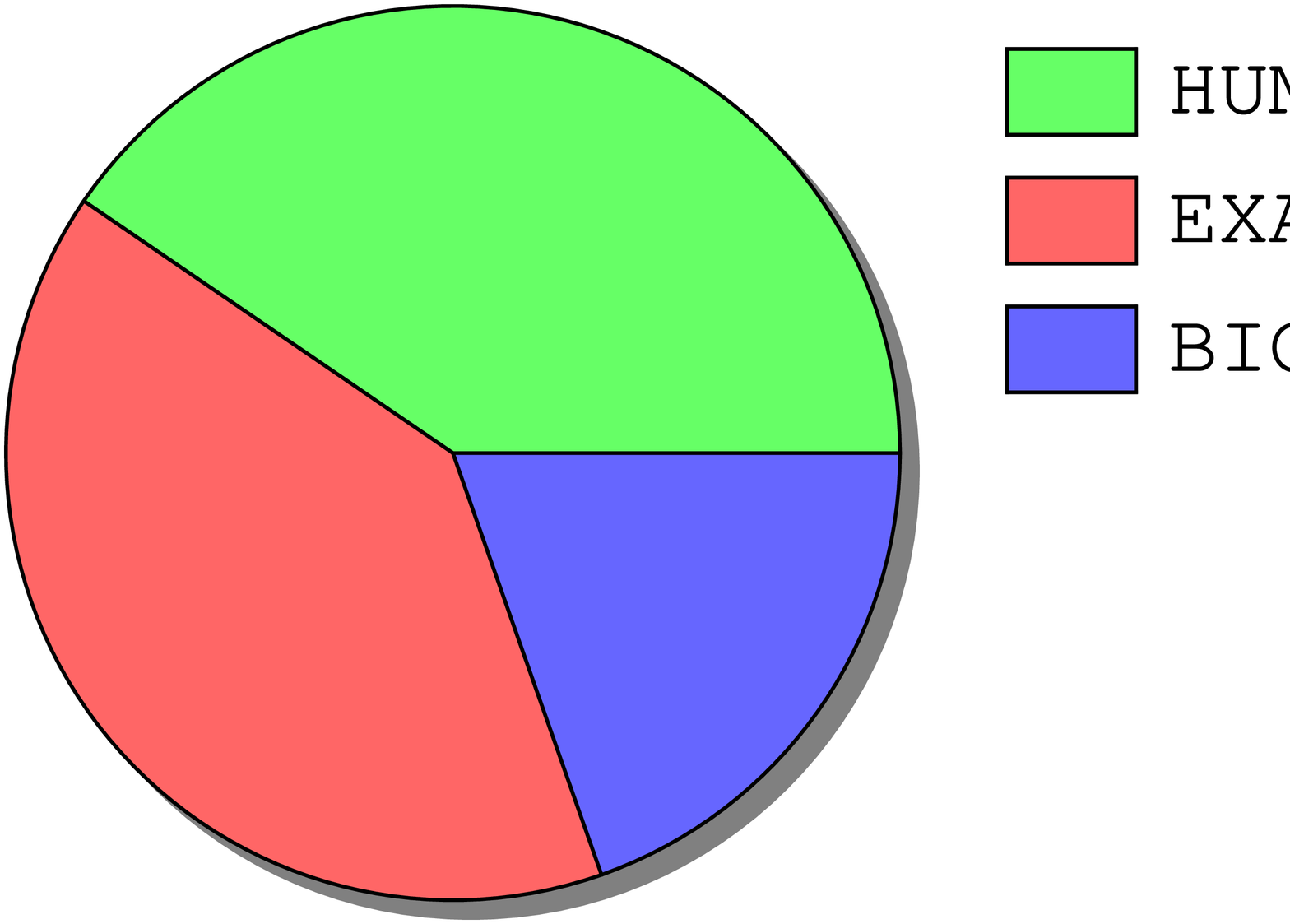}}
\put(290,525){H\includegraphics[scale=0.1]{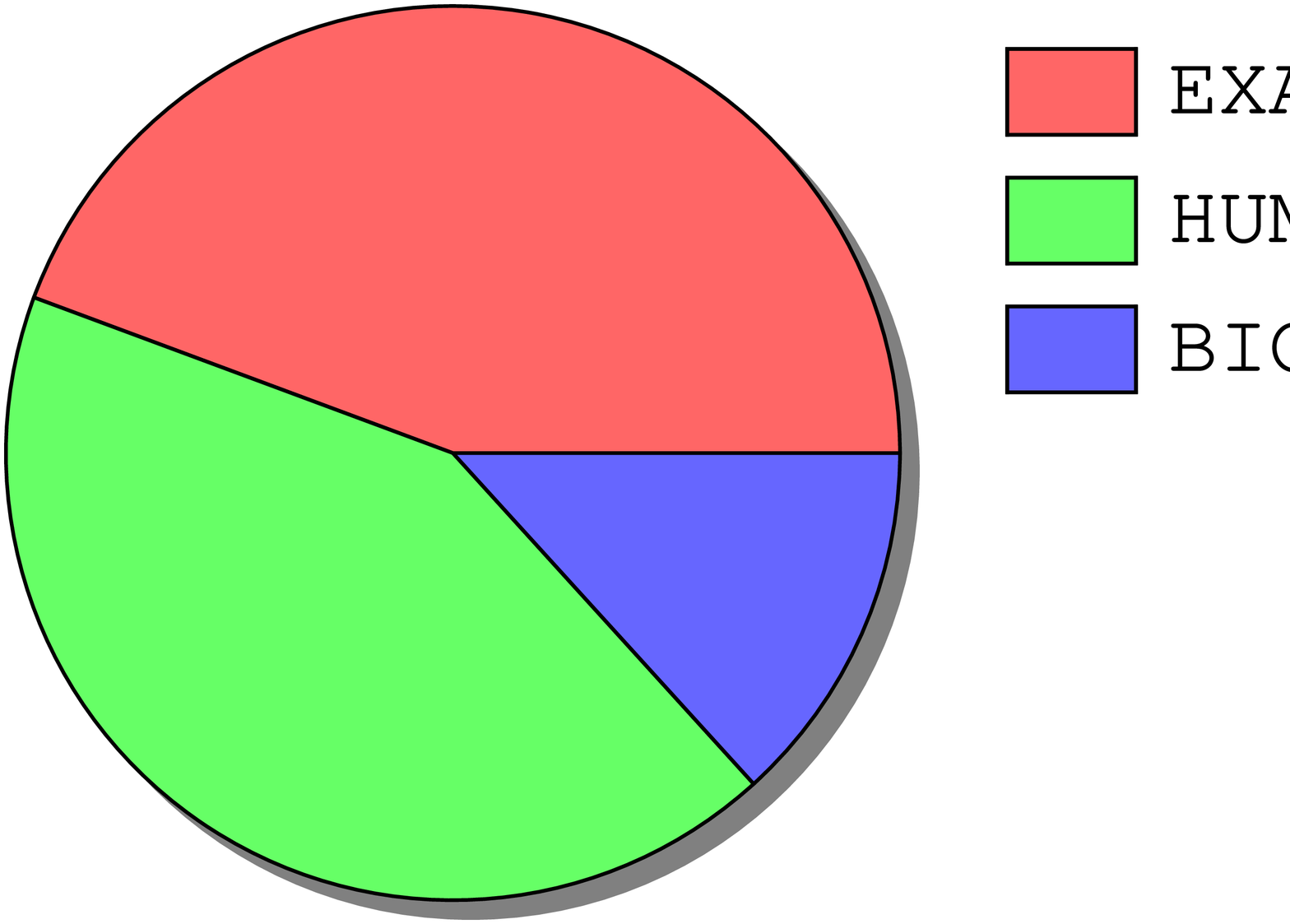}}
\put(290,445){I \includegraphics[scale=0.1]{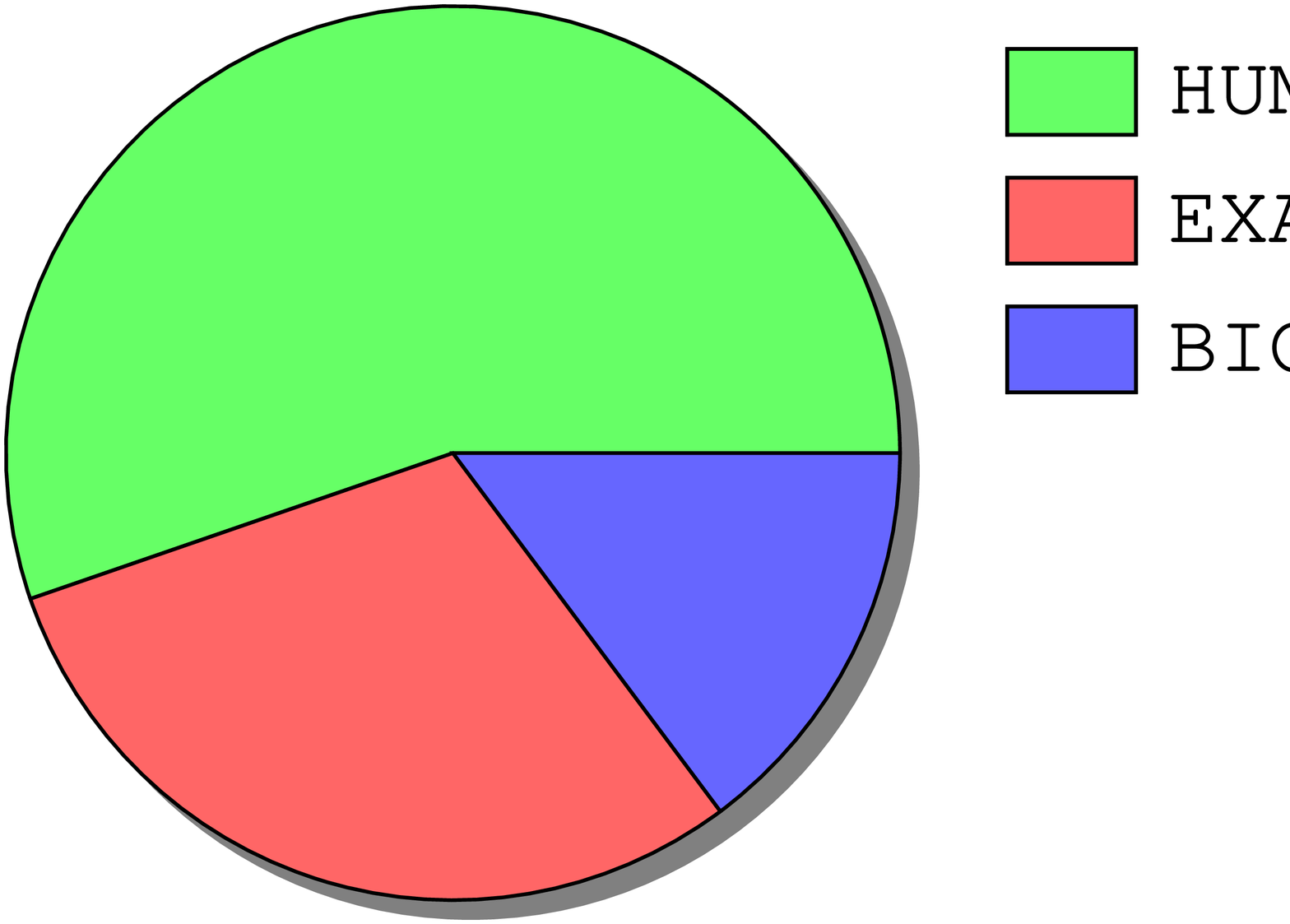}}
\put(290,370){J \includegraphics[scale=0.1]{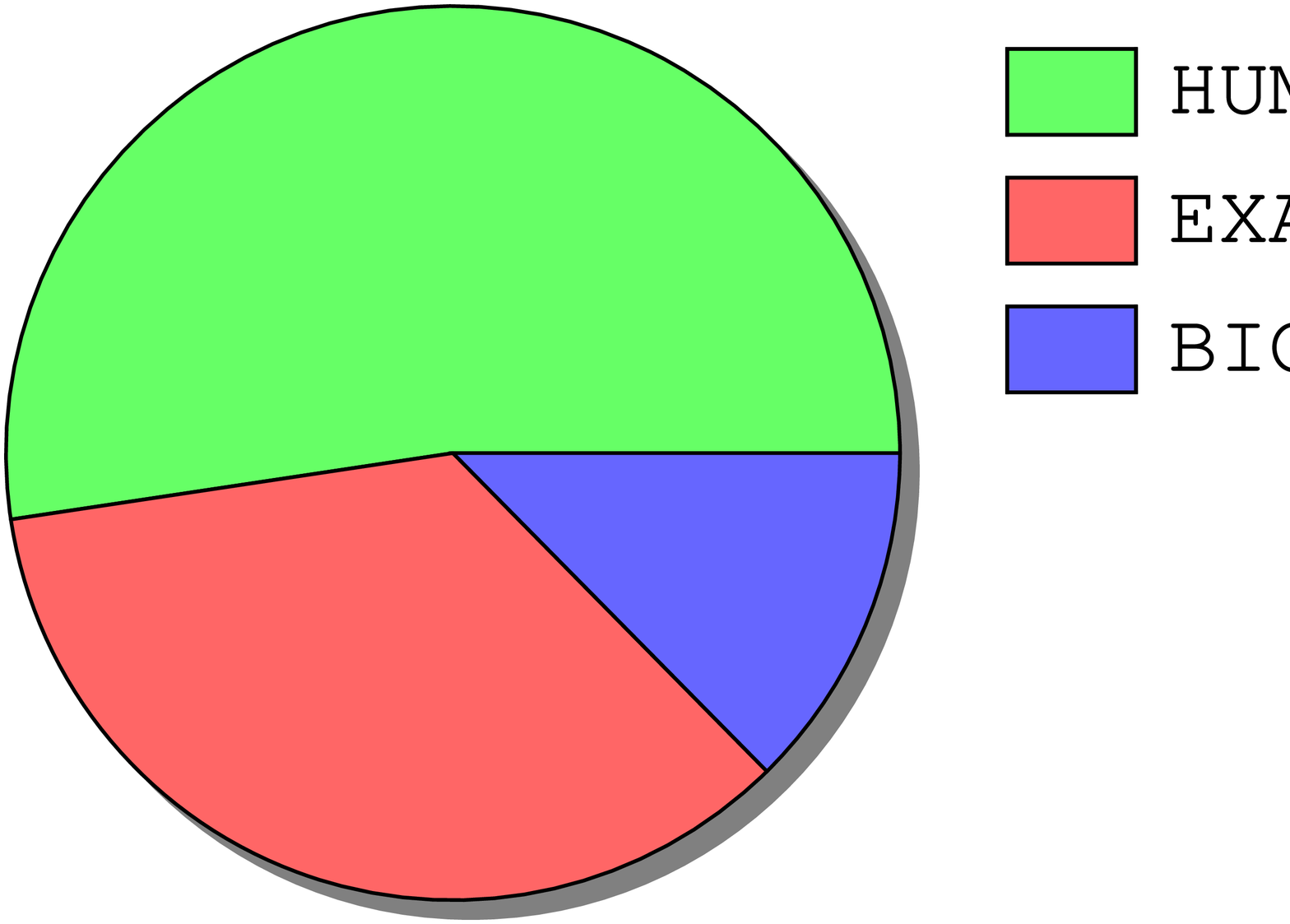}}
\put(290,290){K\includegraphics[scale=0.1]{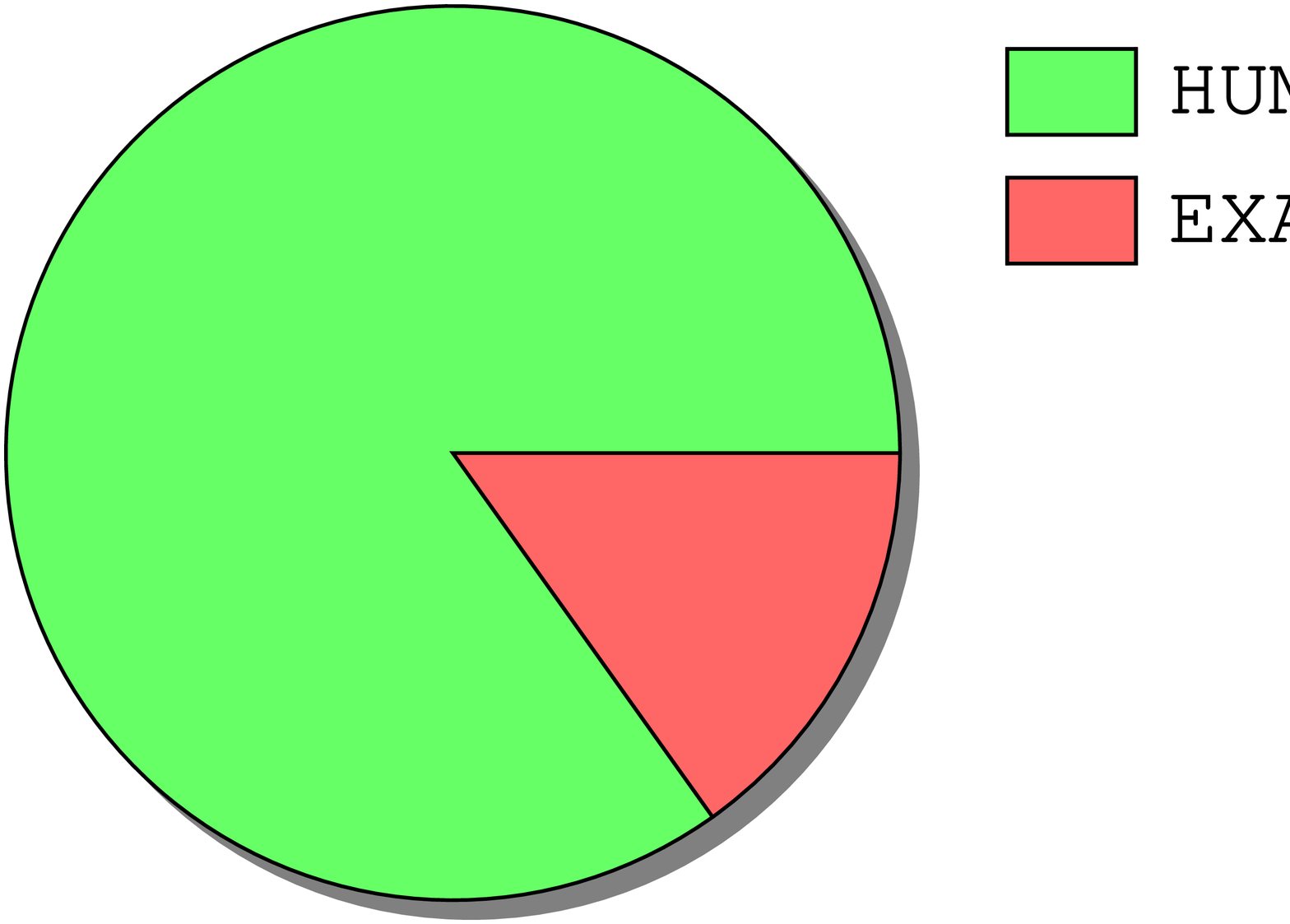}}
\put(290,215){L \includegraphics[scale=0.1]{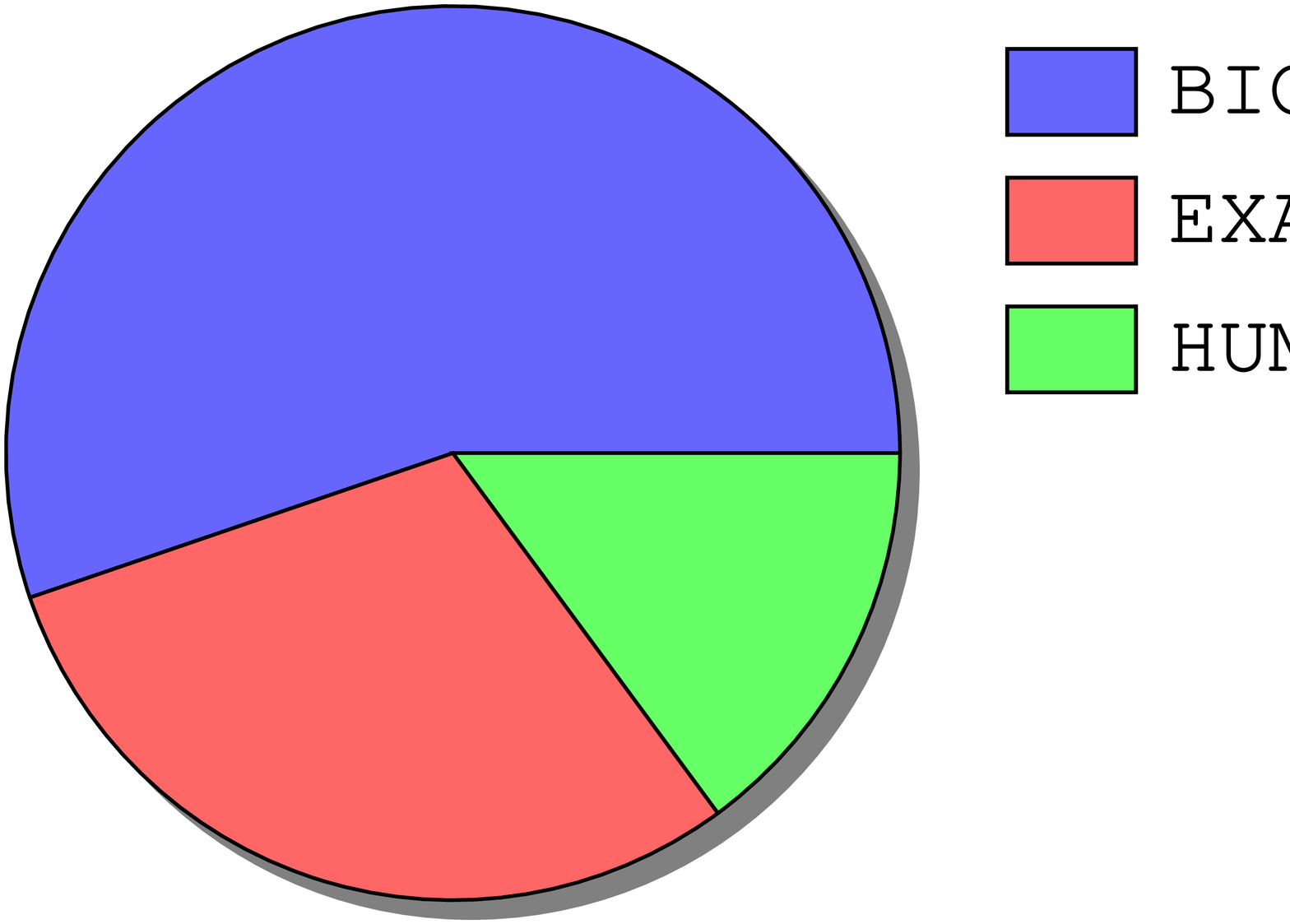}}
\put(290,135){M\includegraphics[scale=0.1]{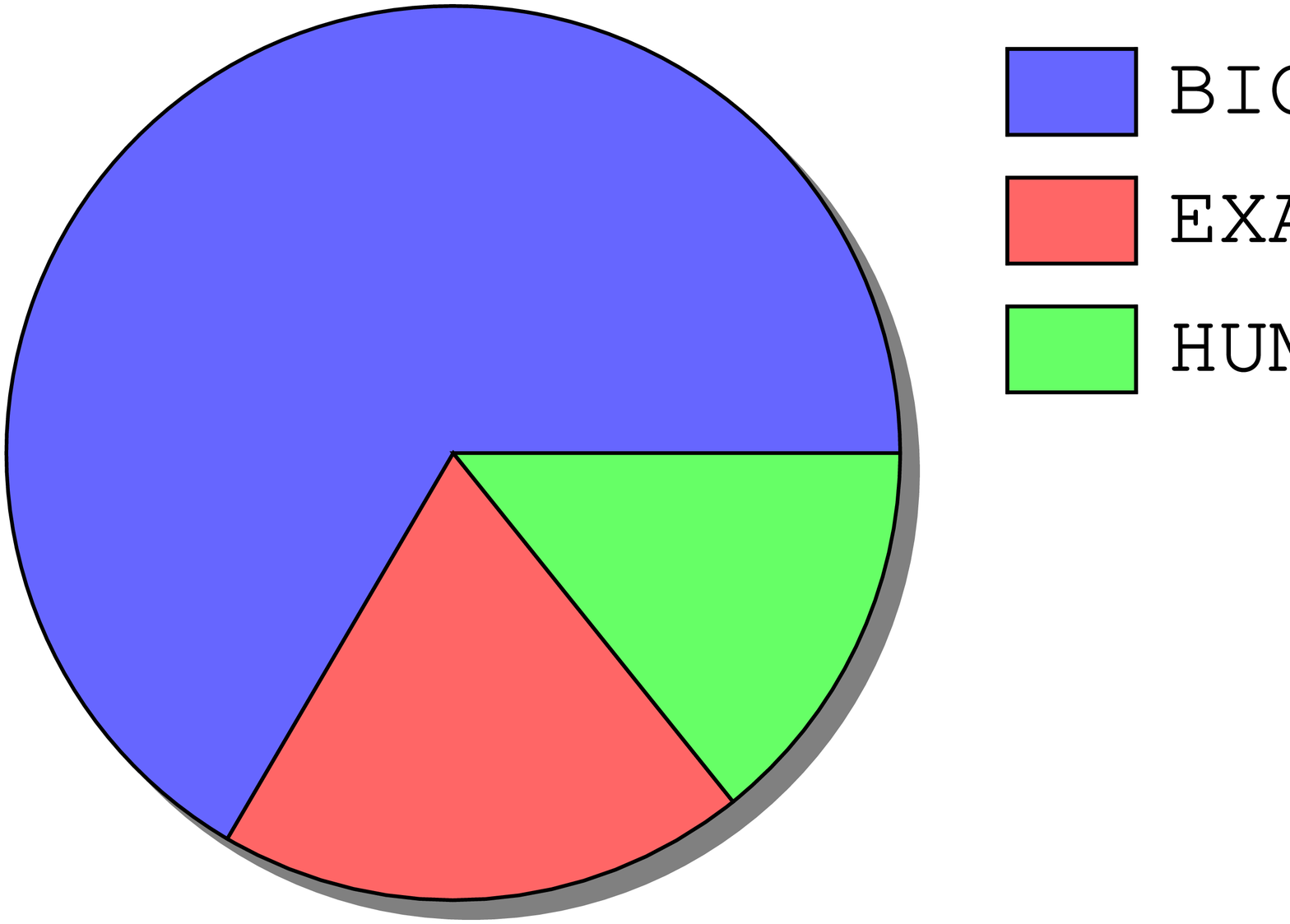}}
\put(290,060){N\includegraphics[scale=0.1]{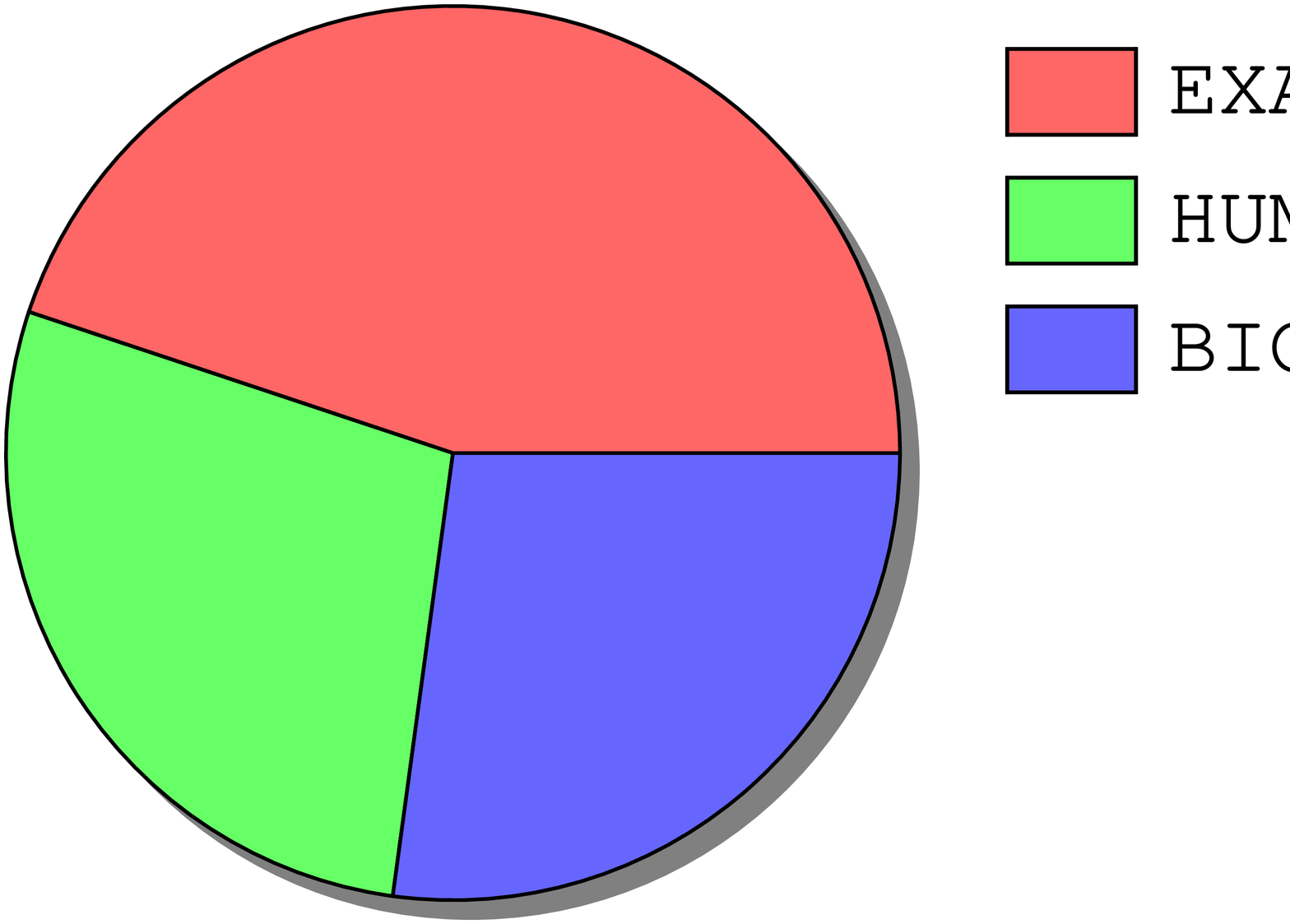}}
\put(290, -20){O\includegraphics[scale=0.1]{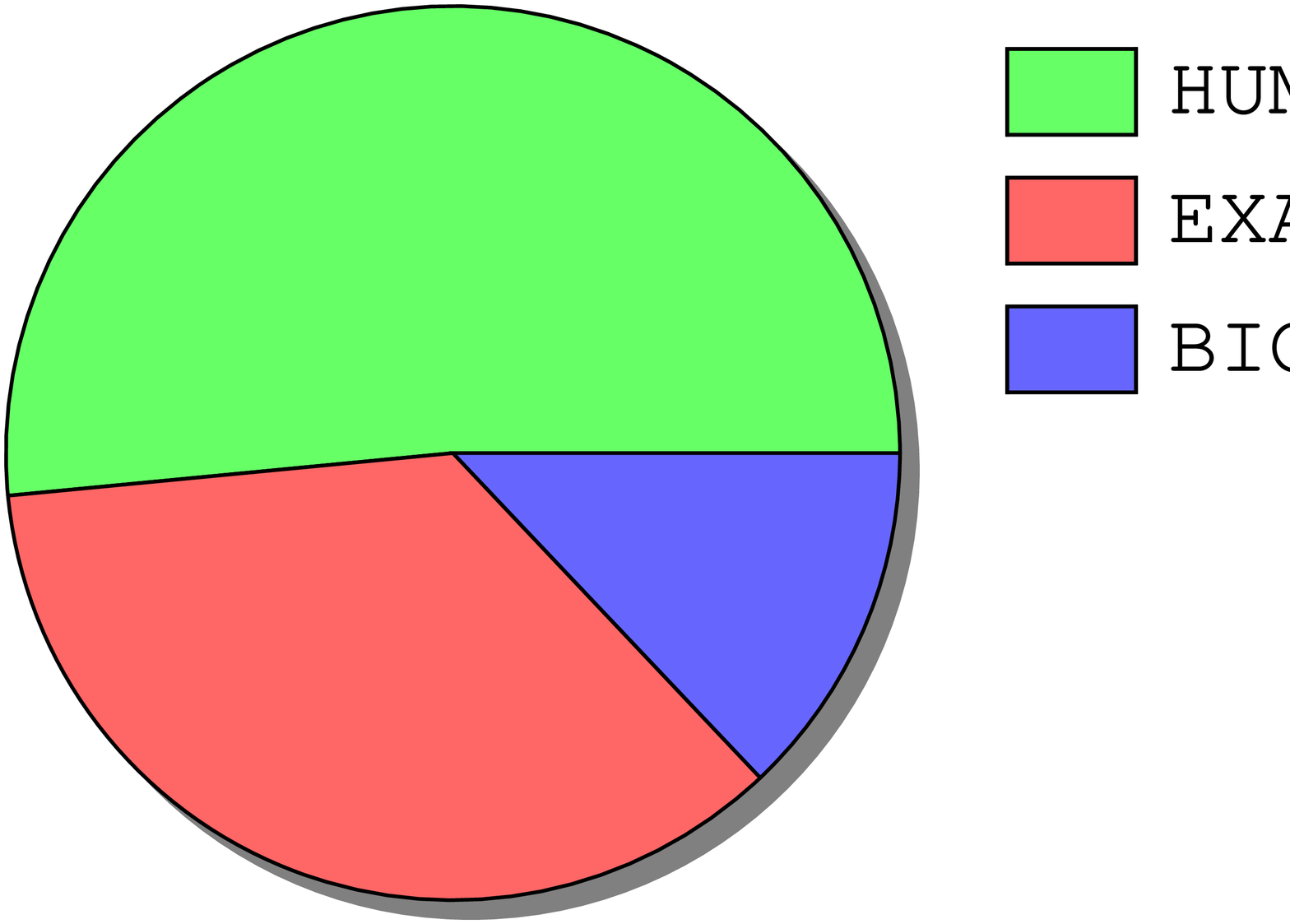}}
\end{picture}
\end{picture}
\end{center}
\caption{Pie charts with the percentage of nodes in each area of knowledge for each
cluster obtained by using the Clustering Coefficient.}
\label{fig:tree4}
\end{figure*}

Because every node in the collaborative network are labeled with the
author department, the percentage of nodes belonging to each
department is given by each obtained group. The results can be seen in
Figures ~\ref{fig:tree3} and ~\ref{fig:tree4} for concentric
clustering coefficient and convergence ratio, respectively. Only the
seven most representative departments of a cluster are shown, other
departments are merged into a single section of the pie chart. Note
that the most representative clusters for department segregation are
located in the second blanch points.

Figure~\ref{fig:tree3} shows the pie charts of the sets of nodes
obtained by hierarchical clustering and considering the clustering
coefficient. Each pie chart includes a respective legend showing the
most highly represented institutes and their relative percentage
considering each level.  For example, considering the branching level
1 (i.e. groups D to G), the sum of the percentages of each institute
for all pie charts should add to 1.  This is the case of the institute
H1 at the branching level 1, which presents a participation of
$17.1\%$ in pie chart D, $3.2\%$ in chart E and $75.3\%$ in chart G.

For most levels, the pie charts do not present marked homogeneity as
far as the nature of the institutes is concerned (i.e. human, exact
and biological areas). The departments are namely according its
knowledge area followed by a number, where H\# stands for human, E\#
for exact and B\# for biological areas.  However, we found the
remarkable result that a large part of the most representative
institutes in chart C were not only from the biological area, but also
located in a same city in the countryside of S\~ao Paulo State.
Therefore, these two factors seem to have implied a distinctive
pattern of collaborations.  In addition, by taking into account the
respective clustering coefficient measurements in
Figure~\ref{fig:tree1}, it becomes clear that these collaborations
involve a peak of clustering coefficient at the hierarchical level 1.
This provides further indication that the collaborations in pie chart
C indeed takes place at a more localized level, implied by the
geographical position of those institutes.  This more localized
collaboration pattern remains in the next branching, i.e. in pie chart
F.  The institutes in the sister chart, i.e. G, have a more widespread
collaboration pattern as indicated in the wider hierarchical
clustering coefficient signature in Figure~\ref{fig:tree1}.  As far as
the subdivision of the chart B is concerned, one of the sister charts
(i.e. D) contains a substantially higher overall percentage of
institute than the chart E.  Therefore, we will not consider the
latter and its respective subdivisions J and K in the following
discussion.  Charts D and E are characterized by the absence of the
secondary peak in the clustering coefficient signature (compared to
charts F and G).  The institutes in chart D are heterogeneous as far
as the scientific area is concerned, but are all located in the
capital (except for E5).

Figure~\ref{fig:tree4} shows the distribution of the scientific area
for each respective pie chart as in Figure~\ref{fig:tree3}.  Most of
the cases in the upper branching (i.e. B, D, E, I, J, K) have a
predominance of the human area.  Contrariwise, only the cases G, O
resulted with predominance of the human area in the lower branch.  The
biological area is over-represented in the branch C, F, L, M.  This is
in agreement with the above discussion.  The exact area is more
uniformly distributed among all cases, predominating only in cases H
and N.

\section{Concluding Remarks}

One of the interesting applications of complex networks has been for
the investigation of patterns of authorship and collaborations in
scientific production (e.g.~\cite{Newman:2001mz,Newman:2001zr,
Newman:2004gf,Newman:2001fr,Cardillo:2006ly, Newman:2007ca}).  The
current work has extended such investigations by considering
concentric measurements, which are capable of providing additional
information about the connectivity around each node
(e.g.\cite{Costa:2004fj, Costa:2006lr}), as well as the organization
of the results by using pattern recognition methods (more specifically
hierarchical clustering).  We applied such a methodology to real data
related to the scientific collaborations between authors from the
several institutes that compose the University of S\~ao Paulo (USP).
A number of interesting results have been obtained.  First, we found
that the geographical position of institutes tended to produce
well-defined groups, characterized by more localized clustering
coefficient.  This suggests that the collaborations are more intense
about the institutions in the same city.  We also found that the three
main scientific areas tended to be differently represented in the
obtained groups, with the exact sciences tending to appear more
uniformly in the majority of groups.  This indicates that this area is
characterized by a less uniform pattern of collaborations.  Future
works could address additional measurements and clustering methods, as
well as the study of co-authorships with external institutions.

\vspace{0.3cm}
{\bf Acknowledgment:} Luciano da F. Costa is grateful to FAPESP
(05/00587-5) and CNPq (301303/06-1) for financial support. Filipi
Nascimento Silva is grateful to CNPq (133256/2007-3) for sponsorship.

\bibliography{bibliotecarefs}

\end{document}